# RANDOM GEOMETRY
# &
# QUANTUM SPACETIME

FROM SCALE-INVARIANT RANDOM
GEOMETRIES AND ASYMPTOTIC SAFETY
TO RANDOM HYPERBOLIC SURFACES
AND JT GRAVITY

# Random Geometry and Quantum Spacetime

## From scale-invariant random geometries and asymptotic safety to random hyperbolic surfaces and JT gravity

Dissertation to obtain the degree of doctor
from Radboud University Nijmegen
on the authority of the Rector Magnificus prof. dr. J.M. Sanders,
according to the decision of the Doctorate Board
to be defended in public on

Thursday, January 25, 2024
at 14:30

by

**Blanca Alicia Castro Bermudez**

# Contents

# 1

# **Introduction**

The quest to understand the fundamental nature of our universe has led us, physicists, to explore two outstanding theories: Quantum Mechanics (QM), which describes the behavior of elementary particles, and General Relativity (GR), which explains how space and time couple to energy and matter. According to the theory of General Relativity, the force of gravity that governs the dynamics of our universe arises from the geometrical properties of spacetime itself. Within this framework, the gravitational interactions between masses and energy are consistently described by the curvature of a pseudo-Riemannian four-dimensional manifold, whose metric satisfies Einstein's field equations. Furthermore, it states that this and all laws of physics must remain unchanged under coordinate transformations and this is achieved by demanding diffeomorphism invariance in General Relativity. In the domain of quantum mechanics and quantum field theory (QFT), our foundational understanding of the microscopic world undergoes a radical change, revealing its non-deterministic nature and its probabilistic interpretation. For example, in quantum mechanics, elementary particles, like electrons, exhibit both particle-like and wave-like behaviors and they are described by a probabilistic wave function, which evolves via the Schrödinger equation. Observables, such as position and momentum, yield probabilistic outcomes upon measurement. QFT extends this non-determinism, characterizing fundamental particles as excitations of quantized fields filling spacetime, and their interactions are represented through Feynman diagrams, emphasizing a universe profoundly governed by uncertainty and probability.

These two very different theories have revolutionized our understanding of the universe and successfully explain a vast array of phenomena, from the outcomes of proton collisions in the Large Hadron Collider (LHC) to the gravitational wave signals coming from black holes merging at light-years from Earth. However, they reach their limits when faced with extreme conditions, such as the origin of the universe and the interior of black holes, where quantum effects and gravitational interactions intertwine, developing a new domain beyond the scope of either theory. In this context, one of the primary challenges in contemporary physics is the formulation of a unified quantum theory of gravity, suitably named Quantum Gravity (QG). Furthermore, since our world operates fundamentally on quantum principles, general relativity is widely regarded as representing merely the low-energy limit of a more fundamental quantum theory of gravity.

## 1.1. Quantum Gravity and quantum spacetime

When it comes to gravity, the traditional methods for quantizing classical physical systems, which have proven successful in other areas, encounter significant challenges. For example, given Einstein-Hilbert's action in $d$ dimensions

$$I_{EH} = \frac{1}{16\pi G} \int_{\mathcal{M}} \mathrm{d}^d x \ \sqrt{-g} \ (R - 2\Lambda),$$ (1.1)

where $G$ is the $d$-dimensional Newton's constant and $\Lambda$ is the cosmological constant. In order to obtain a Hamiltonian and quantize in a canonical fashion, a preferred spacetime direction is required to define conjugate momenta and Hamiltonian evolution, but this choice explicitly breaks diffeomorphism invariance. Moreover, the path integral formalism poses its own set of challenges. In order to compute the path integral

$$\mathcal{Z}(G, \Lambda) = \int \mathcal{D}g \ e^{\frac{i}{\hbar} I_{EH}[g]},$$ (1.2)

one needs to integrate over the space of all possible $d$-dimensional spacetime geometries weighted by an oscillating exponential of the Einstein-Hilbert action. This presents considerable mathematical complexities, like determining the appropriate measure $\mathcal{D}g$, the ambiguity surrounding the well-defined imposition of physical constraints like topology change and the relation between Lorentzian and Euclidean signatures [10]. Furthermore, the factor $e^{\frac{i}{\hbar} I_{EH}}$ introduces rapidly changing oscillations of the integrand, resulting in potential cancellations, interference patterns, and difficulties in accurate evaluation. In the context of QFT, the later issues are often addressed by performing a Wick rotation, which transforms time from the real to the imaginary axis and the metric from Lorentzian to Euclidean signature. However, this approach becomes notably intricate in the context of QG, where the fluctuations of spacetime geometry come into play.

Even if one assumes the well-posedness of QG in the framework of quantum field theory in four dimensions, nonrenormalizability emerges as a critical challenge when treating the metric field perturbatively, losing its predictive power at microscopic scales due to increasingly significant quantum fluctuations, particularly at and beyond the Planck scale. These difficult challenges highlight that the basic principles of both theories do not work together well, suggesting that we might need to rethink the core ideas of at least one of them.

### 1.1.1. Lorentzian vs. Euclidean

In the exploration of quantum gravity, a critical point is the choice between Euclidean and Lorentzian formulations, fundamentally influencing our comprehension of spacetime dynamics and its quantum behavior. The Lorentzian approach, while aligned with the causal unfolding of events in our physical reality, poses considerable mathematical challenges due to the intricate nature of Lorentzian metrics. Conversely, the Euclidean perspective offers a Riemannian manifold framework, which is more comprehensible, and transforms oscillatory path integral terms into damped exponentials, simplifying the mathematical treatment of these integrals.

The Euclidean perspective goes beyond mere mathematical convenience [78]. It establishes links with statistical mechanics and field theory techniques and tools such as partition functions and correlation functions. These facilitate the transfer of insights and methodologies to QG. When recasting the path integral formulation of quantum gravity in Euclidean spacetime, the transition to a probability measure becomes intuitive. This parallels the way statistical mechanics describes systems by assigning probabilities to various equilibrium configurations. Similarly, in Euclidean quantum gravity, the path integral generates a probability distribution over distinct geometries, assigning relative weights to different spacetime configurations.

Even though the map from Lorentzian to Euclidean metrics is not straightforward and there are many caveats [151], the probabilistic interpretation of Euclidean path integrals for QG finds resonance with the concept of random geometries. In this framework, a random geometry is selected from an en-semble with a well-defined probability Boltzmann distribution given by the Euclidean gravitational action. In this thesis, I make use of this powerful relation to study Euclidean QG using mathematics from random geometry and explain how these insights can help us understand our Lorentzian universe.

### 1.1.2. Dimensionality

Although our universe is observed as four-dimensional, investigating quantum gravity in lower dimensions offers valuable insights and tools for comprehending spacetime at a quantum level. In lower dimensions, the dynamics of the gravitational field are simpler since there are no propagating degrees of freedom in vacuum. Additionally, it provides the advantage of having more mathematical tools available, such as uniformization, an infinite-dimensional conformal symmetry group, and discretization and combinatorial techniques in two and three dimensions. These tools facilitate both analytical and numerical computations.

Furthermore, the mathematical terrain of four-dimensional random geometries remains largely beyond the reach of current mathematical techniques. This makes 2D and 3D quantum gravity natural topics to explore. In this thesis, I will concentrate on 2- and 3-dimensional Euclidean quantum gravity and explore its connection to the 4-dimensional Lorentzian case.

If one considers a compact, closed, connected, and orientable manifold $\mathcal{M}$ in 2-dimensions and performs a formal Wick rotation, the path integral (1.2) becomes a partition function and can be written as

$$\mathcal{Z}_{2d}^{eu}(G, \Lambda) = \int \mathcal{D}\tilde{g} \; e^{-\frac{1}{\hbar} I_{EH}[\tilde{g}]}, \tag{1.3}$$

where $\tilde{g}$ is a Euclidean metric and the corresponding Euclidean action is

$$I_{EH} = \frac{\Lambda}{8\pi G} \int_{\mathcal{M}} d^2x \sqrt{\tilde{g}} - \frac{1}{16\pi G} \int_{\mathcal{M}} d^2x \sqrt{\tilde{g}} R. \tag{1.4}$$

The first term in this expression can be identified with the volume $V_{\tilde{g}}$ of $\mathcal{M}$, and according to the Gauss-Bonnet Theorem, the second term is a topological invariant known as the *Euler characteristic*,

$$\int_{\mathcal{M}} d^2x \sqrt{\tilde{g}} R = 4\pi \chi(\mathcal{M}) = 4\pi(2 - 2g), \tag{1.5}$$

where $g$ is the genus of the manifold. Therefore, (1.3) can be written as

$$\mathcal{Z}_{2d}^{eu}(G, \Lambda) = \sum_{g \geq 0} e^{\frac{2-2g}{4G\hbar}} \mathcal{Z}(\Lambda/G),$$ (1.6)

where $\mathcal{Z}(\Lambda/G)$ is

$$\mathcal{Z}(\Lambda/G) = e^{-\frac{\Lambda}{8\pi\hbar G} V_{\bar{g}}}.$$ (1.7)

Thus, this theory depends solely on the genus and volume of the manifold and is organized as a genus expansion. For a fixed genus, the Euclidean path integral of 2-dimensional quantum gravity takes the familiar form of a Boltzmann distribution $\mathcal{Z} = e^{-\beta H}$. As a result, we can think of (1.3) as describing an ensemble of geometries, each assigned a Boltzmann weight. This distinctive nature aligns seamlessly with the concept of random geometries, forming the foundational focus of this thesis.

## 1.2. Quantum Gravity Approaches

In light of the challenges to reconcile the principles of quantum mechanics with the fundamental nature of gravity, diverse approaches have been developed, spanning from canonical approaches such as Loop Quantum Gravity [18], which introduces quantized discrete geometries at the minutest scales, and String Theory [86], which envisages particles as oscillating strings within higher-dimensional spacetime, to more conservative approaches such as Asymptotic Safety (AS) [144, 128, 134], discrete and lattice-based approaches, such as (Causal) Dynamical Triangulations [104, 105], Matrix Models [88, 56], Group Field Theory [125] and Causal Sets [138], or indirect frameworks such as the AdS/CFT correspondence [109, 148]. In this thesis, we will explore the connection of various of these approaches with random geometries.

### 1.2.1. Asymptotic Safety

Let us start by describing asymptotically safe gravity, this has been suggested as a potential scenario for restoring the predictive power of GR as a quantum field theory while maintaining the pseudo-Riemannian metric structure as an effective description of spacetime geometry at arbitrarily short length scales. This is achieved by studying the gravitational path integral using nonperturbative functional renormalization group (FRG) methods while maintaining a continuous description of spacetime itself.

The FRG is based on the Wilsonian concept of a step-by-step integration of high-energy fluctuations as we move from lower scales up to an ultraviolet (UV) cutoff. The scale-dependent effective action $\Gamma_k$ accounts for all quantum fluctuations at energies larger than $k$. A central equation in this context is the Wetterich equation [146],

$$k\partial_k\Gamma_k = \frac{1}{2}\text{Tr}\left(\frac{k\partial_k R_k}{R_k + \Gamma_k^{(2)}}\right), \tag{1.8}$$

which describes how $\Gamma_k$ evolves as the energy scale $k$ changes given an infrared (IR) regulator $R_k$. In the context of gravity, this flow equation offers a bridge between the fundamental microscopic constituents of spacetime and the effective macroscopic geometry described by the metric.

In practice, analyzing the renormalization group flow involves truncating the Wetterich equation to facilitate a more practical analysis. This process entails approximating the infinite tower of scale-dependent couplings in $\Gamma_k$ with finitely many dimensionless couplings $\bar{\lambda}_i(k)$ corresponding to the same number of operators,

$$\Gamma_k[g] = \sum_i^P \bar{\lambda}_i(k)\mathcal{O}_i[g]. \tag{1.9}$$

The asymptotic safety scenario proposes the existence of a nontrivial UV fixed point of the renormalization group flow with a finite number of attractive directions. Fixed points are determined by zeros of the beta functions $\beta_i = \partial_k\bar{\lambda}_i$ and attractive directions by negative eigenvalues of its hessian. In the context of asymptotically safe gravity, the FRG consistently suggests the potential existence of a fixed point within the space of couplings associated with higher-order curvature operators. This implies the existence of a quantum field theory of gravity that remains well-defined and predictive up to the Planck scale.

## 1.2.2. Dynamical Triangulations

Moving on to (Causal) Dynamical Triangulations ((C)DT), a strength of this approach lies in the presence of a coordinate-independent cutoff, facilitating the study of quantum gravity as a lattice field theory similar to the non-perturbative and numerical investigations conducted in Quantum Chromodynamics (QCD). Although unlike QCD where the lattice is fixed, in QG the lattice represents spacetime itself and one needs to sum over all possible lattices. This approach envisions spacetime as being composed of elementary building blocks (triangles in 2D, tetrahedra in 3D, etc.), effectively acting as a cutoff for gravitational degrees of freedom. In this framework, spacetime emerges through the arrangement of these building blocks, with their configurations governed by the DT action. This is obtained by evaluating the Regge action (the discrete version of the Einstein-Hilbert action) [127] on a triangulation with all sidelengths unity. The DT action is given by

$$I_{DT}[T] = k_d N_d(T) - k_{d-2} N_{d-2}(T), \tag{1.10}$$

in $d$-dimensions, where the coupling constants are $k_d \sim \frac{\Lambda}{G}$ and $k_{d-2} \sim \frac{1}{G}$, and $N_d$, $N_{d-2}$ are the number of $d$ and $(d-2)$-dimensional building blocks in the triangulation $T$, respectively. For example, in 2D, the 2-dimensional building blocks are triangles, and curvature is captured by deficit angles, $\epsilon_e$, concentrated on 0-dimensional objects: vertices (See Figure 1.2).

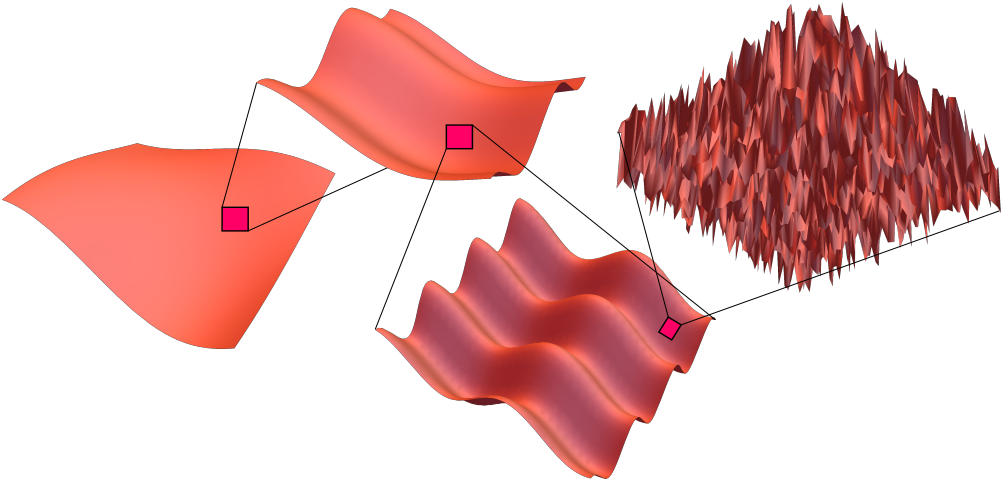

**Figure 1.1.:** Artistic impression of zooming in and out in quantum gravity.

The central concept in (Causal) Dynamical Triangulations is replacing the purely formal measure $\int \mathcal{D}g$ with a discrete measure over all possible fixed-length triangulations of manifolds. These can be thought of as the gluing of $d$-dimensional building blocks along their $(d-1)$-dimensional boundaries where each edge has a constant length. Thus, the DT measure accounts for all possible ways to glue these building blocks up to symmetry factors. Then, the partition function can be computed assuming either that the triangulations are Euclidean (DT) or that a well-defined Wick rotation exists (which holds true for CDT). In both scenarios,

$$\mathcal{Z}(k_{d-2}, k_d) = \sum_{T \in \mathcal{T}} \frac{1}{s(T)} e^{-I_{DT}[T]}, \qquad (1.11)$$

where $\mathcal{T}$ is the space of all possible fixed-length $d$-dimensional triangulations and $s(T)$ represents the order of the symmetry group of the triangulation, and its finiteness will be shown in the upcoming chapter. Given this condition, the partition function (1.11) defines an ensemble of triangulations where each inequivalent triangulation contributes with a Boltzmann weight determined by (1.10). This resonates with the concept of random discrete geometries, or more specifically random triangulations. Moreover, given this probability distribution on the space of triangulations, one can ask what the properties of a 'typical' element of the ensemble are. For DT, these are non-smooth fractal configurations that allow the proliferation of baby universes [10].

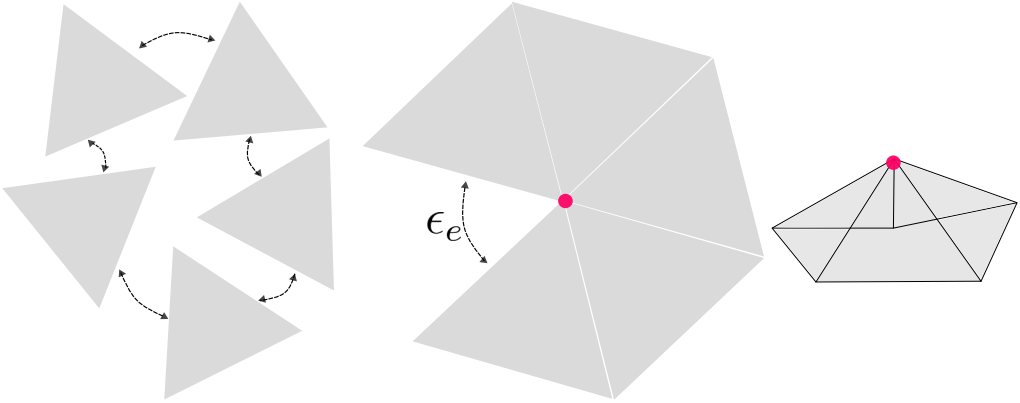

**Figure 1.2.:** Example of a 2-dimensional disk constructed by gluing triangles together. The curvature is captured by the deficit angle $\epsilon_e$.

In two dimensions, (1.11) offers a well-defined framework for Euclidean quantum gravity. Moreover, we can establish an explicit correspondence between 2-dimensional Euclidean quantum gravity path integral (1.6) and the DT partition function by fixing the genus for both, and expressing $\mathcal{Z}(\Lambda/G)$ in terms of the lattice parameter $a$ and the number of triangles in each triangulation, denoted as $n$. Subsequently, a shift to the canonical ensemble is made, where one works with a fixed number of building blocks. Then, we can write the DT canonical partition function

$$\mathcal{Z}_n = \sum_{T \in \mathcal{T}_n} \frac{1}{s(T)}, \tag{1.12}$$

where $\mathcal{T}_n$ is the set of all triangulations of $n$ triangles.

However, this discretization procedure does not completely eliminate the challenges linked to the continuum approach. To bridge the gap between the discrete world of Dynamical Triangulations and a quantum spacetime geometry that describes physics at arbitrary scales, we need to take a continuum limit by removing the lattice cutoff. If we consider the total volume of the triangulation, denoted as $V \sim N_d \ a^d$, where $a$ is the lattice spacing, the continuum limit at fixed $V$ is achieved by tuning the coupling constants $k_d$ and $k_{d-2}$ to a point in phase space where $a \rightarrow 0$ and the number of building blocks $N_d$ diverges. This resonates with the concepts of critical phenomena and universality in statistical physics. Here, we are looking for a critical point where by effectively refining the mesh, the discretized spacetime closely approximates a continuous geometry. Furthermore, universality arises because we expect that this continuous geometry is independent of the types of building blocks we use (triangles, squares, circles, etc.).

Despite the many results in (C)DT [3], the status of four-dimensional (C)DT remains purely as a model for various reasons. For example, if one does not impose a well-defined Wick rotation, the continuum limit is dominated by fractal geometries called branched polymers that do not resemble a manifold structure, and its fractal dimension is 2 [8]. Moreover, even if one is in the CDT case, it is still a challenge to study the coarse-grained properties of quantum spacetime and to construct physical observables, like curvature [106].

## 1.2.3. Matrix Models

Matrix models provide a method to approach the combinatorics of 2-dimensional Euclidean DT using discretizations of manifolds. Let us start by considering the partition function

$$\mathcal{Z}_N(q_4) = \int \mathrm{d}A \ e^{-N \, I_{MM}[A]}, \tag{1.13}$$

with an action of the form

$$I_{MM} = \frac{1}{2}\mathrm{Tr}(A^2) - \frac{q_4}{4}\mathrm{Tr}(A^4), \tag{1.14}$$

where $A$ is a Hermitian matrix of dimensions $N \times N$ and $dA$ is the $U(N)$ invariant Lebesgue measure on the space of $N \times N$ hermitian matrices. The Feynman diagram expansion of this model has a geometric interpretation involving tessellations of piece-wise linear manifolds, often referred to as fatgraphs or ribbon graphs (see Figure 1.3b). This can be seen by recalling that Feynman diagrams are the graphic representation of Wick's theorem, which states that expectation values of products of fields (in this case $A_{ab}$) are obtained by summing over all possible ways to contract their indices.

The quadratic term in (1.14) can be thought of as the inverse propagator, we denote it by

$$\langle A_{ab} A_{cd} \rangle = N^{-1} \delta_{ad} \delta_{bc}, \tag{1.15}$$

where the factor of $N$ comes from the overall normalization in (1.13). Its graphical representation is two double-line half edges, one for each of the two indices of the matrix $A$ (see Figure 1.3a). The 4-valent interaction term is the quartic term in (1.14),

$$\langle A_{ab} A_{cd} A_{ef} A_{gh} \rangle = N q_4 \; \delta_{ad} \delta_{cf} \delta_{eh} \delta_{gb}, \tag{1.16}$$

this can be drawn as four of these double lines joined at a vertex.

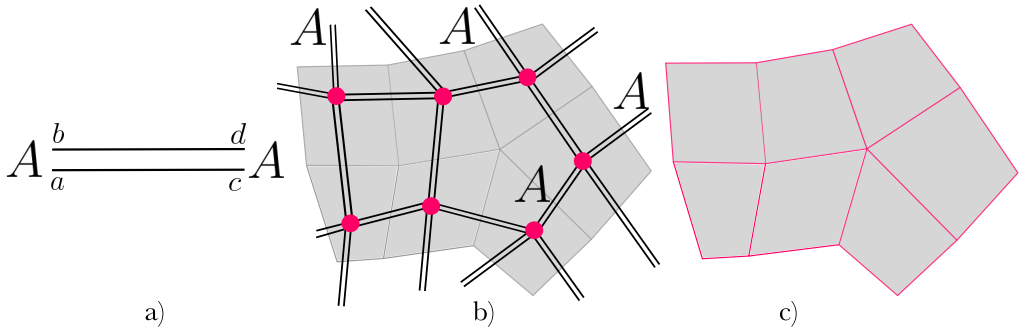

a)             b)             c)

**Figure 1.3.:** a) Propagator of the matrix model action (1.14). b) Fatgraph with vertices with 4 legs obtained from the Feynman diagram expansion of the matrix model action (black double-lines) and its dual quadrangulation (in grey). c) Dual quadrangulation.

Given this propagator and interaction, we can compute expectation values of products of $A$ using Wick's theorem, this accounts for all possible ways of gluing double edges. The resulting Feymann graphs possess a dual correspondence with quadrangulations of a surface, where each vertex of the fatgraph corresponds to a face of the quadrangulation (see Figure 1.3).

Using these observations we can deduce the form of the partition function (1.13). Each propagator or edge of the fatgraph corresponds to an edge in the quadrangulation, so according to (1.15), it contributes a factor of $1/N$. According to (1.16), each vertex of the fatgraph (face of the quadrangulation) contributes a factor of $N$ $q_4$ to the amplitudes. Each face of the fatgraph, which involves tracing over the indices of a matrix, corresponds to a vertex in the dual quadrangulation, and it contributes a factor of $N$ to the amplitudes.

Given this counting, the partition function of connected fatgraphs is

$$\ln \mathcal{Z}_N(q_4) = F_N(q_4) = \sum_{\mathcal{Q}} \frac{N^{2-2g}}{s(\mathcal{Q})} q_4^{v(\mathcal{Q})}. \tag{1.17}$$

Here, $2 - 2g = V(\mathcal{Q}) + F(\mathcal{Q}) - E(\mathcal{Q})$ represents the Euler characteristic of a connected (fat)graph $\mathcal{Q}$ with vertices with 4 legs, where $F$ is the number of faces, $E$ is the number of edges and $V$ is the number of vertices. The term $s(\mathcal{Q})$ stands for the symmetry factor associated with the number of automorphisms of the fatgraph. From this point, it becomes evident that (1.17) can be expressed as a genus expansion

$$F_N(q) = \sum_{g \geq 0} N^{2-2g} F_g(q). \tag{1.18}$$

This topological expansion organizes the terms $F_g$ (called the genus-$g$ partition function) according to the specific genus of surfaces they are associated with.

Until now, we have focused on an action that expands fatgraphs dual to quadrangulations instead of triangulations. However, universality suggests that the continuum limit of the model should be independent of the specific fundamental polygon used to construct a discrete manifold. As a result, our formulation closely resembles the statistical model outlined in (1.11), highlighting a correspondence

$$k_2 \sim -\ln(q_4), \quad k_0 \sim \ln(N). \tag{1.19}$$

In a more general framework, one can consider a matrix action with a potential of the form

$$I_{MM} = \frac{1}{2} \text{Tr}(A^2) + \sum_{k=2}^{L} \frac{q_{2k}}{2k} \text{Tr}(A^k). \tag{1.20}$$

In this case, the genus-$g$ partition function is

$$F_g(\mathbf{q}) = \sum_{\mathcal{G}} \frac{1}{s(\mathcal{G})} q_4^{v_4(\mathcal{G})} q_6^{v_6(\mathcal{G})} \dots q_{2L}^{v_{2L}(\mathcal{G})} \tag{1.21}$$

where $\mathbf{q} = (q_4, q_6, \ldots, q_{2L})$, $\mathcal{G}$ is a connected fatgraph of genus $g$ and $v_{2i}(\mathcal{G})$ is the number of vertices of the fatgraph with $2i$ legs or, equivalently, the number of faces of the dual tesselation. Remarkably, this description accommodates analytical outcomes in both pure gravity and gravity coupled with matter [100].

Given the correspondence between matrix models and DT, one expects that the continuum limit of DT corresponds to critical points in the large $N$ behavior of matrix models. Specifically, in order to take the continuum limit of matrix models, one needs to take $N \to \infty$ and tune $q_{2k}$ to a critical value, simultaneously. In this way, matrix models offer a valuable mathematical framework for exploring 2D quantum gravity. These models generate connected graphs that carry the interpretation of random geometries. This correspondence between matrix models and random geometries serves to bridge the divide between discrete and continuous portrayals of spacetime. Notably, the FRG approach finds application in the analysis of matrix models describing Dynamical Triangulations [64] and Causal Dynamical Triangulations [47], further enhancing the understanding of these concepts.

Matrix models also have a profound connection with string theory, serving as a bridge between the discrete world of lattice quantum gravity and the continuous spacetime emerging in string theory. For example, in [20] it was demonstrated that certain large $N$ limits of matrix models describe the dynamics of strings in a background spacetime. In this context, the Hermitian matrix $A$ in the matrix model corresponds to the world-sheet embedding of a string, and the Feynman diagrams of the matrix model become world-sheet surfaces over which strings propagate. This correspondence has opened up new avenues for understanding the underlying structures of both 2-dimensional quantum gravity and string theory. Notably, matrix models have been instrumental in exploring the non-perturbative aspects of string theory, such as D-brane dynamics [71] and black hole microstates [95], as we will see in the next section.

## 1.2.4. Holography: AdS/CFT and JT Gravity

Holography refers to the idea of encoding the physics of a theory with gravity in a boundary of spacetime. It is not a recent development and traces its origins back to the time of Bekenstain and Hawking's computation of black hole entropy [22, 87]. Moreover, the examination of supersymmetric black holes in String Theory offered valuable insights that ultimately laid the foundation for a QG approach that incorporates the idea of holography: the AdS/CFT correspondence [109]. This is a framework that posits a dictionary between certain quantum field theories in $d$ dimensions and gravitational theories in $(d + 1)$ dimensions.

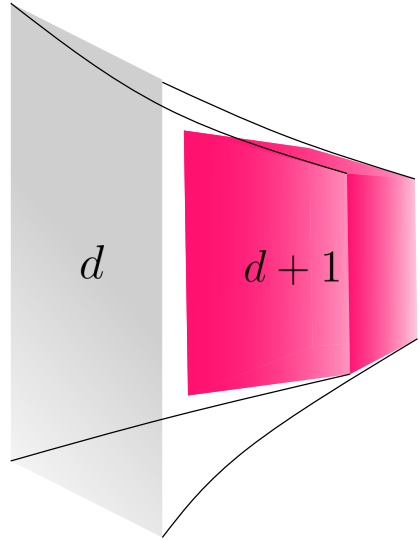

**Figure 1.4.:** Artistic impression of a holographic system. The physics of the $d + 1$ dimensional spacetime with gravity can be described by a system living in $d$-dimensions without gravity.

Although AdS/CFT does not provide a comprehensive theory of quantum gravity per se, it offers a valuable perspective on the interplay between gravity and quantum field theory. For example, in the well-known case of $AdS_5$ and $\mathcal{N} = 4$ super Yang-Mills in four dimensions [148]. In lower dimensions, a notable instance within this correspondence, known as "JT/SYK", has emerged as particularly instrumental for the investigation of black hole evaporation in two and four dimensions. SYK is a model of $N$ Majorana fermions with random couplings [133, 99, 107]. The gravitational side, represented by Jackiw-Teitelboim (JT) gravity [91, 139], serves as an arena to explore quantum gravity phenomena with implications that extend to higher-dimensional spacetimes.

JT gravity is a 2-dimensional theory of gravity defined, in Euclidean signature, by the action

$$I_{JT} = -\frac{S_0}{2\pi} \left( \frac{1}{2} \int_{\mathcal{M}} \sqrt{g} R + \int_{\partial\mathcal{M}} \sqrt{h} K \right) - \left( \frac{1}{2} \int_{\mathcal{M}} \sqrt{g} \phi(R + 2) + \int_{\partial\mathcal{M}} \sqrt{h} \phi(K - 1) \right),$$

(1.22)

where the first term is the Einstein-Hilbert action with a boundary term and $S_0 = \frac{1}{4G}$. Similarly to (1.4), this part of the action is topological in 2D and is proportional to the Euler characteristic of the surface $\mathcal{M}$, $\chi(\mathcal{M}) = 2 - 2g - n$, where $g$ is the genus of the surface and $n$ is the number of boundaries. The second term contains a scalar field $\phi$, called *dilaton*, which acts as a Lagrange multiplier that imposes a negative curvature $R = -2$ in the bulk, meaning the geometry is locally isometric to the hyperbolic plane, or Euclidean two-dimensional AdS space. This is equivalent to imposing $\mathcal{M}$ to be a hyperbolic surface of genus $g$ and a boundary $\partial\mathcal{M}$ composed by a disjoint union of $n$ boundaries.

Hence, the $n$-boundary JT gravity partition function reduces to a sum over hyperbolic geometries with boundaries of lengths $(\beta_1, \dots, \beta_n)$ which, after integrating out the dilaton in the bulk, can be represented as [132]

$$\mathcal{Z}(\beta_1, \dots, \beta_n) = \int \mathcal{D}g \, \mathcal{D}\phi \, e^{-I_{JT}} = \int \mathcal{D}g \, e^{S_0\chi(\mathcal{M})}\delta(R+2) \int \mathcal{D}\phi \, e^{\int_{\partial\mathcal{M}} \sqrt{h}\phi(K-1)}. \tag{1.23}$$

Note that, on the right-hand side, $\delta(R+2)$ represents a delta function at every point $x \in \mathcal{M}$ and the dilaton is evaluated at the boundary $\partial\mathcal{M}$.

The last expression suggests that the partition function admits a topological expansion and a split between bulk and boundary dynamics. This can be made precise by identifying the existence of a geodesic closed curve of length $b_i$ per boundary $\beta_i$ that splits a hyperbolic surface into two types of parts: bulk and *trumpets*. The bulk part of (1.23) is an integral over the moduli space of hyperbolic surfaces of genus $g$ and $n$ geodesic boundaries, and it is given [132] by the Weil-Petersson volume $V_{g,n}^{WP}(b_1, \dots, b_n)$ [120]. The trumpet is the part of the geometry that connects the bulk with the boundaries (See Figure 1.5).

The boundary term in (1.23) comes from $n$ asymptotic boundaries with extrinsic curvature $K$ [132]. The term "asymptotic boundary" accounts for the fact that we are not working in the full AdS spacetime, which is infinite in extent and does not have a traditional boundary that restricts or limits the spacetime itself, but in a very large region of it that has a nearly-conformal boundary located near infinity. This is why asymptotic boundaries are also known as *nearly AdS boundaries*. Their extrinsic curvature is given by the Schwarzian derivative of the near-boundary coordinates. From the holographic point of view, it effectively describes a one-dimensional quantum mechanical system given by the so-called Schwarzian action. The trumpet partition function is the partition function for a geometry with the Schwarzian action in the asymptotic

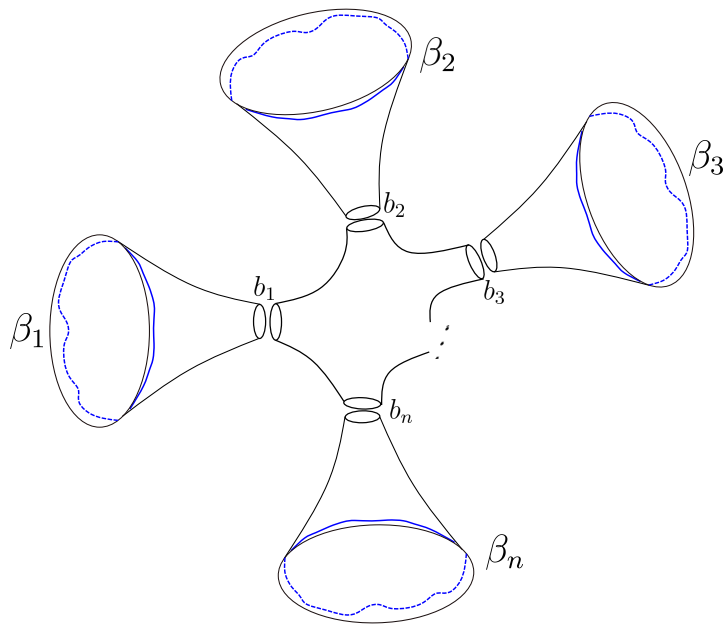

**Figure 1.5.:** Example of a genus-0 hyperbolic surface with $n$ asymptotic boundaries ($\beta_1, \ldots, \beta_n$) (shown in blue). The surface can be split into the bulk and trumpets by cutting along geodesic boundaries ($b_1, \ldots, b_n$).

boundary ending on a geodesic boundary, and it is given by [132]

$$\mathcal{Z}_{tr}(\beta, b) = \frac{e^{-b^2/4\beta}}{2\sqrt{\pi\beta}}. \tag{1.24}$$

Therefore, (1.23) can be written as

$$\mathcal{Z}(\beta_1, \dots, \beta_n) = \sum_{g=0}^{\infty} e^{S_0(2-2g-n)} \int_0^{\infty} b_1 db_1 \cdots \int_0^{\infty} b_n db_n \tag{1.25}$$

$$\cdot\, V_{g,n}(b_1, \dots, b_n) \mathcal{Z}_{tr}(\beta_1, b_1) \dots \mathcal{Z}_{tr}(\beta_n, b_n). \tag{1.26}$$

Let us focus on the case of one asymptotic boundary for now. From (1.26), we can see that

$$\mathcal{Z}(\beta) = \frac{1}{e^{S_0}} \sum_{g=0}^{\infty} e^{S_0(2-2g)} \mathcal{Z}_g(\beta), \tag{1.27}$$

where

$$\mathcal{Z}_g(\beta) = \int_0^{\infty} b\, db\; V_{g,1}(b) \mathcal{Z}_{tr}(\beta, b). \tag{1.28}$$

This resembles the matrix model genus expansion (1.18) with $N = e^{S_0}$[1] which is usually identified with the inverse of the Planck's constant $\hbar = e^{-S_0}$.

The key observation that AdS/CFT brings is this partition function captures the statistical behavior of the holographic system at a given Euclidean temperature $T$ via the relation $\beta \sim \frac{1}{T}$ and should be expressed in terms of the holographic theory Hamiltonian, $H$,

$$\mathcal{Z}(\beta) = \text{Tr}(e^{-\beta H}). \tag{1.29}$$

Recent advancements [132, 137] have unveiled a precise connection between JT gravity and Matrix Models. Within this context, the partition function of JT gravity can be associated with a matrix integral involving a random Hamiltonian $H$ and a potential term $V(H)$ in the following manner

$$\langle \mathcal{Z}(\beta) \rangle = \frac{1}{Z} \int dH e^{-N \text{Tr}V(H)} \text{Tr}(e^{-\beta H}), \tag{1.30}$$

where $Z$ is a normalization factor. This connection between JT gravity and Matrix Models is significant because it provides another computational framework to study quantum gravity using matrix integrals. The matrix integral formulation allows us to explore the statistical behavior of JT gravity. Moreover, this connection highlights a potential link between JT gravity and discrete surfaces and lattice approaches to QG.

---

[1]The extra factor of $\frac{1}{e^{S_0/\hbar}}$ comes from the marked geodesic boundary that attaches the trumpet to the bulk and it also appears in MM when there is a marked face.

## 1.3. Outline

As we have discussed thus far, a central challenge in the development of a theory of quantum gravity is to find an alternative to the classical smooth metric concept of spacetime geometry within the quantum regime. Within the extensive realm of quantum gravity approaches, lattice field approaches offer valuable insights, especially in scenarios like Asymptotic Safety, thanks to the presence of a coordinate independent cutoff. Dynamical Triangulations and matrix models introduce the notion of spacetime as a composition of elementary building blocks. The Euclidean gravitational path integral assumes a probabilistic interpretation, and within this framework, random geometries emerge as a fundamental cornerstone. Within this landscape, the connection between JT gravity and matrix models adds depth to this exploration, providing a pathway to explore quantum gravitational phenomena through the lens of random geometries.

This thesis is driven by a central question: "What can we learn from random geometries about the structure of quantum spacetime?" In Chapter 2, we will provide a partial review of the mathematical foundation of this thesis, random geometry. In Chapter 3, we use a construction coming from random geometry called *Mating of Trees* to build scale-invariant random geometries that have the potential to implement the UV fixed point predicted by Asymptotic Safety in two and three dimensions. In Chapter 4 we explore how our understanding of random critical maps yields the discovery of a new family of deformations of JT gravity. Furthermore, the connection between JT gravity and matrix models leads us to delve deeper into the link between discrete geometry and hyperbolic surfaces, building upon insights from metric maps and irreducible metric maps in Chapter 5.

# RANDOM GEOMETRY

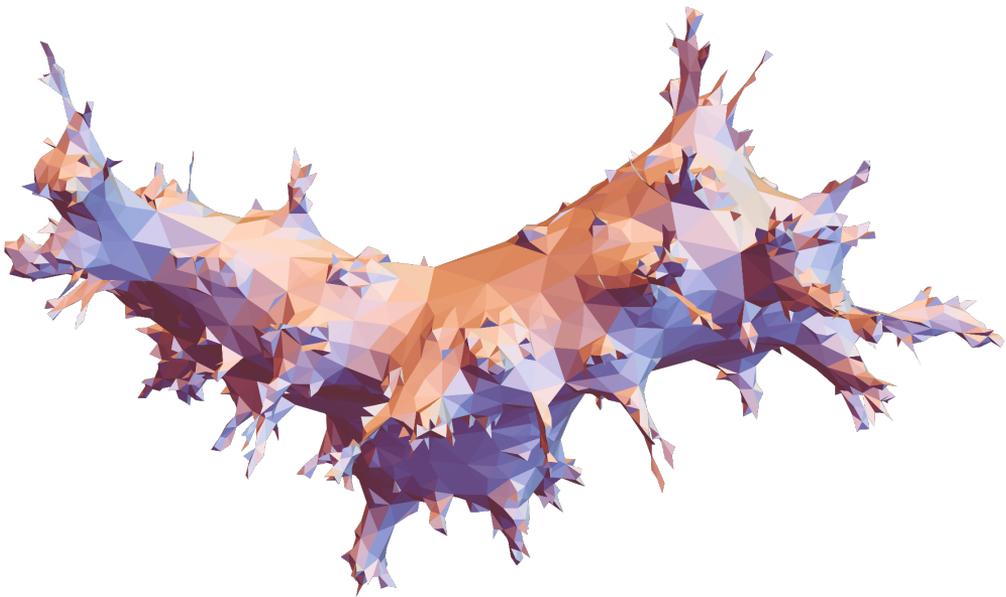

**Figure 2.1.:** Example of a random triangulation of the sphere with 3,000 vertices.

As we have seen in Chapter 1, the challenge of quantizing gravity in 2-dimensional Euclidean quantum gravity aligns with the task of enumerating graphs or discrete representations of manifolds. Random geometry serves as the mathematical foundation of this thesis and the results presented in different approaches to quantum gravity. In this chapter, we review essential terms for describing random geometries. For further details, we refer to [66] and [37] for an updated review.

A ***multigraph*** is a finite set of vertices, edges, and an incidence map $(V, E, \mathbb{I})$,

where *V* and *E* are sets of vertices and edges, respectively, and $\mathbb{I} : E \longrightarrow V \times V$ associates two vertices with an edge. A graph is *connected* if for any two vertices of the graph, there is a path connecting them (see Figure 2.2, Left). The *degree* of a vertex is the number of edges incident to it. The *graph distance* between two vertices in a graph is the shortest number of edges between them.

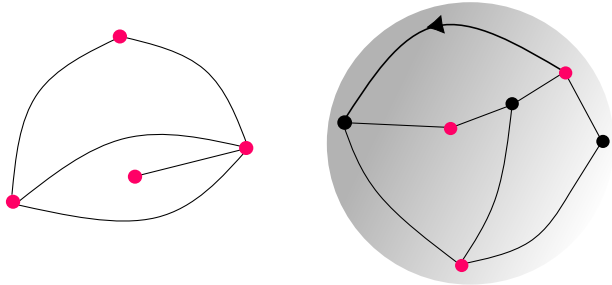

**Figure 2.2.: Left**: Example of a connected graph with four vertices and five edges. **Right**: Example of a rooted pointed planar bipartite quadrangulation, i.e. all faces of length 4. The root is marked with an arrow, the marked vertex is the end of the root.

A ***genus-g map*** (m) is an equivalence class of connected *multigraphs* embedded in a surface of genus *g*, such that its edges do not cross other than at their endpoints, modulo orientation-preserving homeomorphisms of the surface and such that all regions in the surface bounded by the edges of the map are topological disks. Each of these disks is called a ***face*** (*f* ∈ *F*). The *length of a face* is the sum of the lengths of its edges, where each edge has length 1. A map's *girth* is the length of the shortest closed path that starts and ends at the same vertex without retracing any edge or vertex in between. To see that a genus-g map describes a discrete surface, it is sometimes useful to think of the faces as identical unit polygons equipped with the Euclidean metric and the incidence relations of the map as prescriptions on how to glue these polygons along their sides in order to obtain a piecewise flat metric on the sphere for genus 0 and torus for genus 1, etc. The *Euler-characteristic* of a genus-g map is $2 - 2g = |V(\mathfrak{m})| + |F(\mathfrak{m})| - |E(\mathfrak{m})|$.

In particular, at some points, it will be useful to talk about genus-*g* maps with specific characteristics. A genus-*g* map with one oriented edge (called the root), is called a rooted map. A genus-*g* map with a distinguished vertex is called a pointed map. Considering genus-*g* maps with an even number of edges per face will also be useful. In the case $g = 0$, an even map is also bipartite, i.e. its vertices can be colored black and pink such that the endpoints of each vertex have a different color. Figure 2.2 shows a rooted pointed planar bipartite map.

## 2.1. Random trees

Let us start by showing how to count the simplest planar maps: trees. A ***tree*** is a planar map with one face. In particular, a rooted plane tree is a rooted planar map with one face. In broad terms, a ***random tree*** is a tree that is generated randomly according to some specified probability distribution or process. For example, one might generate a random tree by starting with a single vertex and adding new vertices one by one, connecting each new vertex to an existing vertex with a certain probability. The resulting tree's structure and characteristics will depend on the probabilities and rules set for this process.

In this case, we will focus on a specific type of tree called *rooted binary tree*. This is a tree with a marked vertex where each vertex in the tree can either be an *internal* vertex or a *leaf*. Each internal vertex branches into two vertices which can be internal, leaves or a combination of them (Figure 2.3).

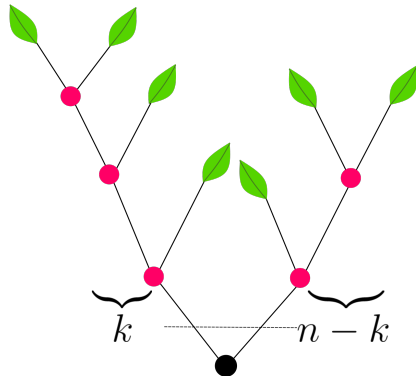

**Figure 2.3.:** Rooted binary tree with 6 internal vertices (pink) and 7 leaves (green). The root is shown in black and removing it erases the edges crossed by the horizontal dashed line. This results in left and right subtrees with $k = 3$ and $n - k = 2$ internal vertices, respectively.

To enumerate rooted binary trees with $n$ internal nodes, denoted as $b_n$, it is essential to recognize that $b_0$ equals 1. Then, we decompose our tree as follows: start with a rooted binary tree with $n+1$ internal nodes, subsequently eliminate the root, and consider the resultant left and right subtrees (as depicted in Figure 2.3). These resultant trees will have $k$ and $n-k$ internal nodes, which implies that

$$b_{n+1} = \sum_{k=0}^{n} b_k \ b_{n-k}. \tag{2.1}$$

Next, these values are organized into a generating function denoted as $B(x) = \sum_{n \geq 0} b_n x^n$. Utilizing the recursion relation (2.1), we can effectively determine the coefficients of $B^2(x)$, which results in

$$B(x)^2 = x \ B(x) + 1. \tag{2.2}$$

Solving this quadratic equation and using the fact that $B(0) = b_0 = 1$, we obtain

$$B(x) = \frac{1 - \sqrt{1 - 4x}}{2x}. \tag{2.3}$$

Then, we can extract the coefficient of $x^n$ of this power series. This results in

$$b_n = -\frac{1}{2} \binom{1/2}{n+1} (-4^{n+1}) = \frac{1}{n+1} \binom{2n}{n} = \text{Cat}(n), \tag{2.4}$$

where $\text{Cat}(n)$ are the Catalan numbers. Generating functions are part of the field of analytic combinatorics and they are powerful tools for the enumeration of maps [72]. In the following section, we will elaborate further on this point.

The last point we want to make is that once we count the number of rooted binary trees, we can define the partition function of the rooted binary trees of $n$ internal vertices as $\mathcal{Z}_n^{\text{bin-tree}} = \text{Cat}(n)$. In particular, we can explore its asymptotic behavior as the number of vertices becomes very large using known characteristics of the Catalan numbers. In this case, we have

$$\mathcal{Z}_n^{\text{bin-tree}} \overset{n \to \infty}{\sim} \frac{1}{\sqrt{\pi}} \ 4^n n^{-3/2}. \tag{2.5}$$

We define a uniform random rooted binary tree with $n$ vertices as a tree sampled uniformly, i.e. with probability $1/\mathcal{Z}_n^{\text{bin-tree}}$. The *height* of a vertex in a tree is the distance in the tree to its root. Since at each height $n$ we can have up to $n^2$ vertices, the random height of a random rooted binary tree grows as $\sqrt{n}$. Furthermore, in section 2.5, we will see that this is a general feature of many types of trees.

## 2.2. Random maps

Let $\mathcal{M}_{g,n}(L_1, \ldots, L_n)$[1] be the set of genus-$g$ maps with $n$ faces of half-lengths[2] $\{L_i \in \mathbb{N}\}_{1 \le i \le n}$. The cardinality of this set is the total number of genus-$g$ maps including a symmetry factor $\text{Aut}(\mathfrak{m})$ counting orientation-preserving automorphisms that preserve the face labels,

$$\|\mathcal{M}_{g,n}(L_1, \ldots, L_n)\| = \sum_{\mathfrak{m} \in \mathcal{M}_{g,n}(L_i)} \frac{1}{\text{Aut}(\mathfrak{m})}. \tag{2.6}$$

### 2.2.1. Generating function

For fixed values of $g$ and $n$, these are nothing more than positive real numbers and they can be encoded in a generating function

$$F_g[\mu] = \sum_{k=1}^{\infty} \frac{1}{k!} \sum_{L_1=1}^{\infty} \mu_{L_1} \cdots \sum_{L_k=1}^{\infty} \mu_{L_k} \mu_0^{2-2g-k+\sum_i L_i} \|\mathcal{M}_{g,k}(L_1, \ldots, L_k)\|, \tag{2.7}$$

where $\mu_{L_i} \ge 0$ are called weights[3] and the powers of $\mu_0$ account for the number of vertices with weight $\mu_0$[4] in a map of genus-$g$ with $k$ faces of half-length $\sum_i L_i$ according to Euler's characteristic.

The generating function is a formal power series in the infinite number of variables $\mu = (\mu_0, \mu_1, \ldots)$. Therefore, we can write

$$\|\mathcal{M}_{g,n}(L_1, \ldots, L_n)\| = \mu_0^{-2+2g+n-\sum_i L_i} \left. \frac{\partial^n F_g[\mu]}{\partial \mu_{L_1} \ldots \partial \mu_{L_n}} \right|_{\mu=0}. \tag{2.8}$$

It can be observed that the weights $\mu_L$ control the likelihood of having faces of half-length $L$ in the random map, e.g. if $(\mu_0 = 1, \mu_1 = 0, \mu_2 \ne 0, \mu_3 = 0, 0, 0, \ldots)$, then only faces of length $2 \cdot 2$ are allowed and the genus-$g$ map is a *quadrangulation* of a genus-$g$ surface.

---

[1]The notation and conventions are chosen to align with other families of geometries presented later.

[2]We are working with maps with faces of an even length $2L_i$.

[3]Non-trivial weights can arise naturally when integrating out decorations/matter systems living on the map.

[4]For practical reasons, we will usually set $\mu_0 = 1$, which can be done without loss of generality (e.g. see [37]).

## 2.2.2. String equation

Coming back to the generating function $F_g[\mu]$, in order to analyze it, it is convenient to write it in terms of another power series, the *string equation*

$$\tilde{x}(u) = u - \sum_{k=0}^{\infty} \mu_k \binom{2k-1}{k} u^k. \tag{2.9}$$

In particular, the genus-0 generating function is proven to be [141, 66][5]

$$F_0[\mu] = \frac{1}{2} \int_0^{E_\mu} \mathrm{d}u \, \frac{1}{u} \left( \tilde{x}(u)^2 - (u - \mu_0)^2 \mathbb{I}_{u < \mu_0} \right), \tag{2.10}$$

where $E_\mu > 0$ is the smallest solution to the string equation $\tilde{x}(E_\mu) = 0$. Thus, $E_\mu$ is an implicit function of the weight $\mu$. The weight is said to be *admissible* if such a solution exists. Furthermore, other useful quantities we are interested in are the *moment* $M_0$, which is defined as

$$M_0 = \left( \frac{\partial E_\mu}{\partial \mu_0} \right)^{-1} = \frac{\partial \tilde{x}}{\partial u}(E_\mu), \tag{2.11}$$

and the *disk function*, which corresponds to the generating function of genus-0 maps with one marked face and an arbitrary number of unmarked faces. Operationally, this is

$$W_\mu(L) = \frac{\partial F_0[\mu]}{\partial \mu_L} = \int_0^{E_\mu} \mathrm{d}u \, \frac{1}{u} \tilde{x}(u) \binom{2L-1}{L} u^L, \tag{2.12}$$

where, by convention, the marked face is the one to the right of the root (see Figure 2.4). It can also be noted that $\frac{\partial^2 F_0[\mu]}{\partial \mu_0 \partial \mu_1} = E_\mu$ which gives a geometrical meaning to $E_\mu$, it is the generating function of rooted pointed planar bipartite maps [6].

By this point, all our expressions are formal power series, but in order to talk about random geometry and random maps, we need a well-defined probability measure. This can be also understood as a well-defined canonical way of sampling a map. In particular, we would like to define a probability measure[7] on the space of planar maps with $k$ faces with weight $\mu$. This can be done using the generating function in the following way

$$p\left(L_1, ..., L_k\right) = \frac{1}{k!} \frac{\prod_{i=1}^k \mu_{L_i}}{F_0[\mu]} \frac{\mu_0^{2-2g-k+\sum_i L_i}}{\text{Aut}(\mathfrak{m})}, \tag{2.13}$$

where we have set $\mu_0 = 1$. Note that this is only well defined when $\mu$ is such that $F_0[\mu] < \infty$. This is satisfied iff the weight $\mu$ is admissible, i.e. if the string equation has a solution $E_\mu > 0$ [111, 24, 49]. In this way, given an admissible weight, a ***Boltzmann planar map*** is defined as a random planar map with $k$-faces of random lengths sampled with probability (2.13). In general, for $g \geq 0$, $F_g[\mu]$ is called the *partition function of Boltzmann genus-$g$ maps* given its role in 2D Euclidean quantum gravity.

---

[5]For $g > 0$, check [66].

[6]More precisely, half of it, since the orientation of the root is not specified, so $2E_\mu$ gives the actual generating function

[7]On a discrete space, it is given by a probability mass function.

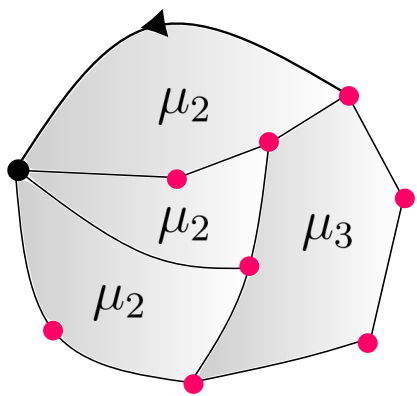

**Figure 2.4.:** A disk of half-perimeter $L = 3$, four unmarked faces, three of them of half-length 2 and one of half-length 3. The distinguished face is shown to the right of the root.

## 2.3. Critical maps

Now that we introduced Boltzmann planar maps, we can study more of their characteristics. In particular, the main point we want to make is that the string equation is the central building block for constructing generating functions and enumerating maps.

### 2.3.1. Criticality

Consider the solution to the string equation $\tilde{x}(E_\mu)/E_\mu = 0$ and its first derivative $\tilde{x}'(E_\mu)$,

$$1 - E_\mu^{-1} = \sum_{k=1}^{\infty} \mu_k \, \frac{1}{2} \binom{2k}{k} E_\mu^{k-1}, \qquad \tilde{x}'(E_\mu) = 1 - \sum_{k=1}^{\infty} \mu_k \, \frac{k}{2} \binom{2k}{k} E_\mu^{k-1}, \qquad (2.14)$$

where we have set $\mu_0 = 1$. Given that $\mu_k, E_\mu > 0$, we can conclude that $0 \leq \tilde{x}'(E_\mu) < 1$. In the mathematics literature [111], if an admissible weight is such that $0 < \tilde{x}'(E_\mu)$, it is called *subcritical*, and if $\tilde{x}'(E_\mu) = 0$, the weight is called **critical**. Furthermore, if a weight is critical, it can be *generic critical* if the string equation behaves as $\tilde{x}(u) \sim (E_\mu - u)^2$ as $u \nearrow E_\mu$ and **non-generic critical** if the string equation behaves as $\tilde{x}(u) \sim (E_\mu - u)^{\alpha - 1/2}$ with $3/2 < \alpha < 5/2$ as $u \nearrow E_\mu$ [111]. In this last case, $E_\mu > 0$ is the radius of convergence of the formal power series $\tilde{x}(u)$.

| Weight $\mu$ | String equation |
|:---:|:---:|
| Subcritical | $\tilde{x}'(E_\mu) > 0$ |
| Generic critical | $\tilde{x}'(E_\mu) = 0, \quad \tilde{x}(u) \sim (E_\mu - u)^2$ |
| Non-generic critical | $\tilde{x}'(E_\mu) = 0, \quad \tilde{x}(u) \sim (E_\mu - u)^{\alpha - 1/2}$ |

**Table 2.1.:** Types of admissible weights.

In the non-generic critical case, it is proven that the weight must behave as $\mu_k \sim h_1^k k^{-\alpha}$ for some $h_1 > 0$ and $k \to \infty$ [75]. The geometrical meaning of this slow decay of critical weights is the proliferation of large faces of half-length $k \gg 1$ (see Figure 2.5, Right).

### 2.3.2. Loop decoration and criticality

To further understand this, we will follow the reasoning used to draw a relation between these maps and the $O(n)$ loop model in [30, 34]. In that case, one considers a planar quadrangulation, $\mathfrak{m}$, dressed with a loop configuration, $\Gamma$, which is a collection of disjoint simple loops drawn on the dual of the quadrangulation. In particular, we will consider the *rigid $O(n)$ loop model*, which means that the loops are only allowed to cross a face of the quadrangulation in opposite edges (see Figure 2.5, Left). The faces crossed by a loop are assigned a weight $h_1$ and the faces that are not crossed are assigned a weight $\mu_2$.

Each of these loops has an arbitrary half-length $k$ and the parameter $n$ controls the weight per loop $n\ h_1^k$. Then, this model is well-defined as a random loop-decorated quadrangulation if its generating function is finite.

The external gasket $E(\mathfrak{m}, \Gamma)$ of a loop-decorated map is obtained by removing from the quadrangulation all edges crossed by a loop or situated within the interior of the loops in $\Gamma$. The faces of the exterior gasket obtained by this procedure are called *holes* (see Figure 2.5, Right). The key result is that the gasket of a critical $O(n)$ loop-decorated quadrangulation is distributed as a non-generic critical Boltzmann planar map with $n = 2\sin\left((\pi(\alpha - 1/2))\right)$ for $\alpha \in (3/2, 5/2)$ [31]. In this construction, the holes are assigned weights $\hat{\mu}_k = \mu_2\delta_{k,2} + n\ h_1^k W_{\hat{\mu}}(k)$, where $W_{\hat{\mu}}(k)$ satisfies the fixed point equation $W_{\mu_2 h_1}^{(n)}(k) = W_{\hat{\mu}}(k)$. $W_{\mu_2 h_1}^{(n)}(k)$ is the disk function of the $O(n)$ loop model with weights $(\mu_2, h_1)$ per face not crossed and crossed by a loop, respectively.

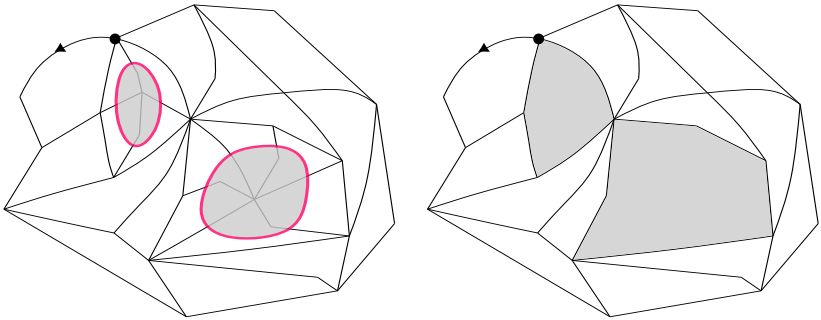

**Figure 2.5.: Left**: Illustrative example of a rigid $O(n)$ loop model configuration with two non-intersecting loops in red. The interiors of the loops are colored grey. **Right**: The corresponding external gasket where the marked faces, or holes, are colored in gray.

We will come back to this topic in Chapter 4 and elaborate on critical hyperbolic surfaces and the hyperbolic $O(n)$ loop model, but interested readers can refer to [30] for proofs concerning the $O(n)$ loop model in the context of maps. It is also worth noting that the $O(n)$ model has been studied in the context of matrix models in [67].

## 2.4. Irreducible maps

Another kind of map that will be of interest in this thesis is irreducible maps. A *2q-**irreducible planar map*** is a planar map of girth (length of the shortest simple cycle) at least 2q and such that every simple cycle of length 2q is a face of degree 2q, or half-length q. This concept can be generalized to genus-$g$ maps by requiring the irreducibility of their universal cover. In the literature, this is called "essential irreducibility" and it only restricts lengths of contractible cycles.

Let $\mathcal{M}_{g,n}^{(\mathfrak{q})}(L_1, \ldots, L_n)$ be the set of genus-$g$ maps with $n$ faces of half-lengths $(L_i \geq \mathfrak{q} \in \mathbb{N})_{1 \leq i \leq n}$ and weights $\mu_{L_i} \geq 0$ per face of half-length $L_i \geq \mathfrak{q}$. We define the *partition function of q-irreducible genus-g maps*, $F_g^{(\mathfrak{q})}[\mu]$, as

$$F^{(\mathfrak{q})}[\mu] = \sum_{k=1}^{\infty} \frac{1}{k!} \sum_{L_1=\mathfrak{q}}^{\infty} \mu_{L_1} \cdots \sum_{L_k=\mathfrak{q}}^{\infty} \mu_{L_k} \mu_0^{2-2g-k+\sum_i L_i} \|\mathcal{M}_{g,n}^{(\mathfrak{q})}(L_1, \ldots, L_k)\|. \tag{2.15}$$

The corresponding string equation is [32, Eq. (3.22)]

$$\tilde{x}^{(\mathfrak{q})}(u) = u \, \mathbb{I}_{\mathfrak{q}=0} - \sum_{k=0}^{\mathfrak{q}} (-1)^{\mathfrak{q}-k} \binom{\mathfrak{q}+k}{2k} \mathrm{Cat}(k) u^{\mathfrak{q}-k} - \sum_{k=\mathfrak{q}+1}^{\infty} \binom{2k-1}{k-\mathfrak{q}} \mu_k u^{\mathfrak{q}+k} - \mu_{\mathfrak{q}}, \tag{2.16}$$

whose solution $\tilde{x}^{(\mathfrak{q})}(E_\mu^{(\mathfrak{q})}) = 0$ defines the power series $E_\mu^{(\mathfrak{q})}$. The disk function can be expressed as [32, Eq. (3.21)]

$$W^{(\mathfrak{q})}(L) = \binom{2L}{L-\mathfrak{q}} \left( \sum_{k=0}^{\mathfrak{q}-1} (-1)^{\mathfrak{q}-k-1} \frac{\mathfrak{q}-k}{L-k} \binom{\mathfrak{q}+k}{2k} \mathrm{Cat}(k) E_\mu^{(\mathfrak{q})L-k} \right. \tag{2.17}$$

$$\left. - \sum_{k=\mathfrak{q}+1}^{\infty} \frac{\mathfrak{q}+k}{L+k} \binom{2k-1}{k+\mathfrak{q}} \mu_k E_\mu^{(\mathfrak{q})L-k} \right) \tag{2.18}$$

for $L > \mathfrak{q}$.

## 2.4.1. Substitution approach

To transition from maps without irreducibility constraints to irreducible maps, a substitution approach has been formulated for the case of $g = 0$ [32], and its extension to $g > 0$ is presented in [35]. In this subsection, we provide an illustrative overview. Let us focus on the planar case and consider the disk function of 2q-irreducible planar maps with a perimeter of half-length $L$, $W_\mu^{(\mathfrak{q})}(L)$, and the disk function of maps with girth at least 2q and perimeter of half-length $L$, $G_\mu^{(\geq \mathfrak{q})}(L)$. The key observations are that: if $L < \mathfrak{q}$, the only map satisfying the irreducibility constraint is a rooted plane tree with $L$ edges, and if $L = \mathfrak{q}$, the only map allowed, besides rooted plane trees, is a rooted map with a single face of half-length q. These relations can be expressed as

$$W^{(\mathfrak{q})}(L)[\mu_{\mathfrak{q}}, \mu_{\mathfrak{q}+1}, \dots] = \text{Cat}(L) + \mu_{\mathfrak{q}} \, \mathbb{I}_{L=\mathfrak{q}} \quad \text{for} \quad 1 \leq L \leq \mathfrak{q}. \tag{2.19}$$

Note that this equation evaluated at $L = \mathfrak{q}$ gives the string equation (2.16).

On the other side, when $L > \mathfrak{q}$: a 2q-irreducible map can be obtained by forbidding inner faces of half-length $\mathfrak{q}-1$ on a $2(\mathfrak{q}-1)$-irreducible map. Rooted maps of girth at least $2\mathfrak{q}$ and perimeter of half-length $L$ can be obtained by substituting each inner face of half-length q with a planar map of girth $2\mathfrak{q}$ and perimeter of half-length q. Finally, any map of the latter kind has girth=$2\mathfrak{q}$ or it is a rooted tree. These three relations can be expressed as

$$G^{(\geq \mathfrak{q})}(L)[\mu_{\mathfrak{q}}, \mu_{\mathfrak{q}+1}, \dots] = W^{(\mathfrak{q}-1)}(L)[0, \mu_{\mathfrak{q}}, \mu_{\mathfrak{q}+1}, \dots], \tag{2.20}$$

$$G^{(\geq \mathfrak{q})}(L)[\mu_{\mathfrak{q}}, \mu_{\mathfrak{q}+1}, \dots] = W^{(\mathfrak{q})}(L)[G^{(\mathfrak{q})}(\mathfrak{q})[\mu_{\mathfrak{q}}, \mu_{\mathfrak{q}+1}, \dots], \mu_{\mathfrak{q}}, \mu_{\mathfrak{q}+1}, \dots], \tag{2.21}$$

$$G^{(\mathfrak{q})}(\mathfrak{q})[\mu_{\mathfrak{q}}, \mu_{\mathfrak{q}+1}, \dots] = G^{(\geq \mathfrak{q})}(\mathfrak{q})[\mu_{\mathfrak{q}}, \mu_{\mathfrak{q}+1}, \dots] - \text{Cat}(\mathfrak{q}), \tag{2.22}$$

and are shown in Figure 2.6.

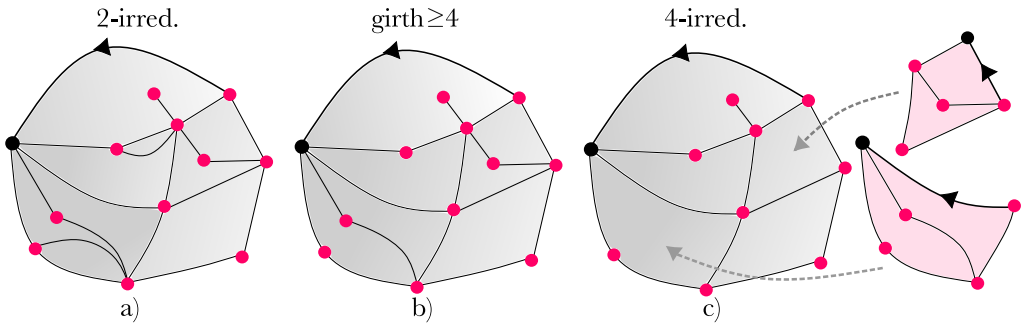

2-irred.    girth≥4    4-irred.

a)    b)    c)

**Figure 2.6.:** a) Example of a 2-irreducible rooted map with perimeter of length 6. b) A rooted map of girth ≥ 4 is obtained by removing all faces of length 2. c) A rooted map of girth ≥ 4 can also be obtained from a 4-irreducible rooted map by substituting each face of length 4 with a rooted disk of girth 4 and perimeter 4.

Together, these relations imply (see [32]) the existence of power series $\mu_k^{(\mathfrak{q})}[\mu_{\mathfrak{q}}, \mu_{\mathfrak{q}+1}, \dots]$ for $1 \leq k \leq \mathfrak{q}$ such that

$$W^{(\mathfrak{q})}(L)[\mu_{\mathfrak{q}}, \mu_{\mathfrak{q}+1}, \dots] = W^{(0)}(L)[\mu_1^{(\mathfrak{q})}, \mu_2^{(\mathfrak{q})}, \dots, \mu_{\mathfrak{q}}^{(\mathfrak{q})}, \mu_{\mathfrak{q}+1}, \dots], \tag{2.23}$$

where $\mu_k^{(\mathfrak{q})}[\mu_{\mathfrak{q}}, \mu_{\mathfrak{q}+1}, \dots]$ are determined by (2.19) for $L < \mathfrak{q}$. Analogous relations for $F_g[\mu]$ $g > 0$ are derived in [35].

The key point lies in the utilization of this weight substitution that facilitates the transition from $2q$-irreducible maps to genus-$g$ maps. The significance of $2q$-irreducible maps becomes pronounced in the continuum limit, where their structure closely parallels that of hyperbolic surfaces. This will be explored in Chapter 5.

## 2.5. Scaling limits

As highlighted in Chapter 1, genus- $g$ maps can be viewed as discretizations of 2-dimensional Euclidean spacetime. These maps hold large significance in the study of quantum gravity in the spirit of lattice field theory, offering a non-perturbative approach to QG. Naturally, a pivotal question arises: How can we systematically explore the transition of discrete maps into the continuum? Thereby bridging the gap between discrete and continuous random geometry. Schematically, this requires a refinement of the lattice structure, taking the number of building blocks to infinity and decreasing their size consistently. The rigorous mathematical avenue for investigating this transition is that of scaling limits of maps. In this section, we will give a short summary of the topic and illustrate the basic example of the Continuous Random Tree (CRT) [2], for further details, refer to [76].

### 2.5.1. Continuum Random Tree (CRT)

To explore scaling limits, we first establish a metric space framework for measuring distances between different maps. In this context, we consider the *Gromov-Hausdorff* metric. For two metric spaces $X$ and $Y$, their Gromov-Hausdorff distance $d_{GH}(X, Y)$ is defined as the infimum of the Hausdorff distances between all possible isometric embeddings of $X$ and $Y$ into a common metric space. Intuitively, a smaller Gromov-Hausdorff distance between $X$ and $Y$ implies a greater similarity between $X$ and $Y$ in terms of their metric structure.

In the context of scaling limits of maps, the Gromov-Hausdorff metric is used to assess the similarity of a map with $n$ vertices, $X_n$, viewed as metric spaces (equipped with their graph distance) with a continuous space $X$, as $n \longrightarrow \infty$. If a sequence of metric spaces $X_n$ converges to a limit metric space $X$ in the Gromov-Hausdorff metric, it indicates that as $n \rightarrow \infty$, $X_n$ becomes increasingly close to $X$ in terms of their metric structure.

We can start by considering a rooted plane tree with $n$ edges together with its graph distance. The *contour function* of this tree is obtained by recording the distance of each vertex from the root starting from the root to its leftmost neighbor and ending from its rightmost neighbor to the root $C_t(k) : [0, 2n] \rightarrow \mathbb{R}_{\geq 0}$ (See Figure 2.7b). This is a discrete positive one-dimensional walk that satisfies $C_t(0) = C_t(2n) = 0$. Therefore, if we consider a uniform random rooted plane tree[8], $C_t$ has the law of a random walk conditioned to stay positive and starting and ending at zero.

The key insight is to recognize a random walk converges to Brownian Motion. In this case,

$$\frac{1}{\sqrt{2n}\sigma} C_t(2n\ t) \xrightarrow[n \rightarrow \infty]{(d_{\mathbb{R}})} \mathbf{e}(t) \tag{2.24}$$

where $0 \leq t \leq 1$ and $\mathbf{e} : [0,1] \rightarrow \mathbb{R}_{\geq 0}$ is a Brownian excursion which is a standard Brownian motion starting and ending at zero and conditioned to stay positive [131] (see Figure 2.7c) and $d_{\mathbb{R}}$ is the norm on the space of real functions on $[0,1]$. Note that the scaling of the contour function and its argument in (2.24) as $n \rightarrow \infty$ means that we are taking the number of steps to infinity while decreasing the size of the steps to be infinitesimal. This resembles exactly the notion of a continuum limit.

Now that we know what the scaling limit of the contour function is, we can map the Brownian excursion, $\mathbf{e}$, back to a tree. We construct a tree by identifying points at the same distance

$$d_{\mathbf{e}}(s, t) = \mathbf{e}(s) + \mathbf{e}(t) - 2 \inf_{u \in [s,t]} \mathbf{e}(u) \qquad 0 \leq s, t \leq 1, \tag{2.25}$$

where the last term restricts the identification when the excursion drops below $\mathbf{e}([s, t])$ (see Figure 2.7c). This object is called the CRT.

In general, any such excursion $X : [0,1] \rightarrow \mathbb{R}_{\geq 0}$ naturally gives rise to a continuous metric space: the *real tree* given by the unit interval $[0,1]$ with metric

$$d(s, t) = X(s) + X(t) - 2 \inf_{u \in [s,t]} X(u), \tag{2.26}$$

where it is understood that we identify $s$ and $t$ whenever $d(s, t) = 0$. In Chapter 3, we will explore how to construct higher dimensional random geometries using these trees.

---

[8]Sampled uniformly in the way explained in Section 2.1

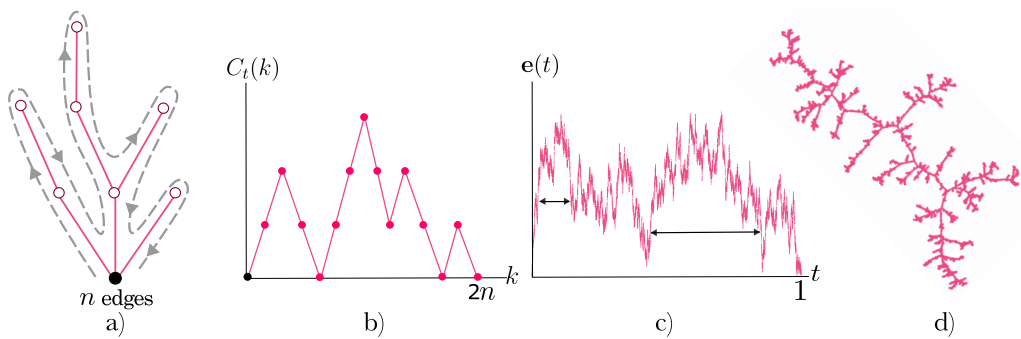

**Figure 2.7.:** a) Rooted plane tree with $n$ edges, the root is shown in black. b) Contour function obtained by recording the distance of each vertex from the root following the grey dotted line. c) Brownian excursion obtained in the scaling limit. An example of the distance identification (2.25) is shown in black horizontal arrows. d) Illustration of the CRT.

### 2.5.2. Universality

With a well-defined concept of a scaling limit in place, we can now explore questions regarding the universal properties of these objects. For instance, in (2.24) we observe that for any random rooted plane tree with $n$ edges, the contour function needs to be rescaled by $n^{-1/2}$ to obtain a well-defined scaling limit. For a general excursion, we will require a rescaling of the form $n^{-1/d_H}$, where $d_H$ is called the *Hausdorff dimension*. Therefore, for the CRT, we have $d_H = 2$. This is one of the exponents that characterize different universality classes, another one is the exponent seen in the asymptotic behavior of the partition function of rooted binary trees in Section 2.1. In this example (2.5), we have that $\mathcal{Z}_n^{\text{bin-tree}} \sim n^{-3/2}$. In general, for a model of random geometry the partition function is expected to scale with $n$ as

$$\mathcal{Z}_n^* \overset{n \to \infty}{\sim} C n^{\gamma_s - 2} \kappa^n, \tag{2.27}$$

where $C$ and $\kappa$ are non-universal constants, and $\gamma_s$ is a critical exponent, known in the physics literature as the *string susceptibility*. For binary trees, we have $\gamma_s = \frac{1}{2}$. The universality class with these values for the Hausdorff dimension is called *branched polymer universality class*.

# Scale-invariant random geometries and Asymptotic Safety

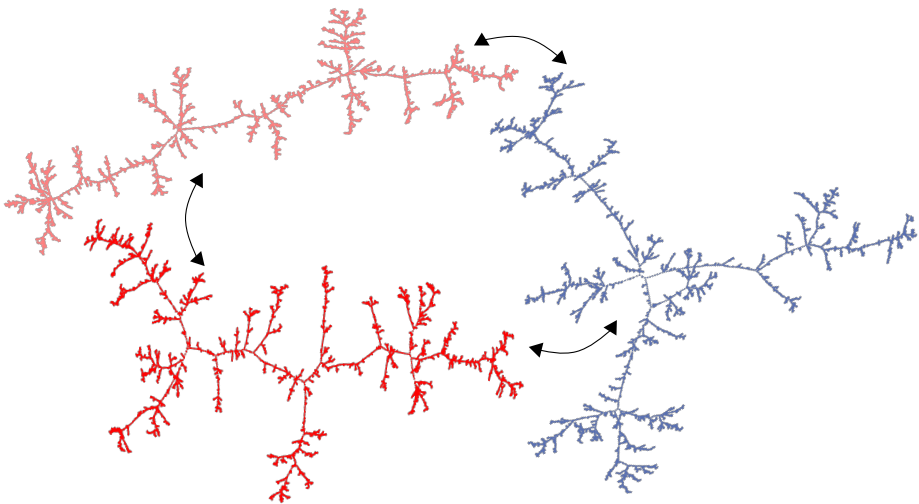

**Figure 3.1.:** Mating of 3 Continuum Random Trees.

In the Asymptotic Safety scenario, the non-perturbative renormalization group flow of the gravitational quantum field theory approaches a UV fixed point at which the dimensionless couplings take finite values and do not change with the energy scale. If these couplings are the ones of geometrical operators, we are led to the conclusion that the quantum laws of the spacetime geometry itself at such a fixed point must be scale-invariant.

Focusing on Euclidean quantum field theories of four-dimensional Riemannian metrics, functional renormalization group methods [128] relying on trun-cations of the renormalization group flow onto finite numbers of couplings have consistently found evidence for the existence of a suitable fixed point, see e.g. [130]. From a mathematical or statistical physics point of view, a Euclidean quantum field theory is ideally understood as a probability measure on an appropriate space of field configurations. This suggests that a mathematical construction of a UV fixed point of Euclidean quantum gravity amounts to the identification of a suitable model of scale-invariant random geometry on the spacetime manifold.

In the toy model of Euclidean two-dimensional quantum gravity this has been realized in the form of the Brownian plane (and its cousins on compact surfaces, like the Brownian sphere), an exactly scale-invariant probability measure on metric spaces with the topology of the plane, that represents a rigorous construction of what is called the pure-gravity universality class in the physics literature. It appears not just as the scaling limit of uniform random triangulations of the 2-sphere [76], but also naturally arises from Liouville Quantum Gravity in the absence of matter coupling [58, 81, 119]. The latter is related to the Liouville Conformal Field Theory [53], which due to the scaling symmetry of conformal field theories makes it clear that we are dealing with a fixed point of the renormalization group flow. As a consequence of this symmetry, the Brownian plane is not a (random) Riemannian manifold, which would require the geometry in any sufficiently small neighbourhood to resemble that of the Euclidean plane, but is genuinely fractal. For instance, its Hausdorff dimension, equal to 4 [74], differs substantially from its topological dimension.

Beyond two dimensions, however, we currently know of no explicit examples of non-trivial scale-invariant random metric spaces with the topology of a three- or four-dimensional manifold. The construction of natural examples is therefore not only an important ingredient for the asymptotic safety scenario but also presents an important open mathematical problem. Essentially there are three ways of approaching the problem, mimicking what we know from two dimensions (where all three ways lead to the same results).

The first way would be to construct a scale-invariant quantum field theory on the spacetime manifold that describes the gauge-fixed degrees of freedom of the Riemannian metric. Subsequently, we would need to figure out how to extract the metric geometry from these. This would be analogous in the two-dimensional setting to first identifying the Liouville Conformal Field theory, which aims to describe two-dimensional Riemannian metrics in conformal gauge, and extract the geometry from there using an appropriate regularization procedure (which has largely been achieved via Quantum Loewner Evo-lution in the case of pure gravity and Liouville First Passage Percolation in the presence of matter fields). However, this procedure is difficult to generalize to higher dimensions, because of the lack of good global coordinate gauges and the challenges involved in constructing non-perturbative interacting quantum field theories.

The second way is to introduce discreteness in the field configurations following the philosophy of lattice field theory. Having a non-zero lattice spacing regularizes the ultraviolet divergences in the path integral while allowing to include field configurations beyond the perturbative regime, as is successfully employed in the numerical investigation of QCD in its strongly-coupled regime. In gravity, the field should describe the geometry itself, and this is naturally achieved by allowing the lattice itself to become dynamical, with the gravitational degrees of freedom entirely contained in the combinatorial data describing the lattice and its geometry. This is done in Dynamical Triangulations [12, 52, 11], where, as we have seen in Section 1.2.2, the lattice is constructed by gluing equilateral simplices. In order to find a scale-invariant random geometry, representing a potential Euclidean quantum field theory at the UV fixed point, it is necessary to take a scaling limit. As is well known from statistical physics, for such a non-trivial scaling limit to exist, the discrete model must be critical, in that it exhibits diverging correlation lengths. Besides criticality, another very important criterion in the case of random geometry is that the manifold structure does not degenerate in the scaling limit. For instance, in Dynamical Triangulations of the 3-sphere, the piece-wise flat geometries built from equilateral tetrahedra have the topology of the 3-sphere and display criticality in the so-called branched polymer phase of the model. However, numerical simulations indicate that shrinking the building blocks leads to the topology degenerating into that of trees, nothing like the manifold structure of the 3-sphere. Apart from the branched polymers, simulations of Dynamical Triangulations in three and four dimensions have not (yet) uncovered critical phenomena that escape this branched-polymer universality. This means that the lattice approach is yet to uncover concrete opportunities to establish scale-invariant random geometries on three- and four-dimensional manifolds. There are however promising avenues in models that restrict the family of triangulations considered. Simulations suggest that four-dimensional *Causal* Dynamical Triangulations [105] feature continuous phase transitions [7, 6] where one expects criticality to be found, while a numerical investigation of a recently proposed model of three-dimensional dynamical triangulations assembled from triples of trees is underway [42].

The third route towards scale-invariant random geometry, the one that we follow in this chapter, also aims to assemble geometries out of simpler building blocks, but instead of relying on criticality and scaling limits to approach scale-invariance with random discrete objects, one takes the building blocks themselves to be scale-invariant. If the assembly does not spoil the scale-invariance and the resulting geometry has the desired topology, this provides a very economical way of uncovering new universality classes. A simple example of such an assembly procedure is that of the Continuum Random Tree out of Brownian motion [2] (see Section 2.5.1). Brownian motion, seen as a random continuous real function on the line (or equivalently as a massless free scalar field in one dimension), is the prime example of a scale-invariant random object. This random real function naturally gives rise to a gluing procedure of the real line into a topological tree with a metric, namely the CRT, that shares the same scaling symmetry (see Section 3.1.5 for a discussion). Since the CRT is not a topological manifold, its relevance for quantum gravity is not obvious, but stems from the possibility of using the CRT itself as building block for larger random geometries.

The hope of assembling manifolds out of random trees may seem far-fetched, but is well established in two-dimensional quantum gravity. At the level of discretized surfaces, bijective encoding of planar maps in treelike combinatorial structures has a long tradition, starting with Mullin's bijection for tree-decorated maps [121] and the Cori–Vauquelin–Scheafer bijection between quadrangulations [48, 135] and labeled trees. The study of the latter bijection in the scaling limit, in which the discrete trees approach the CRT, paved the way for a mathematically rigorous construction of (and convergence to) the scale-invariant Brownian sphere [112, 76, 116]. Generalizations of Mullin's bijection to random planar maps decorated by various critical statistical systems [136, 23, 98, 103] hinted at a different appearance of CRTs in the continuum limit, in a way that ties in closely with Liouville Quantum Gravity and conformally invariant random curves described by Schramm–Loewner Evolution (SLE). In the foundational paper [63] by Duplantier, Miller and Sheffield this approach, going under the name of Mating of Trees, was introduced in its generality. Starting from two independently sampled CRTs, there exists a simple assembly procedure that identifies points in the contours of the two trees, resulting in a scale-invariant random metric space that (almost surely) has the topology of the 2-sphere. This metric space is known to correspond with Liouville Quantum Gravity for a particular value of its coupling constant ($\gamma = \sqrt{2}$). Remarkably, any other value of this coupling $\gamma \in (0, 2)$ associated to a grav-itational universality class in the presence of an arbitrary matter conformal field theory (as long as the matter central charge is below 1), can be achieved by introducing a correlation between the pair of CRTs. As alluded to above, a CRT is assembled from a Brownian motion, meaning that a pair of CRTs is naturally obtained from the two coordinates of a two-dimensional Brownian motion, and this correlation can be understood as the choice of a non-trivial covariance matrix for the latter Brownian motion.

The fact that the CRT provides the universal building block for essentially all scale-invariant random geometries relevant to two-dimensional quantum gravity naturally raises the question of whether higher-dimensional random geometries can be constructed in a similar fashion. One can question whether this is sufficiently motivated from a path integral perspective on quantum gravity, but given that at present we do not know of a single explicit example of a scale-invariant random geometry with three-dimensional manifold topology one should not set too stringent conditions. In this chapter (based on [40]), we propose a rather straightforward generalization of the mating of trees construction, in which the pair of correlated CRTs is replaced by a triple. If the result has a well-defined and scale-invariant random metric structure, something that requires checking, it necessarily gives rise to new universality classes beyond random surfaces. Naturally, the model possesses a three-dimensional parameter space as opposed to the one-dimensional parameter space of mated-CRT surfaces. One of the critical exponents, the string susceptibility, can be calculated (analytically for special regions and numerically elsewhere) and displays a non-trivial dependence on the three parameters, suggesting that they really parameterize an entire family of new universality classes. To start exploring the parameter space we develop a numerical toolbox to simulate the result of mating a triple of trees and measure an important critical exponent related to the metric: the Hausdorff dimension, which governs the relative scaling between volume and radius of geodesic balls in the geometry. Whether the topology induced by the metric really is that of a three-dimensional manifold requires a more refined analysis that is beyond the scope of the current work.

# 3.1. Mating of trees: 2D quantum gravity from Brownian motion

The basic idea of the Mating of Trees approach is that both the geometry and the matter degrees of freedom in two-dimensional quantum gravity on the 2-sphere can be encoded in a single continuous path in the Euclidean plane. For a detailed story of the correspondence, we direct the mathematically-inclined audience towards the foundational paper by Duplantier, Miller and Sheffield [63] and the recent survey by Gwynne, Holden and Sun [79], as well as the many references in the latter. In this section, we provide a high-level introduction to the topic for those that are not entirely comfortable with the probability theory literature. First, we will review two examples of mating-of-trees bijections between discrete surfaces decorated with statistical systems and certain discrete walks in the quadrant. Next, we will explain how this picture extends to the continuum limit and generalizes to the full family of two-dimensional quantum gravity theories coupled to conformal matter.

## 3.1.1. First discrete example: spanning-tree decorated quadrangulations

The simplest example of a model of discrete surfaces that is naturally encoded by a walk in the quadrant, is that of spanning-tree decorated quadrangulations, which goes back to a bijection of Mullin [121] in the sixties.

A *spanning-tree-decorated quadrangulation* is a planar quadrangulation together with a choice of diagonal in each face, such that the graph formed by the diagonals alone has no loops (Figure 3.2b). In this case, the diagonals necessarily take the form of two disjoint trees that together span the vertices of the quadrangulation, hence the name of the model. The two vertices of the root edge belong to the two different trees and mark a root for each of them. If one assigns equal Boltzmann weight to each spanning-tree-decorated quadrangulation (in other words, one samples uniformly), one may think of this decoration as a statistical system coupled to the geometry of the surface described by the quadrangulation, thus as a rather abstract form of matter. Note that the presence of the statistical system has an (entropic) effect on the geometry, as the number of decorations differs from one quadrangulation to the other.

According to Mullin [121] rooted spanning-tree-decorated quadrangulations with $n$ faces are in bijection with excursions in the quadrant of length $2n$ with unit steps parallel to the axes. An *excursion in the quadrant* is a walk $Z_0, Z_1, \ldots, Z_{2n} \in \mathbb{Z}_{\geq 0}^2$ with $2n$ steps that starts and ends at the origin, $Z_0 = Z_{2n} = 0$ (Figure 3.2d). The bijection is rather easy to understand: there exists a unique closed path on the surface starting and ending at the root edge that intersects all edges of the quadrangulation while avoiding all diagonals (Figure 3.2c). The corresponding excursion simply records for the $i$th visited edge the heights $Z_i \in \mathbb{Z}_{\geq 0}^2$ of its left and right extremity in the tree. From the figure it should be clear that between consecutive visits, exactly one of the heights changes by $\pm 1$, so $Z_{i+1} - Z_i \in \{(0, \pm 1), (\pm 1, 0)\}$ and one indeed obtains the desired excursion in the quadrant. It is straightforward to check that any such excursion can be obtained in this way and that the quadrangulation together with its decoration can be reconstructed from the excursion.

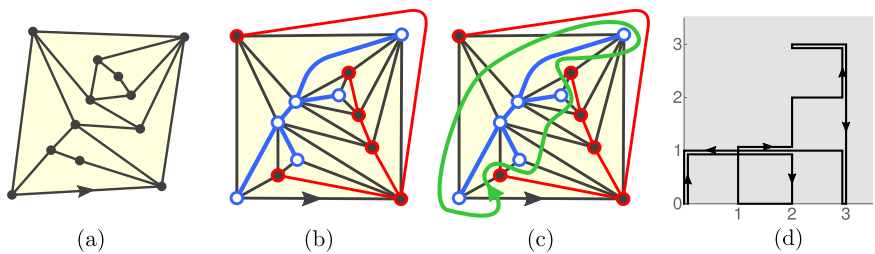

(a)  (b)  (c)  (d)

**Figure 3.2.:** (a) A rooted quadrangulation (note that the white outer region is a face of degree four as well). (b) A spanning-tree-decorated quadrangulation. (c) The space-filling curve. (d) The corresponding excursion $(Z_i)$ in the quadrant. Figure adapted from [21].

In light of what follows, it is useful to think of the reconstruction starting from an excursion as a three step procedure. In the first step one only examines the sequence of horizontal steps of the excursion (and ignoring the vertical coordinate), which encodes a *Dyck path*, i.e. a walk with unit steps on the non-negative integers starting and ending at zero. It encodes a plane tree, which we draw in blue. In the second step, a red tree is constructed similarly from the vertical steps of the excursion. Now every visit of the excursion naturally corresponds to a pair of corners, one on each tree, such that following the excursion corresponds to tracing the contour of the blue tree in counterclockwise direction and the contour of the red tree in clockwise direction. Finally, in the third step the blue and red tree are "mated" into a spanning-tree-decorated quadrangulation by drawing a black edge between each pair of corners.

## 3.1.2. Second discrete example: site-percolated triangulations

The previous example is archetypal, where it is intuitively clear that the surface can be encoded in trees, because the decoration already takes the form of a tree. Admittedly, it is not the most natural statistical system one would think of when trying to couple quantum gravity to a matter field. So let us look at a model that has a simpler interpretation, but for which the trees are well hidden. This is the model of (loopless) triangulations with site percolation [15, 23, 80]. A *triangulation* is a planar map in which all faces have degree three and it is *loopless* if it has no edges starting and ending on the same vertex. A *site percolation* on a rooted triangulation is simply an assignment of one of two colors, say blue and red, to each vertex of the map, with the only requirement that the root edge points from a red to a blue vertex (see Figure 3.4a). If one assigns equal Boltzmann weight to every such rooted site-percolated triangulation with $2n$ triangles, we obtain a very simple example of a statistical system on a random surface. One could think of this system as the high-temperature limit of the standard Ising model living on the vertices of the triangulation. Note that, contrary to the spanning-tree-decorated quadrangulations, each triangulation admits the same number of distinct site percolations, namely $2^n$ because there are precisely $n$ vertices that are not incident to the root edge. This means that the statistical system does not affect the statistics of the geometry, and we are dealing with a model whose geometry lives in the universality class of pure gravity, coupled to a rather trivial type of matter in the form of white noise.

Just like in the previous example, we would like to find a self-avoiding closed curve that in some sense explores the full triangulation. It is natural to consider the partition of the vertex set into monochromatic clusters and examine the cluster interfaces, which naturally correspond to a collection of disjoint closed loops on the graph dual to the triangulation (Figure 3.4a). Since the root edge crosses an interface, it is natural to start the exploration along this interface. Unless the site-percolated triangulation is exactly of the type of Figure 3.2b, with exactly one blue cluster and one red cluster and such that the monochromatic edges in each cluster form a tree, the exploration will return to the root edge before having explored the full map. The rough idea described in [23] is that one can merge all cluster interfaces into a single exploration by following an interface and taking detours into neighbouring interfaces at the very last opportunity before they become inaccessible.

More precisely, one may setup a peeling exploration that visits all $3n$ edges of the triangulation as follows (see [80, Section 2.3]). We start the exploration in the triangle at the right of the root edge and position the tip of the exploration at the other non-monochromatic edge of that triangle (see the top left of Figure 3.4c). Then at each step the edge $e$ at the tip is removed, in such a way that we eventually return to the root edge (the final *target*). There are five different cases (a to e) to be considered, which are summarized in Figure 3.3. If $e$ is adjacent to a triangle containing a non-boundary vertex, then the exploration traces the cluster interface, meaning that it turns left or right depending on the color of the vertex (cases a and b). If however, all vertices of the triangle are on the boundary, one considers the two components that are separated by the triangle, and one implements a detour through the component that does not contain the target. The choice of detour is shown in Figures 3.3c and d by a light green arrow pointing from one component, where it can be regarded as an intermediate target for the exploration, to the other, indicating where the exploration will continue after the first component has been fully explored. Finally, if instead of a triangle the edge $e$ is adjacent to a detour, the detour is followed (case e). An example of a full exploration is shown in Figure 3.4c.

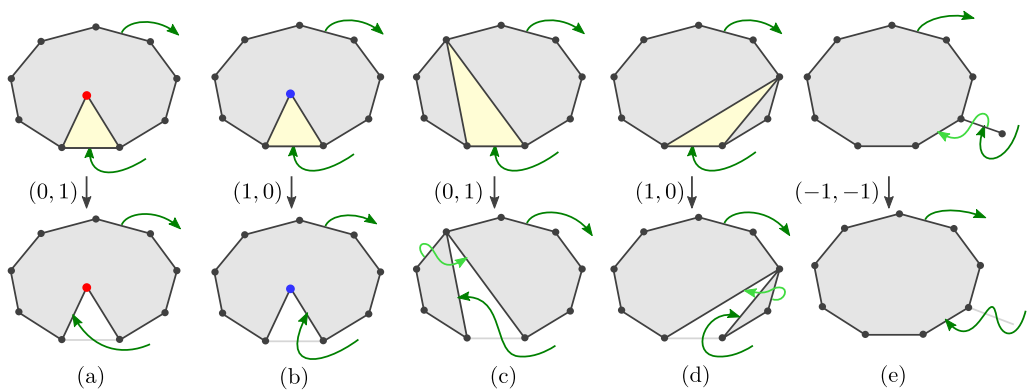

**Figure 3.3.:** The five types of possibilities when peeling away an edge in the exploration. The pair of integers indicates the length change of the contour to the left respectively right of the exploration path.

To obtain a lattice walk, at every step $i = 1, \ldots, 3n$ of the exploration one keeps track of the distances $Z_i \in \mathbb{Z}_{\geq 0}^2$ in clockwise respectively counterclockwise direction along the contour between the tip of the exploration and the root edge. It is easily seen that the only possible changes in these distances are $(0, 1)$, $(1, 0)$ and $(-1, -1)$, so one obtains an excursion with these increments in the quadrant of length equal to the number of edges of the triangulation (Figure 3.4b). These walks are known as Kreweras walks [102]. It is a non-trivial fact that this determines a bijection, see [80, Theorem 2.2] and the earlier references [23, 25].

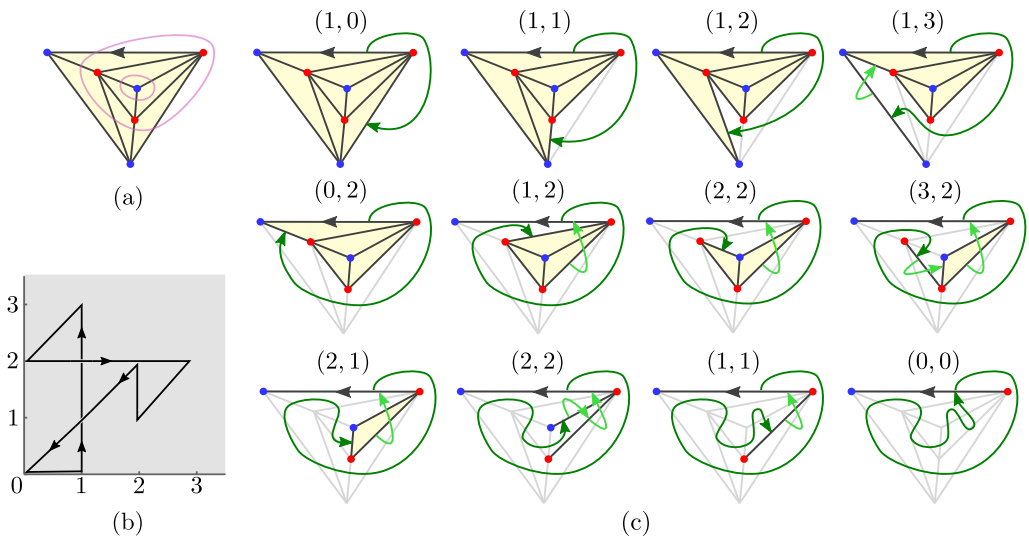

**Figure 3.4.:** (a) A rooted site-percolated triangulation with cluster interfaces indicated in pink. (b) The corresponding excursion in the quadrant with increments $(1,0)$, $(0,1)$ and $(-1,-1)$. (c) The peeling exploration. The dark green curve illustrates the iteratively constructed exploration path, while the lighter green curves indicate the required detours. The integers above each map are contour lengths on the left respectively right of the exploration path.

### 3.1.3. Scaling limit of the walks

A first consequence of these bijections is that one can easily understand the asymptotics of the enumeration. Let us denote by $\mathcal{Z}_n^*$ the canonical partition function with unit Boltzmann weight per configuration, meaning simply the total number of decorated rooted planar maps in the model. In the case of spanning-tree decorated quadrangulations, the number of simple excursions of length $2n$ is easily found to be

$$\mathcal{Z}_n^{\text{spanning-tree}} = \text{Cat}(n)\,\text{Cat}(n+1) \overset{n\to\infty}{\sim} \frac{4}{\pi} 4^{2n} n^{-3}. \tag{3.1}$$

Note that the exponential growth $4^{2n}$ reflects the four different increments available for each of the $2n$ steps of a simple walk. In the case of site-percolated triangulations, the number of Kreweras excursions of length $3n$ is [102, 23]

$$\mathcal{Z}_n^{\text{percolation}} = \frac{4^n}{(n+1)(2n+1)} \binom{3n}{n} \overset{n \to \infty}{\sim} \sqrt{\frac{3}{16\pi}} \, 3^{3n} n^{-5/2}. \tag{3.2}$$

Also here the $3^{3n}$ agrees with the three possible increments of the walk at each of the $3n$ steps. More importantly, the exponents of the power-law correction differ between the two models.

This should not be surprising, because the models feature qualitatively different matter systems and should be expected to belong to different universality classes (see Section 2.5.2). If the universality class corresponds to two-dimensional quantum gravity coupled to a matter conformal field theory with central charge $c \in (-\infty, 1]$, then the KPZ formula predicts [100]

$$\gamma_s = \frac{c - 1 - \sqrt{(c-1)(c-25)}}{12}, \tag{3.3}$$

for the string susceptibility defined in (2.27). We see that spanning-tree-decorated quadrangulations feature the exponent $\gamma_s = -1$ corresponding to $c = -2$. The site-percolated triangulations feature the exponent $\gamma_s = -1/2$ corresponding to $c = 0$, in accordance with the expectation that the latter live in the pure gravity universality class together with binary trees.

Why do we see two different exponents appearing in the enumerations of excursions in the quadrant? Similarly to our tree example in Section 2.5.1, in the large-$n$ limit the random walks in the quadrant, rescaled by $1/\sqrt{n}$, approach a Brownian excursion in the quadrant. In order to specify this process, it is sufficient to know the covariance of the unrestricted Brownian motion, which appears as the limit of the unrestricted random walks. It is precisely in this covariance that the two models differ. Indeed, denoting the $x$ and $y$ components of a walk on $\mathbb{Z}^2$ by $L_t$ and $R_t$ respectively, an unrestricted simple random walk on the square lattice has covariance

$$\text{Var}(L_t) = \text{Var}(R_t) = \frac{t}{2}, \qquad \text{Cov}(L_t, R_t) = 0, \tag{3.4}$$

while an unrestricted Kreweras random walk satisfies

$$\text{Var}(L_t) = \text{Var}(R_t) = \frac{2t}{3}, \qquad \text{Cov}(L_t, R_t) = \frac{t}{3}. \tag{3.5}$$

In general, for a random walk with

$$\text{Cov}(L_t, R_t) = \rho \, \text{Var}(L_t) = \rho \, \text{Var}(R_t), \qquad \rho \in (-1, 1) \tag{3.6}$$

it is known [55] that the number of excursions in the quadrant of length $n$ grows like

$$C \, n^{-1 - \frac{\pi}{\arccos(-\rho)}} \, \kappa^n, \tag{3.7}$$

for some $C > 0$ and $\kappa > 1$. Note that for $\rho = 0$ respectively $\rho = 1/2$ this indeed agrees with (3.1) respectively (3.2). Moreover, it suggests that any other model of random decorated planar maps that admits a bijection with walks in the quadrant and belongs to a universality class with a certain central charge $c$ must satisfy $\rho = -\cos\left(\frac{\pi}{1-\gamma_s}\right)$. Several further examples are indeed known for which this is the case, including bipolar-oriented triangulations ($c = -7$, $\gamma_s = -2$, $\rho = -1/2$) [98] and Schnyder-wood-decorated triangulations ($c = -25/2$, $\gamma_s = -3$, $\rho = -1/\sqrt{2}$) [103].

### 3.1.4. Mating of trees and Liouville Quantum Gravity

The discrete examples make one wonder whether there is a continuum interpretation to the encoding by trees and whether it extends to other universality classes of two-dimensional quantum gravity coupled to conformal matter with $c \in (-\infty, 1]$. This has indeed been shown to be the case in a framework going under the name of mating of trees [63], putting the case $c = 1$ aside with its peculiarities [16]. In order to understand the result, we need to explain first how to describe geometry and space-filling curves in the continuum. We are dealing with quantum gravity on the 2-sphere, which is conveniently represented by the Riemann sphere $\hat{\mathbb{C}} = \mathbb{C} \cup \{\infty\}$. Let $\hat{g}_{ab}$ be some fixed conformal[1] background metric on $\hat{\mathbb{C}}$ of unit area. Then, we are after a random real field $\phi$ on $\hat{\mathbb{C}}$, that we informally interpret as describing a random Riemannian metric $g_{ab} = e^{\gamma\phi}\hat{g}_{ab}$ of unit volume, and independently a random continuous space-filling curve $\eta_{\hat{g}} : [0,1] \longrightarrow \hat{\mathbb{C}}$ such that $\eta_{\hat{g}}(0) = \eta_{\hat{g}}(1)$ and $\eta_{\hat{g}}([s,t])$ has volume $t - s$ with respect to the background measure $\sqrt{\hat{g}} \, \mathrm{d}^2 z$ for $0 < s < t < 1$.

---

[1]Conformal in the sense that $\hat{g}_{ab}$ preserves the angles of $\hat{\mathbb{C}}$, and thus corresponds to a metric of the form $f(z)(\mathrm{d}x^2 + \mathrm{d}y^2)$ for $z = x + iy \in \hat{\mathbb{C}}$ for some function $f : \hat{\mathbb{C}} \longrightarrow [0, \infty)$. A natural choice, corresponding to constant curvature $\hat{R} = 8\pi$ and unit area, is $f(z) = \frac{1}{\pi}(1 + |z|)^{-2}$.

The former is provided by Liouville Quantum Gravity (LQG) and the latter by Schramm–Loewner Evolution (SLE), which we both briefly discuss.

By the uniformization theorem, any two-dimensional Riemannian metric on the sphere is isometric to a conformal rescaling $g_{ab} = e^{\gamma\phi}\hat{g}_{ab}$ of the background metric $\hat{g}_{ab}$. Liouville Quantum Gravity with coupling constant $\gamma \in (0, 2)$ is the path integral quantization of this field $\phi$ (known as the dilaton) with action

$$S_L = \frac{1}{4\pi} \int d^2x \sqrt{\hat{g}} \left( \hat{g}^{ab}\partial_a\phi\partial_b\phi + Q\hat{R}\phi + 4\pi\hat{\mu}e^{\gamma\phi} \right), \tag{3.8}$$

where

$$Q = \frac{\gamma}{2} + \frac{2}{\gamma}, \tag{3.9}$$

$\hat{R}$ is the scalar curvature of $\hat{g}_{ab}$ and $\hat{\mu} > 0$ a parameter known as the cosmological constant. If gravity is coupled to a conformal matter field with central charge $c \in (-\infty, 1)$, then the parameter $Q \in (2, \infty)$ is related to $c$ via

$$c = 25 - 6Q^2. \tag{3.10}$$

It is not obvious that one can make sense of a random field with density proportional to $e^{-S_L}$ (where we have taken $\hbar = 1$ for simplicity) in an appropriate space of generalized functions. Luckily $\phi$ is closely related to the Gaussian Free Field (GFF) on $\hat{g}_{ab}$, the free massless real scalar field with action $S_L$ in which $Q$ and $\hat{\mu}$ are set to zero, which has an unambiguous probabilistic interpretation if we restrict the constant mode, for instance, by requiring zero mean (see [145] for a recent introduction to the GFF). Accounting for the residual Möbius symmetry by marking three points, say $z_1, z_2, z_3 \in \hat{\mathbb{C}}$, it can be shown that $\phi$ is obtained from the GFF by a deterministic position-dependent shift [53, 17].

The resulting random field $\phi$ is not defined point-wise but should be viewed as a generalized function. In order to make sense of the rescaling $g_{ab} = e^{\gamma\phi}\hat{g}_{ab}$, it is thus necessary to consider a suitable regularization. For instance, we could look at the circle average $\phi_\epsilon(z)$, taken to be the average of $\phi(z)$ over a circle of radius $\epsilon$ around $z$. Then, a normalized *quantum area* measure on $\hat{\mathbb{C}}$ can be defined via

$$\mu_\phi = \lim_{\epsilon \to 0} \frac{e^{\gamma\phi_\epsilon(z)}\sqrt{\hat{g}}\, d^2z}{\int_{\hat{\mathbb{C}}} e^{\gamma\phi_\epsilon(z)}\sqrt{\hat{g}}\, d^2z}, \tag{3.11}$$

that is independent of $\hat{g}_{ab}$ and is such that $z_1, z_2, z_3$ are uniform points for $\mu_\phi$[2].

---

[2] More precisely, this means that if we sample three new independent points $z_1', z_2', z_3'$ with probability distribution $\mu_\phi$ and look at the transformation of $\mu_\phi$ under the unique Möbius transformation that sends $z_i'$ to $z_i$, $i = 1, 2, 3$, then the result has the same law as $\mu_\phi$.

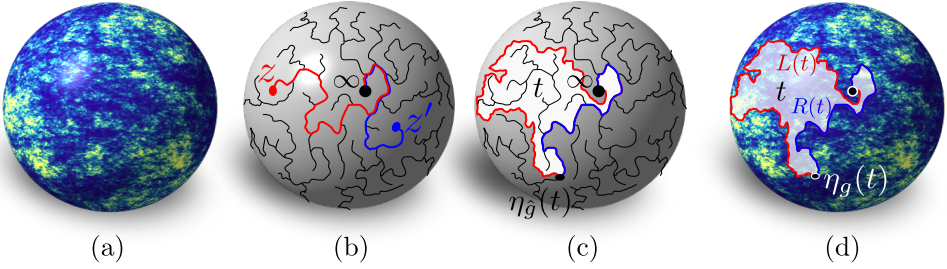

(a)          (b)          (c)          (d)

**Figure 3.5.:** (a) LQG$_\gamma$: simulation of the random measure $\mu_\phi$ on the round 2-sphere (lighter regions contain more quantum area than darker regions). (b) Space-filling SLE$_{\kappa'}$: an illustration of the imaginary geometry flow lines from $z$ and $z'$ to $\infty$. Here $\eta_z$ (in red) meets $\eta_{z'}$ (in blue) from the left, and thus $z$ precedes $z'$ in the space-filling curve. (c) $\eta_{\hat{g}}([0, t])$ is a closed region (in white) of $\hat{g}$-area $t$ in $\hat{\mathbb{C}}$ with $\infty$ and $\eta_{\hat{g}}(t)$ on its boundary. (d) LQG$_\gamma$ + SLE$_{\kappa'}$: the region $\eta_g([0, t])$ has quantum area $\mu_\phi(\eta_g([0, t])) = t$ and $L(t)$ and $R(t)$ are the left and right boundary lengths of this region measured by $\nu_\phi$ respectively.

See Figure 3.5a for an illustration. Similarly, one may introduce a *quantum length* measure via

$$\nu_\phi = \lim_{\epsilon \to 0} \frac{e^{\gamma \phi_\epsilon(z)/2} \sqrt[4]{\hat{g}} \, |\mathrm{d}z|}{\sqrt{\int_{\hat{\mathbb{C}}} e^{\gamma \phi_\epsilon(z)} \sqrt{\hat{g}} \, \mathrm{d}^2 z}}. \tag{3.12}$$

Given a region $U \subset \hat{\mathbb{C}}$ and a curve $\Gamma : [0, 1] \to \hat{\mathbb{C}}$, $\mu_\phi(U)$ and $\nu_\phi(\Gamma([0, 1]))$ should be interpreted as rigorous definitions for the usual area $\int_U \sqrt{g} \, \mathrm{d}^2 z$ and length $\int_0^1 \sqrt[4]{g} |\Gamma'(t)| \mathrm{d}t$ as measured by the Riemannian metric $g_{ab} = e^{\gamma \phi} \hat{g}_{ab}$. The Riemann sphere $\hat{\mathbb{C}}$ equipped with the random measures $\mu_\phi$ and $\nu_\phi$ is called the *unit-area $\gamma$-quantum sphere.*

Next, we describe the random space-filling curve $\eta_{\hat{g}}$ arising from SLE$_{\kappa'}$ with $\kappa' = 16/\gamma^2 \in (4, \infty)$. A concise way to introduce this curve is via *imaginary geometry*. If $h : \mathbb{C} \to \mathbb{R}$ is a smooth real function, one can consider the flow lines of the complex vector field $e^{ih/\chi}$ where $\chi = \frac{\sqrt{\kappa'}}{2} - \frac{2}{\sqrt{\kappa'}}$. More precisely, to $z \in \mathbb{C}$ we associate the curve $\eta_z$ determined by

$$\eta_z'(t) = e^{\frac{i}{\chi} h(\eta_z(t))}, \quad \eta_z(0) = z, \quad t \geq 0. \tag{3.13}$$

Remarkably these flow lines are still well-defined if we take $h$ to be the (far from smooth) whole-plane GFF $h$ on $\hat{\mathbb{C}}$ (independently but similar to the one for LQG). It can be shown, see [117], that for each $z \in \mathbb{C}$ the flow-line $\eta_z(t)$ does not self-intersect and approaches $\infty$ as $t \longrightarrow \infty$, and that for two distinct starting points $z, z' \in \mathbb{C}$ the flow lines $\eta_z$ and $\eta_{z'}$ almost surely eventually meet and stay together before reaching $\infty$ (Figure 3.5b). One may use this to associate an order to the points in the complex plane: $z$ precedes $z'$ if $\eta_z$ meets $\eta_{z'}$ from the left. The space-filling curve $SLE_{\kappa'}$ is the continuous non-self-crossing path $\eta_{\hat{g}} : [0, 1] \longrightarrow \hat{\mathbb{C}}$ starting and ending at $\infty$ that visits the points of $\mathbb{C}$ in this order, parametrized such that $\eta_{\hat{g}}([s, t])$ has area $t - s$ for $0 \le s < t \le 1$.

Liouville Quantum Gravity for $\gamma \in (0, 2)$ and the space-filling $SLE_{\kappa'}$ for $\kappa' \in (4, \infty)$ are intimately related to each other when $\kappa' = 16/\gamma^2$. To formulate this let us consider a unit-area $\gamma$-quantum sphere $\mu_\phi$, $\nu_\phi$ and an independently-sampled space-filling curve $\eta_{\hat{g}}$. It is then natural to consider the reparametrization $\eta_g$ of $\eta_{\hat{g}}$ such that it explores the quantum area at unit rate, meaning that $\mu_\phi(\eta_g([s, t])) = t - s$ for $0 < s < t < 0$. Now for every $t \in (0, 1)$, the traces $\eta_g([0, t])$ and $\eta_g([t, \infty])$ are closed subset of $\hat{\mathbb{C}}$. The boundary $\eta_g([0, t]) \cap \eta_g([t, \infty])$ at which they meet consists of two continuous curves starting at $\eta_g(t)$ and ending at $\infty$. Let $Z(t) = (L(t), R(t)) \in \mathbb{R}^2_{>0}$ be the $\nu_\phi$-lengths of these curves (Figure 3.5d). The crucial insight of the mating of trees approach [63] is that the process $Z(t)$ has a very simple law. To be precise, according to [118, Theorem 1.1] (and [14, Theorem 1.3] for the precise normalization) it has the law of a two-dimensional Brownian motion $(L(t), R(t))$ started from $(0, 0)$ with covariance

$$\text{Var}(L(t)) = \text{Var}(R(t)) = \frac{2}{\sin\left(\frac{\pi\gamma^2}{4}\right)}|t|, \text{ and } \text{Cov}(L(t), R(t)) = -2\cot\left(\frac{\pi\gamma^2}{4}\right)|t|$$

(3.14)

and conditioned to stay in $[0, \infty)^2$ and to return to $(0, 0)$ at time $t = 1$. Moreover, both the quantum sphere and the space-filling curve are almost surely determined by this process. This means that, at least in principle, one can reconstruct the measures $\mu_\phi$ and $\nu_\phi$ as well as the curve $\eta_{\hat{g}}$ simply by looking at the Brownian excursion $Z(t)$. In the next subsection, we will discuss an explicit procedure.

At this point, one should recognize the analogy with the discrete mating-of-trees bijections that we described above. The quantum surface is the continuum analogue of the random planar map, while the space-filling $SLE_{\kappa'}$ is the analogue of the exploration path determined by the statistical system living on the planar map. In the discrete case the bijections with lattice walks show that the random discrete surface is completely determined by a corresponding ran-dom walk, which starts at the origin and is conditioned to stay in the positive quadrant $\mathbb{Z}_{\geq 0}^2$ before returning to the origin after $n$ steps. Upon rescaling the walk by $1/\sqrt{n}$ and normalizing the time to run over $[0, 1]$, its law converges in a probabilistic sense to that of the Brownian excursion $(L(t), R(t))$. Comparing (3.6) to (3.14) we observe that $\rho = -\cos(\pi\gamma^2/4)$ and therefore the string susceptibility $\gamma_s$ must be related to $\gamma$ by

$$\gamma_s = 1 - \frac{4}{\gamma^2}, \tag{3.15}$$

which is easily checked to be consistent with the relations between $\gamma_s$, $\gamma$, $c$, $Q$ given in (3.3), (3.9) and (3.10).

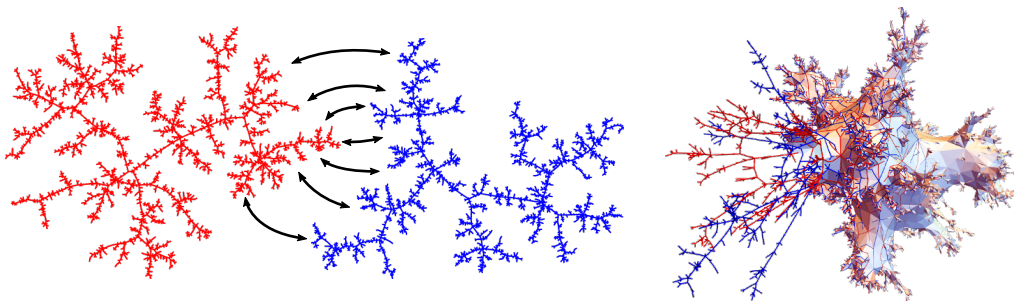

**Figure 3.6.:** Illustration of the mating of trees construction. The excursions $L(t)$ and $R(t)$ in the positive half line each encode a real tree (in red and blue respectively). Upon pairwise identification of points in their contours a topological sphere emerges. The right figure illustrates an intermediate state in which only part of the contour is identified.

To wrap up, mating of trees provides a procedure to go back and forth between a unit-area $\gamma$-quantum sphere together with a space-filling $\text{SLE}_{\kappa'}$ curve on one side and a pair or correlated continuum random trees encoded by a Brownian excursion in the quadrant on the other, and each side (almost surely) determines the other.

### 3.1.5. Mated-CRT maps

As should be clear from Figure 3.5b, the union of the flow lines $\eta_{z_1}, \dots, \eta_{z_k}$ of points $z_1, \dots, z_k \in \mathbb{C}$ has the structure of a tree spanning $z_1, \dots, z_k$ and $\infty$. If we increase the number $k$ of points this tree approaches a tree that spans the whole sphere, and the space-filling curve $\eta_g$ can be understood as tracing the contour of this tree. The quantum length measure $\nu_\phi$ assigns a metric structure to the tree, and one can interpret the process $R_t$ as the distance in the tree between $\eta_g(t)$ and $\infty$. Similarly, $L_t$ is the distance to $\infty$ along a complementary tree, which informally one can think of as what is left of the surface after the first tree is removed.

The reconstruction of the quantum sphere from the two-dimensional Brownian excursion can also be understood from the perspective of the trees. Both coordinates $L(t)$ and $R(t)$ describe a continuous excursion in the positive real line starting and ending at 0. As we stated in Section 2.5.1, any such excursions $X : [0, 1] \longrightarrow \mathbb{R}_{\geq 0}$ naturally gives rise to a continuous metric space: the real tree, given by the unit interval $[0, 1]$ with metric

$$d(s, t) = X(s) + X(t) - 2 \inf_{u \in [s,t]} X(u), \tag{3.16}$$

where we identify $s$ and $t$ whenever $d(s, t) = 0$. If $L(t)$ and $R(t)$ are uncorrelated, which happens for $\gamma = \sqrt{2}$, each is an independent Brownian excursion on the line and the corresponding real tree is the CRT. In general, they encode a pair of correlated random trees very similar to the CRT. It is relatively straightforward (see [63, Section 1.3]) to see that pairwise identification of points in the contours of the two trees leads to a space that is (almost surely) topologically equivalent to the 2-sphere. See Figure 3.6 for an illustration. What is not at all obvious is that the result has a natural conformal structure, let alone a natural metric. A convenient way to see that it does is by considering successively finer discretizations of the surface as follows.

Consider an excursion $X : [0, 1] \longrightarrow \mathbb{R}_{\geq 0}$ in the positive real line such that $X(0) = X(1) = 0$. For any positive integer $n$ one may associate to $X$ a triangulation of the $n$-sided polygon with vertices labeled from 1 to $n$ as follows [83]. We divide the interval $[0, 1]$ into $n$ equal parts $[0, \frac{1}{n}], [\frac{1}{n}, \frac{2}{n}], \dots, [1 - \frac{1}{n}, 1]$, one for each vertex. For any $1 \leq x < y \leq n$ such that the vertices with labels $x$ and $y$ are not neighbours, we draw a diagonal connecting these vertices if there is a horizontal segment below the graph of $X$ connecting the intervals $[\frac{x-1}{n}, \frac{x}{n}]$ and $[\frac{y-1}{n}, \frac{y}{n}]$ in the graph, i.e. if there is an $s \in [\frac{x-1}{n}, \frac{x}{n}]$ and a $t \in [\frac{y-1}{n}, \frac{y}{n}]$ such that $X(s) = X(t)$ and $X(u) \geq X(s)$ for all $u \in [s, t]$. For generic $X$, for instance when $X$ is a Brownian excursion, the result is a triangulation.

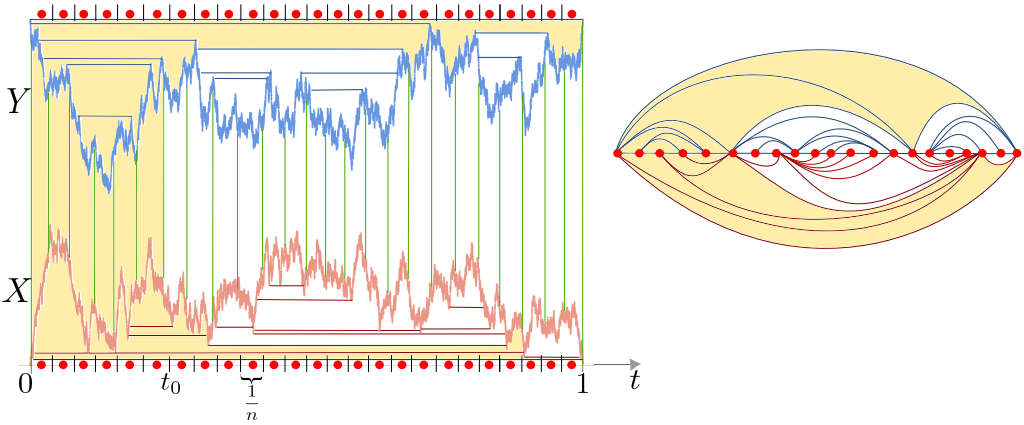

**Figure 3.7.:** **Left**: The components $(X, Y)$ of a 2D Brownian excursion are drawn with $Y$ drawn upside down for illustration purposes. The interval $[0, 1]$ is divided into $n$ equal parts and each of them corresponds to a vertex. The horizontal segments that lead to edges between vertices are indicated. **Right**: The triangulation resulting from gluing the pair of triangulated $n$-gons (the one associated to $Y$ in blue on top and the one associated to $X$ on the bottom). Note that the top and bottom arc are yet to be identified. The shaded region represents the region explored by the space-filling curve up to time $t_0$.

In the case of a two-dimensional Brownian excursion $(L(t), R(t))$ we thus naturally obtain a pair of triangulated polygons by applying the construction to both coordinates. Gluing these two polygons together produces a triangulation of the 2-sphere with $2n - 4$ triangles, called the *Mated-CRT map* [83]. It is naturally equipped with a Hamiltonian cycle, i.e. a simple closed path on the triangulation that visits all vertices (in this case in order of their labels $1, \ldots, n$), and rooted on the edge that connects the vertices with label $1$ and $n$. See Figure 3.7 for an example. This random triangulation is very much analogous to the discrete models in sections 3.1.1 and 3.1.2, except that it exists for any of the universality classes parametrized by $\gamma \in (0, 2)$. For any $n$ we obtain a unit-area Riemannian metric $g_{ab}^{(n)}$ on the 2-sphere by interpreting each triangle as an equilateral Euclidean triangle of area $1/(2n - 4)$. Informally, the metric associated to the mating of trees should be obtained as the large-$n$ limit of these metrics $g_{ab}^{(n)}$. More precisely, one can work with the canonical Tutte embedding of the triangulation in the sphere and it is shown in [83, Theorem 1.1 and Remark 3.7] that the corresponding measure $\sqrt{g^{(n)}}\mathrm{d}^2 z$ converges (in a weak sense) to $\mu_\phi$ as $n \rightarrow \infty$ and the Hamiltonian cycle to the space-filling curve $\eta_g$.

So far we have not discussed geodesic distances in the quantum sphere. Naively, one would expect to find a metric structure, i.e. a distance between $x, y \in \hat{\mathbb{C}}$, by minimizing the quantum length $\nu_\phi(\Gamma([0, 1]))$ of a curve $\Gamma$ from $x$ to $y$. But due to the fractal nature of the geometry, this limit is identically zero. Instead one should consider a different regularization [81], namely there exists a deterministic positive real number $d_\gamma > 2$ depending only on $\gamma$ such that the regularized distance

$$D_\epsilon(x, y) = \frac{\inf_\Gamma \int_\Gamma e^{\frac{\gamma}{d_\gamma} \phi_\epsilon(z)} (\hat{g})^{\frac{1}{2d_\gamma}} |\mathrm{d}z|}{\left( \int_{\hat{\mathbb{C}}} e^{\gamma \phi_\epsilon(z)} \sqrt{\hat{g}} \, \mathrm{d}^2 z \right)^{1/d_\gamma}} \tag{3.17}$$

possesses a well-defined limiting metric $D_\phi(x, y)$ (in probability) as $\epsilon \rightarrow 0$ when appropriately rescaled (by a factor of order $\epsilon^{-1+\frac{2}{d_\gamma}}$ in $\epsilon$). The value $d_\gamma$ is precisely the Hausdorff dimension of this metric [85], which informally is saying that the $\mu_\phi$-quantum area of a geodesic ball of radius $r$ around any point is of order $r^{d_\gamma}$ when $r \rightarrow 0$. The exact value of $d_\gamma$ is only known for $\gamma = \sqrt{8/3}$, corresponding to the pure gravity universality class, where $d_{\sqrt{8/3}} = 4$. For $\gamma \neq \sqrt{8/3}$, rigorous bounds are known [60, 13] as well as numerical estimates [21]. Moreover, as $\gamma \rightarrow 0$ the dimension $d_\gamma$ approaches 2 (see [59] for bounds on the convergence rate) in accordance with the constant curvature solution $g_{ab}$ to the classical Liouville action at $\gamma = 0$.

It is widely expected that this random metric $D_\phi(x, y)$ agrees with the large-$n$ limit of the graph distance within the $n$-vertex Mated-CRT map when normalized by $n^{-1/d_\gamma}$, and also with the geodesic distance as measured by $g_{ab}^{(n)}$ with the same normalization. A proof is still out of reach, but it is known [60, Theorem 1.6] that the number of vertices in a ball of radius $r$ around a randomly chosen vertex in the limit $n \rightarrow \infty$ grows like $r^{d_\gamma}$ with increasing radius $r$. Therefore the simple model of Mated-CRT maps can be used to estimate the Hausdorff dimension of Liouville Quantum Gravity for any $\gamma \in (0, 2)$. We will pursue this avenue in the next section.

### 3.1.6. Baby universes

The string susceptibility $\gamma_s$ can also be interpreted at the level of the Mated-CRT maps in terms of the distribution of sizes of *minimal-neck baby universes* (minbus) within the geometry. Since the Mated-CRT map is a loopless triangulation, the minimal length of a simple closed cycle is two. We let a *minbu of size $k$* for $2 \leq k \leq n-2$ be a connected region of $2k-2$ triangles not containing the root edge that is separated by a cycle of length two from the remaining $2(n-k)-2$ triangles. The string susceptibility is often introduced [92] as the exponent featuring in the expected number $E_{n,k}$ of minbus of size $k$,

$$\lim_{n \to \infty} \frac{E_{n,k}}{n} = C\, k^{\gamma_s - 2} + o(k^{\gamma_s - 2}) \qquad \text{as} \qquad k \to \infty. \tag{3.18}$$

Let us verify that this definition agrees with the relation (3.15).

Note from the construction of the Mated-CRT map that a minbu of size $k$ is associated to any $x = 1, \ldots, n-k$ for which both triangulated polygons have a diagonal connecting $x$ to $x+k$. Therefore $E_{n,k}/n$ is the probability of this event when $x$ is sampled uniformly. Denoting $X(t) = (X_1(t), X_2(t)) = (L(t), R(t))$, this happens precisely when

$$\min_{u \in [\frac{x}{n}, \frac{x+k-1}{n}]} X_i(u) > \max\left( \min_{s \in [\frac{x-1}{n}, \frac{x}{n}]} X_i(s), \min_{t \in [\frac{x+k-1}{n}, \frac{x+k}{n}]} X_i(t) \right) \qquad \text{for } i = 1, 2.$$

In the limit $n \to \infty$ the probability is the same as that for an unrestricted correlated two-dimensional Brownian motion $(\tilde{X}_1(t), \tilde{X}_2(t))$, such that

$$\lim_{n \to \infty} \frac{E_{n,k}}{n} = \mathbb{P}\left( \min_{u \in [x, x+k-1]} \tilde{X}_i(u) > \max\left( \min_{s \in [x-1, x]} \tilde{X}_i(s), \min_{t \in [x+k-1, x+k]} \tilde{X}_i(t) \right)_{\text{for } i=1,2} \right). \tag{3.19}$$

But this is essentially the probability that a two-dimensional correlated Brownian motion started close to the origin, remains in the quadrant for time at least $k$ and is close to the origin again at time $k$. This can be estimated using the heat kernel of the Brownian motion [19, Lemma 1], and scales with $k$ as $k^{-1-\gamma^2/4}$ (see discussion about Brownian motion in the wedge below). We thus find

$$\lim_{n \to \infty} \frac{E_{n,k}}{n} = C\, k^{-1-\gamma^2/4} + o(k^{-1-\gamma^2/4}) \qquad \text{as } k \to \infty. \tag{3.20}$$

This is clearly in agreement with (3.15) and (3.18).

# 3.2. Mated-CRT graphs from multi-dimensional Brownian excursions

## 3.2.1. Mated-CRT graphs

Now that we know how to read metric properties and the string susceptibility from the combinatorial data of a Mated-CRT map, let us introduce a natural generalization. Let $d = 2, 3, \ldots$ and $\mathbf{C}$ be a real positive-definite symmetric $d \times d$ matrix. Then we may consider $d$-dimensional Brownian motion $X(t) = (X_1(t), \ldots, X_d(t))$ started at the origin in $\mathbb{R}^d$ with covariance matrix $\mathrm{Cov}(X_i(t), X_j(t)) = \mathbf{C}_{ij}|t|$. A *Brownian excursion with covariance* $\mathbf{C}$ is then such a Brownian motion for $t \in [0, 1]$ that is conditioned to start and end at the origin and stay in the octant $\mathbb{R}^d_{>0}$ for $t \in (0, 1)$. We can associate to this Brownian excursion a random (multi-)graph $G_n^{\mathbf{C}}$ on $n$ vertices with a distinguished Hamiltonian cycle by gluing the $d$ triangulated $n$-gons associated to the $d$ excursions $X_1(t), \ldots, X_d(t)$ along their boundary (Figure 3.8). For $d = 2$ this graph is planar and $G_n^{\mathbf{C}}$ corresponds to the graph underlying the Mated-CRT map, while for $d \geq 3$ the graph is generally non-planar.

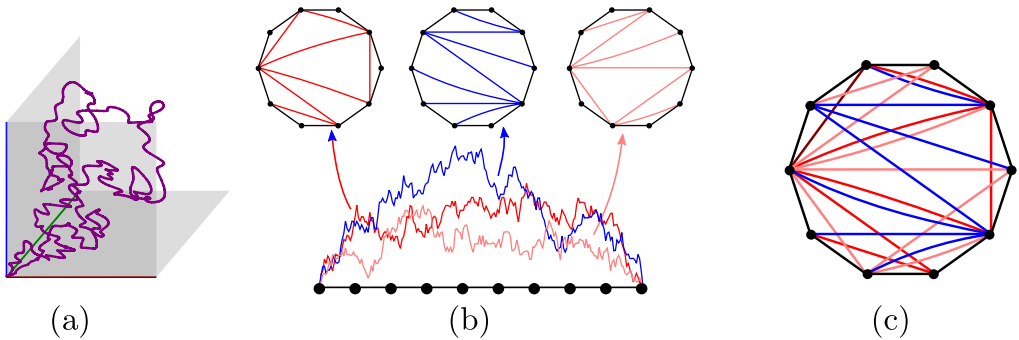

(a)             (b)             (c)

**Figure 3.8.:** (a) Illustration of a three-dimensional Brownian excursion. (b) To each of the $d = 3$ components of the excursion we may associate a triangulation of the $n$-gon. (c) The resulting Mated-CRT graph $G_n^{\mathbf{C}}$ with the Hamiltonian cycle appearing in black.

The central question is whether the graph $G_n^{\mathbf{C}}$, seen as a metric space induced by the graph distance, possesses a scaling limit, meaning that there exists some real number $d_H^{\mathbf{C}} > 0$ for which the rescaled metric space $n^{-1/d_H^{\mathbf{C}}} G_n^{\mathbf{C}}$ has a continuous limit as $n \to \infty$ (in a Gromov–Hausdorff sense). A posi-tive answer for $d \geq 3$ would give rise to new families of universality classes of random geometries, which based on the two-dimensional case one would expect to depend on the covariance matrix $\mathbf{C}$. Note that the construction of $G_n^{\mathbf{C}}$ is invariant under scaling of the coordinate axes, and its law therefore is invariant under coordinate-wise rescaling of the matrix $\mathbf{C}$. We may thus assume unit diagonal entries of $\mathbf{C}$ without loss of generality, and we are left with a $d(d-1)/2$-dimensional phase space of models. In $d = 2$ this 1-dimensional phase space is parametrized by the Liouville coupling $\gamma \in (0, 2)$.

As a first indication of a non-trivial scaling limit for $d \geq 2$, we will compute the string susceptibility of $G_n^{\mathbf{C}}$. The definition of a minbu (minimal-neck baby universe) is easily extended to $G_n^{\mathbf{C}}$: a minbu of size $k$ in $G_n^{\mathbf{C}}$ is a pair of vertices with label $x$ and $x + k$, such that removing $x$ and $x + k$ and all incident edges from $G_n^{\mathbf{C}}$ one is left with two connected components with $n - k - 1$ and $k - 1$ vertices respectively. Following the same reasoning as in the two-dimensional case the expected number $E_{n,k}^{\mathbf{C}}$ of minbus of size $k$ satisfies

$$\lim_{n \to \infty} \frac{E_{n,k}^{\mathbf{C}}}{n} = C k^{\gamma_s - 2} + o(k^{\gamma_s - 2}),$$

if the heat kernel $P_t^{\mathbf{C}}(x, y)$ of the $d$-dimensional Brownian motion with covariance matrix $\mathbf{C}$ on the octant with absorbing boundary conditions falls off like $P_t^{\mathbf{C}}(x, y) = c\, t^{\gamma_s - 2} + o(t^{\gamma_s - 2})$ as $t \to \infty$. Let us take a closer look at this process to see whether this is realized.

## 3.2.2. Brownian Excursions in a cone

Instead of dealing with correlated Brownian motion in the octant $\mathbb{R}_{\geq 0}^d$, it is often more convenient to work with uncorrelated Brownian motion in an appropriate cone $W \subset \mathbb{R}^d$. If $\mathbf{C}$ is a positive-definite symmetric matrix, then we can find an invertible real $d \times d$ matrix $R$ such that $\mathbf{C} = RR^T$ (in fact $R$ may be taken to be lower-triangular with positive entries on the diagonal, in which case it is called the Cholesky decomposition of $\mathbf{C}$). Let $W = R^{-1} \mathbb{R}_{\geq 0}^d$ be the preimage of the octant by the linear map $R$. Then the standard $d$-dimensional Brownian motion in the cone $W$ is mapped by $R$ to a Brownian motion with covariance matrix $\mathbf{C}$ in the octant.

The corresponding heat kernel $P_t^{\mathbf{C}}(x, y) \mathrm{d}^d y$ measures the probability density that a standard Brownian motion started at $x \in W$ remains within $W$ for at least time $t$ and is located at $y \in W$ at time $t$. By separation of radial and an-gular motion, it can be explicitly expressed in terms of the orthonormal eigenmodes of the spherical Laplace-Beltrami operator $L_{\mathbb{S}^{d-1}}$ on the spherical region $W \cap \mathbb{S}^{d-1} \subset \mathbb{R}^d$ with Dirichlet boundary conditions,

$$
\begin{cases}
L_{\mathbb{S}^{d-1}} m_i(\tilde{x}) = -\lambda_i m_i(\tilde{x}) & \text{for } \tilde{x} \in W \cap \mathbb{S}^{d-1}, \\
m_i(\tilde{x}) = 0 & \text{for } \tilde{x} \in \partial W \cap \mathbb{S}^{d-1}.
\end{cases}
\tag{3.21}
$$

Namely [19, Lemma 1]

$$
P_t^{\mathrm{C}}(x, y) = \frac{e^{-\frac{|x|^2 + |y|^2}{2t}}}{|x|^{\frac{d}{2}-1} |y|^{\frac{d}{2}-1}} \sum_{j=1}^{\infty} \frac{1}{t} I_{\alpha_j} \left( \frac{|x||y|}{t} \right) m_j \left( \frac{x}{|x|} \right) m_j \left( \frac{y}{|y|} \right),
\tag{3.22}
$$

where $I_\alpha(r)$ is a modified Bessel function satisfying $I_\alpha(r) \sim r^\alpha$ as $r \to 0$ and

$$
\alpha_j = \sqrt{\lambda_j + \left( \frac{d}{2} - 1 \right)^2}.
\tag{3.23}
$$

It follows that for fixed $x, y \in W$,

$$
P_t^{\mathrm{C}}(x, y) = c \, t^{-\alpha_1 - 1} + o(t^{-\alpha_1 - 1})
\tag{3.24}
$$

as $t \to \infty$, where the exponent depends on the fundamental eigenvalue $\lambda_1$ of $L_{\mathbb{S}^{d-1}}$ on $W \cap \mathbb{S}^{d-1}$. Hence the string susceptibility of the mated-CRT graph with covariance matrix $\mathbf{C}$ is

$$
\gamma_{\mathrm{s}} = 1 - \alpha_1 = 1 - \sqrt{\lambda_1 + \left( \frac{d}{2} - 1 \right)^2}.
\tag{3.25}
$$

In the two-dimensional case, the appropriate linear transformation $R$ is

$$
R = \begin{pmatrix} \sin \alpha & -\cos \alpha \\ 0 & 1 \end{pmatrix}, \qquad \mathbf{C} = RR^T = \begin{pmatrix} 1 & -\cos \alpha \\ -\cos \alpha & 1 \end{pmatrix}.
\tag{3.26}
$$

Then $R^{-1}\mathbb{R}_{\geq 0}^2$ is a cone of opening angle $\alpha$, with the right boundary ray along the positive x-axis (Figure 3.9). The corresponding fundamental eigenmode is $m_1(\theta) = \sin(\pi\theta/\alpha)$ with eigenvalue $\lambda_1 = \pi^2/\alpha^2$. We see that $\gamma_{\mathrm{s}} = 1 - \pi/\alpha = 1 - 4/\gamma^2$ is consistent with (3.15).

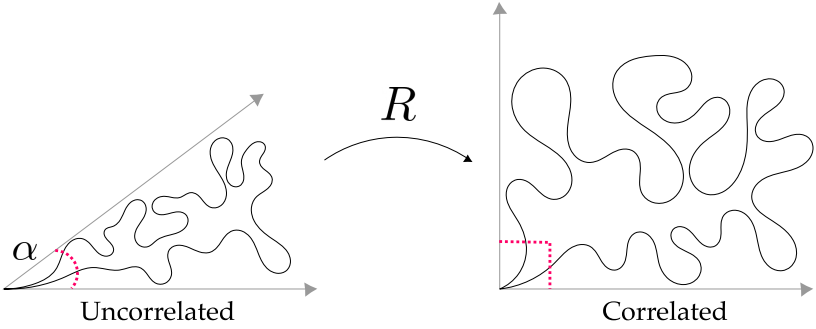

**Figure 3.9.:** An uncorrelated BE in a cone of opening angle $\alpha$ is mapped to a Correlated BE in $\mathbb{R}^2_{\geq 0}$ by (3.26).

In the three-dimensional case the most general positive-definite symmetric matrix $\mathbf{C}$ with unit diagonal entries is given by [29]

$$\mathbf{C} = \begin{pmatrix} 1 & -\cos(\alpha) & -\cos(\gamma) \\ -\cos(\alpha) & 1 & -\cos(\beta) \\ -\cos(\gamma) & -\cos(\beta) & 1 \end{pmatrix} \quad \text{with } \alpha, \beta, \gamma \in (0, \pi), \ \alpha + \beta + \gamma > \pi. \tag{3.27}$$

The corresponding cone $W_{\alpha,\beta,\gamma} = R^{-1}\mathbb{R}^3_{\geq 0}$ intersects $\mathbb{S}^2$ in a spherical triangle $T_{\alpha,\beta,\gamma}$ with angles $\alpha$, $\beta$ and $\gamma$ (Figure 3.10) and its corresponding 3-dimensional phase space is the interior of a rounded tetrahedron (Figure 3.11). The eigenvalue $\lambda_1$ of the fundamental eigenmode $m_1$ of $T_{\alpha,\beta,\gamma}$ is only known for special values of the angles. For example, for birectangular spherical triangles ($\beta = \gamma = \pi/2$) the fundamental eigenvalue is known to be [142]

$$\lambda_1 = \left(1 + \frac{\pi}{\alpha}\right)\left(2 + \frac{\pi}{\alpha}\right), \tag{3.28}$$

corresponding to a string susceptibility of $\gamma_s = -\frac{1}{2} - \frac{\pi}{\alpha}$. On the other hand, for spherical triangles with very small area $\alpha + \beta + \gamma - \pi$ the fundamental eigenvalue is well approximated by that of the Laplacian on a Euclidean triangle with angles $\alpha, \beta, \pi - \alpha - \beta$ and area $\alpha + \beta + \gamma - \pi$. Denoting the fundamental eigenvalue of the unit-area Euclidean triangle with these angles by $\lambda_{\text{Eucl}}(\alpha, \beta)$, we thus have

$$\lambda_1 = \frac{\lambda_{\text{Eucl}}(\alpha, \beta)}{\alpha + \beta + \gamma - \pi} + o(1) \qquad \text{as } \gamma \to \pi - \alpha - \beta. \tag{3.29}$$

In the case of the equilateral triangle, it is a classical computation that $\lambda_{\text{Eucl}}(\frac{\pi}{3}, \frac{\pi}{3}) = \frac{4}{\sqrt{3}}\pi^2$. Hence, for the equilateral case we find the string susceptibility

$$\gamma_s = -\frac{2\pi}{3^{3/4}\sqrt{\alpha - \frac{\pi}{3}}} + 1 + o\left(\sqrt{\alpha - \frac{\pi}{3}}\right). \qquad (\alpha = \beta = \gamma) \qquad (3.30)$$

We see that $\gamma_s \to -\infty$ as $\alpha \to \frac{\pi}{3}$ analogous to the behaviour of the string susceptibility in 2D when the cone becomes very narrow. The maximum is reached near the maximal area region in both cases. For more general regions we use the Finite Element Method (FEM) to determine the solutions numerically[3]. See Figure 3.11.

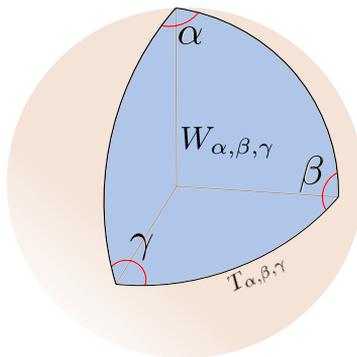

**Figure 3.10.:** The cone $W_{\alpha,\beta,\gamma}$ corresponds to the solid spherical region delimited by the spherical triangle $T_{\alpha,\beta,\gamma}$ in the unit sphere.

---

[3]Recently, more precise methods to compute these fundamental eigenvalues have been developed in [51].

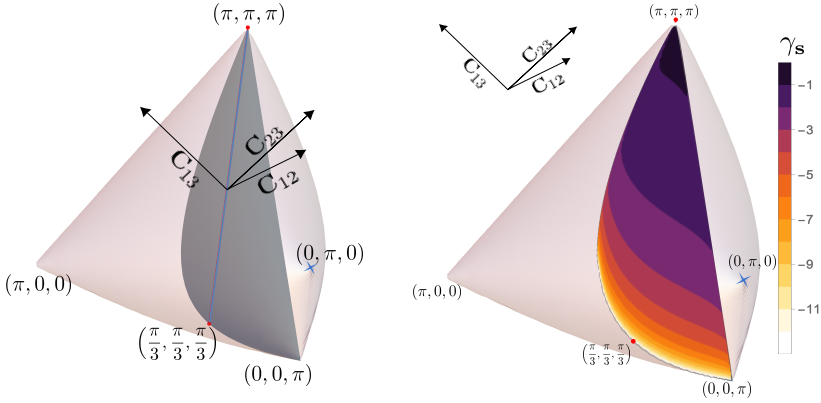

**Figure 3.11.:** Left: The phase space of covariance matrices $\mathbf{C}$, parametrized by $\mathbf{C}_{12} = -\cos(\alpha)$, $\mathbf{C}_{23} = -\cos(\beta)$, $\mathbf{C}_{13} = -\cos(\gamma)$, spans the "tetrahedral" region $\mathbf{C}_{12}^2 + \mathbf{C}_{13}^2 + \mathbf{C}_{23}^2 - 2\mathbf{C}_{12}\mathbf{C}_{13}\mathbf{C}_{23} < 1$. The corners are labeled with their spherical angles $(\alpha, \beta, \gamma)$. The blue diagonal corresponds to equilateral spherical triangles ($\alpha = \beta = \gamma$), while the gray plane indicates the isosceles spherical triangles ($\beta = \gamma$). The red dots are the smallest $(\frac{\pi}{3}, \frac{\pi}{3}, \frac{\pi}{3})$ and largest $(\pi, \pi, \pi)$ equilateral triangles, which will be of particular interest in the results. Right: estimates of the string susceptibility $\gamma_s$ for isosceles triangles obtained from Finite Element Methods.

### 3.2.3. A simpler biased Brownian excursion

Unit-time Brownian excursions in a non-trivial cone $W$ are challenging objects to simulate efficiently. For this reason, we introduce a slightly biased version of the Brownian excursion, which is easier to simulate. As we will see in a minute, for a unit-time Brownian excursion the integral $\int_0^1 |X(t)|^{2\alpha_1-2} dt$ has finite expectation value $C_{\alpha_1} = 2^{\alpha_1-1}\Gamma(\alpha_1)/\alpha_1 > 0$. We may thus introduce the Brownian excursion $\hat{X}(t)$ obtained from $X(t)$ by biasing its law by the value of this integral, meaning that the probability measure of $\hat{X}(t)$ is that of $X(t)$ multiplied by $\int_0^1 |X(t)|^{2\alpha_1-2} dt/C_{\alpha_1}$. In probabilistic terms, the new random excursion $\hat{X}(t)$ is absolutely continuous with respect to the unit-time Brownian excursion and therefore displays the same critical exponents. In particular, the Mated-CRT graph $\hat{G}_n^C$ associated to the excursion $\hat{X}(t)$ will display the same local geometry as the unbiased Mated-CRT graph $G_n^C$ when $n \to \infty$, and therefore agree on the Hausdorff dimension $d_\gamma$ (if it exists) and the string susceptibility $\gamma_s$.

This is quite useful, because we claim that $\hat{X}(t)$ can be more easily sampled than $X(t)$. Let $S$ be a random point chosen from the spherical region $W \cap \mathbb{S}^{d-1}$ with probability distribution $m_1(x)^2$ (recall that this eigenmode is assumed to be normalized and thus $m_1(x)^2$ integrates to one on $W \cap \mathbb{S}^{d-1}$). Independently, we sample two independent Brownian motions $X^1(t)$ and $X^2(t)$ started at $S$ and conditioned to touch the boundary $\partial W$ at the origin. If $T_1$ and $T_2$ are the hitting times of the origin, then we make the identification

$$\hat{X}(t) = \frac{1}{\sqrt{T_1 + T_2}} \begin{cases} X^1(T_1 - t(T_1 + T_2)) & 0 \le t \le \frac{T_1}{T_1 + T_2} \\ X^2(t(T_1 + T_2) - T_1) & 1 \ge t \ge \frac{T_1}{T_1 + T_2} \end{cases}. \tag{3.31}$$

In words, we concatenate the reversal of the first curve with the second to produce an excursion from the origin, which is then rescaled to have unit duration (taking into account the usual Brownian scaling relations).

To understand why this works, let us first compute the distribution of the hitting time $T_1$ (which is the same as that of $T_2$). From the $y \to 0$ limit of the heat kernel (3.22) with $|x| = 1$,

$$P_t^C(x, y) \overset{y \to 0}{\sim} t^{-\alpha_1 - 1} e^{-\frac{1}{2t}} |y|^{\alpha_1 + 1 - \frac{d}{2}} m_1\left(\frac{y}{|y|}\right) m_1(x), \qquad (|x| = 1)$$

it follows that the hitting time $T_1$ is independent of the starting position $S$ and distributed as an *inverse gamma distribution* with index $\alpha_1$ and scale $1/2$, i.e. has density

$$\frac{1}{2^{\alpha_1} \Gamma(\alpha_1)} t^{-\alpha_1 - 1} e^{-\frac{1}{2t}} \mathrm{d}t \qquad \text{on} \quad \mathbb{R}_{>0}. \tag{3.32}$$

Next, we determine the density of the Brownian excursion $X(t)$ in $W$ at a fixed time $s \in (0, 1)$, which is obtained from the heat kernel (3.22) via the limit

$$\lim_{y_1, y_2 \to 0} \frac{P_s(y_1, x) P_{1-s}(x, y_2)}{P_1(y_1, y_2)} = \frac{|x|^{2\alpha_1 + 1}}{s^{\alpha_1 + 1}(1 - s)^{\alpha_1 + 1}} e^{-\frac{|x|^2}{2}\left(\frac{1}{s} + \frac{1}{1-s}\right)} |x|^{1 - |d|} m_1\left(\frac{x}{|x|}\right)^2. \tag{3.33}$$

It follows that $X(s)/|X(s)|$ for any $s$ is distributed like the point $S$ above and that the distance $|X(s)|$ to the origin has probability density

$$\rho_s(r)\mathrm{d}r = \frac{1}{s^{\alpha_1 + 1}(1 - s)^{\alpha_1 + 1}} \frac{r^{2\alpha_1 + 1}}{2^{\alpha} \Gamma(\alpha + 1)} e^{-\frac{r^2}{2}\left(\frac{1}{s} + \frac{1}{1-s}\right)} \mathrm{d}r. \tag{3.34}$$

Integrating this expression against $r^{2\alpha_1-2}$ yields the previously claimed expectation value

$$C_{\alpha_1} := \mathbb{E}\left[\int_0^1 |X(t)|^{2\alpha_1-2}\right] = \int_0^1 ds \int_0^\infty dr \, r^{2\alpha_1-2}\rho_s(r) = 2^{\alpha_1-1}\frac{\Gamma(\alpha_1)}{\alpha_1}. \quad (3.35)$$

Suppose now that $\hat{X}(t)$ is the biased Brownian excursion and, conditionally on $\hat{X}(t)$, let $U \in [0,1]$ be a random variable sampled with density proportional to $|\hat{X}(u)|^{2\alpha_1-2}du$. Then we let

$$T_1 = \frac{U}{|\hat{X}(U)|^2}, \qquad T_2 = \frac{1-U}{|\hat{X}(U)|^2}, \qquad S = \frac{\hat{X}(U)}{|\hat{X}(U)|}. \quad (3.36)$$

We will demonstrate that $T_1$, $T_2$ and $S$ are independent and $T_1$ and $T_2$ are distributed precisely as the inverse gamma distribution mentioned above.

From (3.34) it follows that the joint distribution of the pair $(|\hat{X}(U)|, U)$ has probability density

$$\frac{1}{C_{\alpha_1}}\rho_u(r)r^{2\alpha_1-2}drdu \qquad \text{on } \mathbb{R}_{>0} \times (0,1). \quad (3.37)$$

Since $T_1$ and $T_2$ are bijectively related to $|\hat{X}(U)|$ and $U$ via (3.36), the joint density of the pair $(T_1, T_2) \in \mathbb{R}_{>0}^2$ is obtained from this by the transformation $r = 1/\sqrt{\tau_1 + \tau_2}$, $u = \tau_1/(\tau_1 + \tau_2)$, with Jacobian $2r^{-5}drdu = d\tau_1 d\tau_2$, which yields

$$\frac{1}{C_{\alpha_1}}\rho_u(r)r^{2\alpha_1-2}dsdr = \frac{1}{(2^{\alpha_1}\Gamma(\alpha_1))^2}\tau_1^{-\alpha_1-1}e^{-\frac{1}{2\tau_1}}\tau_2^{-\alpha_1-1}e^{-\frac{1}{2\tau_2}}d\tau_1 d\tau_2. \quad (3.38)$$

Comparing with (3.32), we observe that $T_1$ and $T_2$ are independent and distributed with the desired inverse gamma distribution.

Finally, conditionally on $T_1$, $T_2$ and $S$, the curves $\hat{X}(U-t)$ and $\hat{X}(U+t)$ are independent $d$-dimensional Brownian motions both started at $\hat{X}(U) = S/\sqrt{T_1+T_2}$ and conditioned to hit to the origin after time $U = T_1/(T_1+T_2)$ and $1-U = T_2/(T_1+T_2)$ respectively. By the scale invariance of the Brownian motion the curves

$$X^1(t) = \sqrt{T_1+T_2}\hat{X}\left(\frac{T_1-t}{T_1+T_2}\right), \qquad X^2(t) = \sqrt{T_1+T_2}\hat{X}\left(\frac{T_1+t}{T_1+T_2}\right) \quad (3.39)$$

are distributed as independent $d$-dimensional Brownian motions started at $S$ and conditioned to hit the origin after time $T_1$ and $T_2$ respectively. But as we computed above, $T_1$ and $T_2$ have precisely the distribution of the hitting time of the origin of a $d$-dimensional Brownian motion started at $S$ and conditioned to tough the boundary $\partial W$ at the origin, so we may lift the latter conditioning. Since this is precisely the inverse of (3.31), we have proven that the identity (3.31) holds for the law of $\hat{X}(t)$.

### 3.2.4. Brownian motion conditioned to hit the origin

We have seen that the biased excursion $\hat{X}(t)$, and therefore also the biased Mated-CRT graph $\hat{G}_n^C$, can be constructed from a pair of $d$-dimensional Brownian motions that are conditioned to hit the boundary $\partial W$ at the origin. Let us discuss these processes in a bit more detail. An important role is played by the harmonic function

$$h(x) = |x|^{-\alpha_1 - \frac{d}{2} + 1} m_1 \left( \frac{x}{|x|} \right). \tag{3.40}$$

It can be recovered from the heat kernel by first computing the $t$-integral

$$\int_0^\infty P_t^C(x, y) \mathrm{d}t = \sum_{j=1}^\infty \frac{1}{\alpha_j} |x|^{1 - \frac{d}{2} - \alpha_j} |y|^{1 - \frac{d}{2} + \alpha_j} m_j \left( \frac{x}{|x|} \right) m_j \left( \frac{y}{|y|} \right) \quad \text{for } |x| > |y|. \tag{3.41}$$

As $y$ tends to zero we thus have

$$\int_0^\infty P_t^C(x, y) \mathrm{d}t \overset{y \to 0}{\sim} h(x) \, h(y) \frac{|y|^{2\alpha_1}}{\alpha_1}, \tag{3.42}$$

which estimates the probability that an unrestricted $d$-dimensional Brownian motion started at $x$ leaves the cone $W$ at a point close to the origin. Even though the Brownian motion has vanishing probability of hitting $\partial W$ at the origin, we may still condition on this event by a so-called Doob's $h$-transform [62] of a standard Brownian motion with respect to this harmonic function. Without diving into the theory of $h$-transforms, we can characterize the Brownian motion $X^i(t)$ conditioned to hit $\partial W$ at the origin as follows. If $A \subset W$ is a closed neighbourhood of $x_0 = X^i(0)$ and we consider the exit time $\tau$ when $X^i(t)$ leaves $A$, then the distribution of the exit point $X^i(\tau)$ is related to that of a standard Brownian motion started at $x_0$ by a factor $h(X^i(\tau))/h(x_0)$.

This characterization gives a simple iterative procedure of constructing $X^i(t)$ started at $x_0 \in W$ from standard Brownian motion (see Figure 3.12). We take the subset $A$ to be the largest Euclidean ball centered at $x_0$ and contained in $W$ and let $x_1$ be a random variable on the sphere $\partial A$ with probability density $h(x)/h(x_0)$. We may then consider a standard Brownian motion started at $x_0$ until it hits the boundary of the ball at time $t_1$. By symmetry this happens at a uniform point $\tilde{x}_1$ on $\partial A$. A $d$-dimensional rotation around $x_0$ that brings $\tilde{x}_1$ to $x_1$ then gives an appropriately sampled path for $X^i(t)$ for $t \in [0, t_1]$. Since $X^i(t)$ is a Markov process, we may iterate this procedure with the new starting point $X^i(t_1) = x_1$ to obtain the path for $X^i(t)$ with $t \in [t_1, t_2]$, etcetera. Of course, infinite iteration is required to reach the origin, but if one is only interested in the path until reaching some small distance $\epsilon > 0$ from the origin, then the number of required iterations can be seen to grow only logarithmically in $1/\epsilon$.

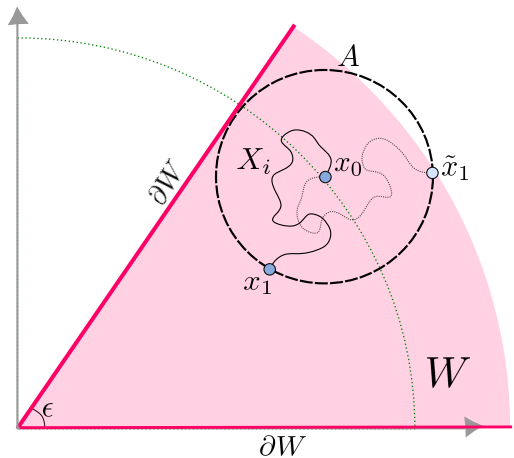

**Figure 3.12.:** The first step in the iterative procedure to produce $X^i(t)$ in the case of a two-dimensional cone $W$: $x_0$ is sampled from the unit circle $W \cap \mathbb{S}^1$ with density $m_1(x)^2 = h(x)^2$; a standard 2-dimensional Brownian motion is run until it exits the disk $A$ at $\tilde{x}_1$, which is then rotated to end at the random position $x_1$ with distribution $h(x)/h(x_0)$. This procedure is to be repeated with a new disk centered at $x_1$.

# 3.3. Simulations and Hausdorff dimension estimates

## 3.3.1. Sampling Mated-CRT graphs numerically

In the previous section we have introduced the random Mated-CRT graph $\hat{G}_n^{\mathbb{C}}$ constructed from the biased Brownian excursion. Let us now turn to the numerical implementation of this construction.

Sampling $\hat{G}_n^{\mathbb{C}}$ with exactly the right probability distribution is challenging, as it relies on continuous Brownian motions. What helps is that the graph $\hat{G}_n^{\mathbb{C}}$ is determined by the ranges of $X^1(t)$ and $X^2(t)$ on intervals of length $(T_1 + T_2)/n$, where $T_1$ and $T_2$ are the time extents that we know to be inverse-gamma distributed. The probability of $T_1$ or $T_2$ being much shorter than their expectation value $1/(2\alpha - 2)$ is very small. Hence, to approximate $\hat{G}_n^{\mathbb{C}}$ well, it suffices to sample the Brownian motions at a time resolution $\epsilon$ that is significantly smaller than $\mathbb{E}[(T_1 + T_2)/n] = 1/(n(\alpha - 1))$. This we do by approximating the Brownian motion by a driftless random walk with increments that are sampled uniformly on the sphere of radius $\sqrt{\epsilon}$ in $\mathbb{R}^d$. In this case, we use $\sqrt{\epsilon} \in [0.0001, 0.001]$. The reason to opt for these increments instead of the potentially more accurate Gaussian increments is that the exit times and exit positions (e.g. from the cone $W$) are more easily controlled with bounded increments.

To be precise, for a desired correlation matrix $\mathbf{C}$ we compute the exponent $\alpha_1$ and fundamental mode $m_1$ (either analytically or numerically if an analytical solution is not available). Then to obtain a single sample of $\hat{G}_n^{\mathbb{C}}$ we perform the following procedure, based on the construction in Section 3.2.3. A random starting point $x_0$ with distribution $m_1(x)^2$ on $W \cap \mathbb{S}^{d-1}$ is chosen using rejection sampling. Two random piece-wise linear curves from $x_0$ to the origin are obtained by running the random walk in an iterative fashion as follows. We find the largest radius $r$ such that the ball $\text{Ball}_r(x_0)$ around $x_0$ is contained in $W$ and choose a point $x_1$ on its boundary with distribution $h(x)/h(x_0)$, again using rejection sampling. Next, we run the mentioned random walk with steps of size $\sqrt{\epsilon}$ until it leaves $\text{Ball}_r(x_0)$, denoting the exit point on the sphere by $\tilde{x}_1$. This random walk is rotated by an orthogonal transformation that only depends on $x_0$, $x_1$ and $\tilde{x}_1$ to produce a piece-wise linear path from $x_0$ to $x_1$ (note that the last segment of this path has to be shortened a bit to end precisely at $x_1$ instead of ending outside $\text{Ball}_r(x_0)$). We iterate this procedure, but now using $x_1$ as the starting point, which extends the piece-wise linear path from $x_1$ to a ran-dom point $x_2$ on the boundary of the largest ball $\text{Ball}_r(x_1)$ around $x_1$, and so on. This is continued until we reach a point within distance $2\sqrt{\epsilon}$ from the origin, after which we add a final segment connecting to the origin. The result is a piece-wise linear path from $x_0$ to the origin that stays strictly in the cone $W$ and approximates the law of the Brownian motion in $W$ conditioned to hit the origin. Concatenating the two paths according to (3.31) leads to a piece-wise linear excursion that approximates the biased unit-time Brownian excursion $\hat{X}(t)$. Finally, the mated-CRT graph of size $n$ is obtained as explained in Section 3.2.1, resulting in an adjacency matrix for the $n$ vertices of the graph.

## 3.3.2. Hausdorff dimension estimates via finite-size scaling

As explained in Section 3.2.1, the central question is whether the metric space induced by the graph distance on $\hat{G}_n^C$ possesses a scaling limit. Does there exist a real number $d_H^C > 0$, which we then call the Hausdorff dimension of the model, such that the metric space $n^{-1/d_H^C} \hat{G}_n^C$ has a limit as $n \longrightarrow \infty$ (in the Gromov–Hausdorff sense)? This statement about the limit is not something one can effectively verify numerically, but there is a necessary condition that is within numerical reach. Let $d_n$ be the graph distance between two uniformly sampled vertices in the random graph $\hat{G}_n^C$. Then for the existence of a sane Gromov–Hausdorff limit it is necessary that $d_n/n^{1/d_H^C}$ converges in distribution as $n \longrightarrow \infty$. Since $d_n$ is relatively easy to measure, this allows us to verify the convergence in distribution and at the same time estimate the value $d_H^C$ through finite-size scaling.

The probability distributions $\rho_n(r) = \mathbb{P}(d_n = r)$ for $r = 0, 1, 2, \ldots$ were estimated for $n = 2^{11}, 2^{12}, \ldots, 2^{19}$ as follows. For each size $n$, the graph $\hat{G}_n^C$ was sampled 80000 times and, in each sampled graph, the graph distances from a uniformly chosen vertex to all other vertices were determined several times. All these distances were stored in a histogram, which upon normalization, provides our best estimate for $\rho_n(r)$ with small statistical errors (see Figure 3.13).

For convenience we extend $\rho_n(r)$ to a continuous function of $r \in \mathbb{R}_{\geq 0}$ via linear interpolation. If $d_n/n^{1/d_H^C}$ converges in distribution to a random variable with density $\rho(x)$, we expect to have the limit

$$\lim_{n \to \infty} n^{1/d_H} \rho_n(n^{1/d_H} x) = \rho(x). \tag{3.43}$$

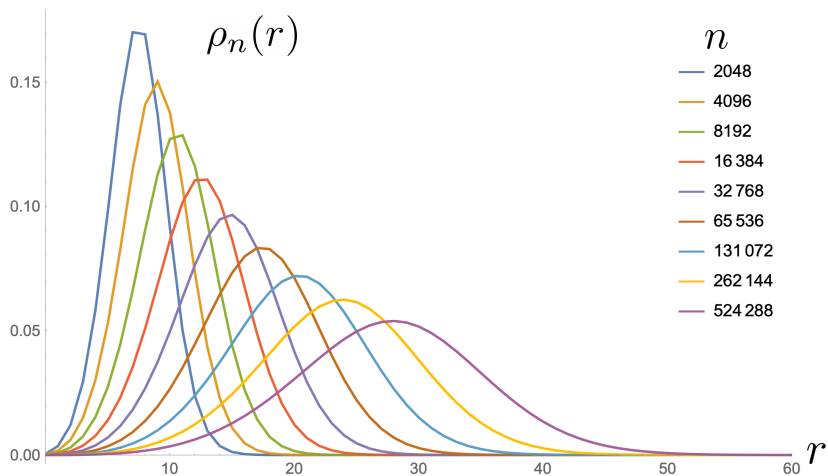

**Figure 3.13.:** Normalized histogram $\rho_n(r)$ for $\alpha = \beta = \gamma = \frac{\pi}{2}$, obtained by sampling $m = 10000$ graphs independently and measured distances $k = 10$ times from a uniformly random chosen vertex.

This is a mildly stronger assumption than what is implied by the Gromov–Hausdorff convergence, but one that is supported by our data. In order to study the limit (3.43), we choose as reference size $n_0 = 2^{19}$ and aim to collapse the curves $\rho_n$ for the other sizes $n$ to $\rho_{n_0}$. More precisely, for each $n = 2^{11}, \dots, 2^{18}$ we determine fit parameters $k_n$ and $s_n$ that minimize the integrated square deviation between $k_n^{-1}\rho_n(k_n^{-1}(x + s_n) - s_n)$ and $\rho_{n_0}$. The shift $s_n$ is included to compensate for discretization effects and is largely independent of $n$. By comparing this expression with (3.43), we see that $k_n \sim C n^{-1/d_H}$, i.e., finding the asymptotic behaviour of $k_n$ is the key to estimating $d_H$.

In order to find more accurate values for $d_H$, we collapse $\rho_n$ two times. In the first one $k_n^{-1}\rho_n(k_n^{-1}(x + s_n) - s_n) \longmapsto \rho_{n_0}$, we extract the values $s_n$ to compute its mean $s$. In the second one, we use $s$ to collapse $k_n^{-1}\rho_n(k_n^{-1}(x + s) - s) \longmapsto \rho_{n_0}$ and we extract $k_n$. Finally, we estimate $d_H$ by fitting $k_n$ to the ansatz

$$\left(\frac{n}{n_0}\right)^{-1/d_H}\left(a + b\left(\frac{n}{n_0}\right)^{-\delta}\right), \tag{3.44}$$

where $a \approx 1$, $\delta$ of order $1/d_H$ and $|b| \ll 1$. This expression takes into account a leading-order correction and has proven to work well for Hausdorff dimension estimations in a similar setting [21]. The fitting procedure was tested

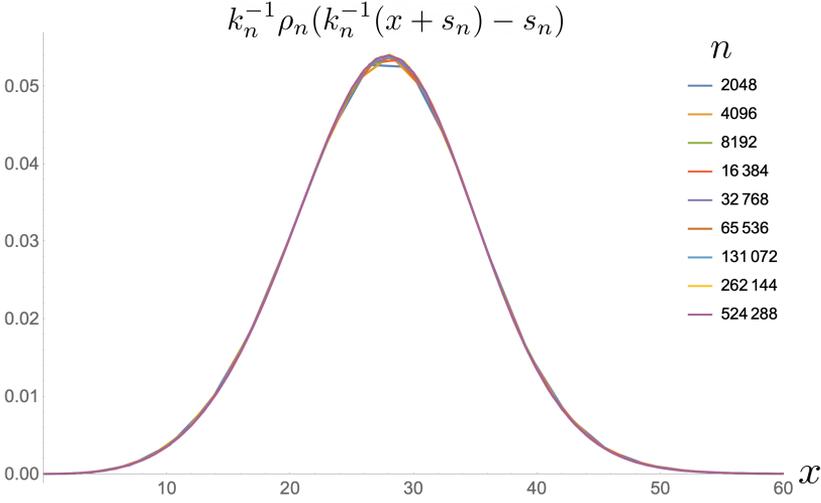

$$k_n^{-1}\rho_n(k_n^{-1}(x + s_n) - s_n)$$

**Figure 3.14.:** Collapsed histograms with optimal shift for $\alpha = \beta = \frac{\pi}{2}$ and $n_0 = 2^{19}$.

by varying the range of volumes included, as well as the values of $\delta$ and $d_H$ while keeping $a \approx 1$ and $|b| \ll 1$. In this way, we determine systematic errors. On the other hand, the statistical errors in the fit parameters were determined using batching (by dividing the data into eight independent batches and applying the analysis each independently). The results are presented in the next subsections.

### 3.3.3. Results for Mated-CRT maps ($d = 2$)

As explained in Section 3.1.5, the Gromov–Hausdorff convergence of the Mated-CRT maps has not been proved, but the number of vertices within a ball of radius $r$ in the graph $\hat{G}_n^C$ with very large $n$ is known to grow as $r^{d_\gamma}$ as $r \to \infty$, where $d_\gamma$ is the Hausdorff dimension of Liouville Quantum Gravity. Here the covariance $C_{12} = -\cos(\alpha)$ is related to $\gamma$ through the relation $\alpha = \pi\gamma^2/4$. This strongly suggests that the convergence (3.43) holds with Hausdorff dimension $d_H^C = d_\gamma$, and thus our methods provide a means of estimating the Hausdorff dimension $d_\gamma$ of Liouville Quantum Gravity. In Figure 3.15 and Table 3.1, we show the numerical values of $d_H$ as a function of $\gamma$.

A distinction between the regions $\gamma < 1$ and $\gamma \geq 1$ is made, since the latter is the domain analysed in [21]. The reason for choosing the four values $\gamma = 1, \sqrt{4/3}, \sqrt{2}, \sqrt{8/3}$ is that they are the values associated to the universality classes of Schnyder-wood-decorated triangulations, bipolar-oriented triangulations,
spanning-tree-decorated quadrangulations and uniform quadrangulations, respectively. For these models, discrete mating-of-trees bijections are available that are at the basis of the high-precision estimates of $d_H$ in [21]. They thus form a good benchmark for the techniques developed in this work. Our results in Table 3.1 are seen to be very well consistent with those in [21, Table 5], although the errors here are significantly larger.

Our current method has the advantage that it can be used to perform simulations at any $\gamma \in (0, 2)$, in particular in the region $\gamma < 1$ where very few numerical estimates for $d_H$ were known (see [21, Section 5] for estimates based on Liouville first-passage percolation). Gaining more accurate estimates for small $\gamma$ is important, because it is in this region that some proposed formulas for the Hausdorff dimension deviate substantially. Two such formulae are the one due to Watabiki [143] (shown in blue in Figure 3.15)

$$d^W = 1 + \frac{\gamma^2}{4} + \sqrt{\left(1 + \frac{\gamma^2}{4}\right)^2 + \gamma^2}, \tag{3.45}$$

and one by Ding and Gwynne [60] (shown in yellow in Figure 3.15)

$$d^{DG} = 2 + \frac{\gamma^2}{2} + \frac{\gamma}{\sqrt{6}}. \tag{3.46}$$

As can be seen in Figure 3.15, our new estimates strengthen the conclusion of [21] that Watabiki's formula is ruled out numerically (in addition to being already inconsistent with the $\gamma \to 0$ bounds in [59]). However, the measurements are still statistically consistent with Ding and Gwynne's formula.

| $\gamma$ | $d_H$ |
|---|---|
| 3/8 | 2.24 ± 0.01 |
| 1/2 | 2.35 ± 0.01 |
| 5/8 | 2.47 ± 0.01 |
| 3/4 | 2.60 ± 0.02 |
| 1 | 2.90 ± 0.04 |
| $\sqrt{4/3}$ | 3.13 ± 0.05 |
| $\sqrt{2}$ | 3.59 ± 0.07 |
| $\sqrt{8/3}$ | 4.07 ± 0.14 |

**Table 3.1.:** Our Hausdorff dimension estimates from simulated mated-CRT maps for different values of $\gamma$. The errors have been determined according to the procedure outlined at the end of Section 3.3.2.

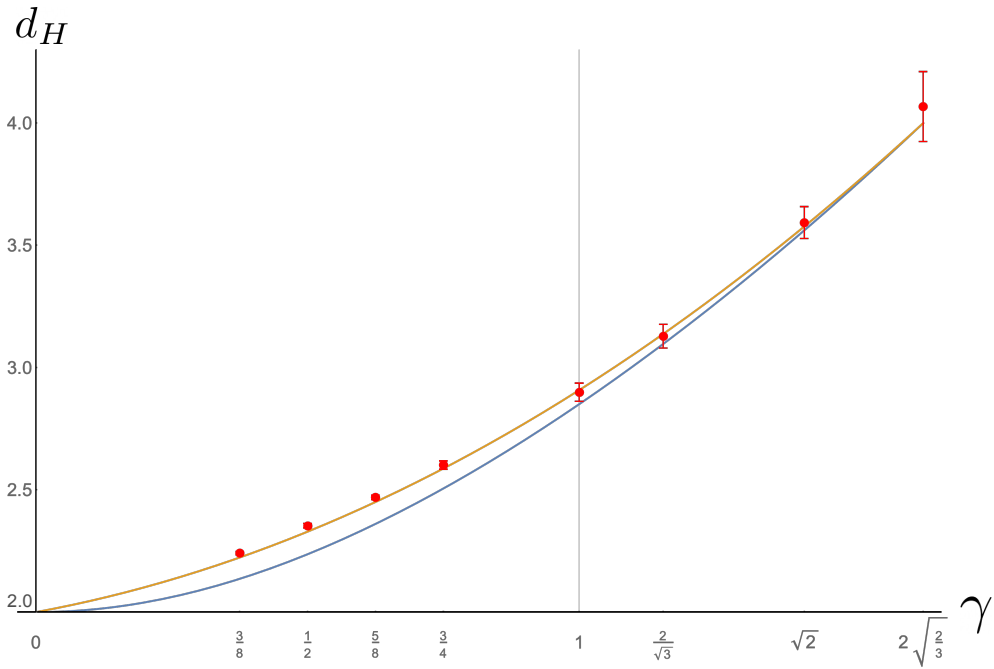

**Figure 3.15.:** Hausdorff dimension estimates from simulated mated-CRT maps for different values of $\gamma$. Watabiki's formula (3.45) is plotted in blue, while Ding and Gwynne's formula (3.46) is plotted in yellow.

### 3.3.4. Results for Mated-CRT graphs in $d$ = 3

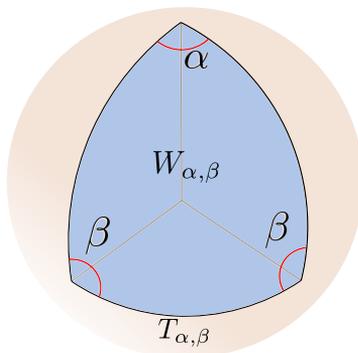

**Figure 3.16.:** The cone $W_{\alpha,\beta}$ is spanned by the isosceles spherical triangle $T_{\alpha,\beta}$ in the unit sphere.

Having benchmarked the numerical methods, we turn to the main numerical results of this work, the Hausdorff dimension estimates of Mated-CRT graphs in $d$ = 3. Since the phase diagram is significantly larger than in $d$ = 2, being three-dimensional instead of one-dimensional, we have chosen to restrict our attention to the two-dimensional subspace corresponding to isosceles spherical triangles (Figure 3.16) in which two of the angles are equal: $\gamma = \beta$ or, equivalently, $\mathbf{C}_{23} = \mathbf{C}_{13}$[4].

The estimates for the Hausdorff dimension $d_H^{\mathbf{C}}$ including error bars for a variety of angle pairs $(\alpha, \beta)$ are presented in Figure 3.17 and listed in Table 3.2. For convenience we record a reasonable fit for $d_H^{\mathbf{C}}$ using a quadratic ansatz in $\mathbf{C}$ that respects the symmetries,

$$d_H \approx 4.83 + 0.42(\mathbf{C}_{12} + \mathbf{C}_{23} + \mathbf{C}_{13}) + 0.37(\mathbf{C}_{12}^2 + \mathbf{C}_{23}^2 + \mathbf{C}_{13}^2) \qquad (3.47)$$

$$- 0.38(\mathbf{C}_{12}\mathbf{C}_{23} + \mathbf{C}_{13}\mathbf{C}_{23} + \mathbf{C}_{12}\mathbf{C}_{23}). \qquad (3.48)$$

Although we have only been able to effectively simulate a limited region of the full phase diagram, several conclusions can be drawn based on the data. First of all, the dependence of $d_H^{\mathbf{C}}$ on $\mathbf{C}$ differs qualitatively from that of the string susceptibility in Figure 3.11, suggesting that we are really dealing with a multi-parameter family of universality classes. Secondly, contrary to the two-dimensional case there appears to be no limit, at least in the isosceles region, where the Hausdorff dimension approaches a "classical" value equal to $d$ itself, in this case 3. Instead, we seem to observe a minimum $d_H \approx 4.1$ when $(\alpha, \beta, \gamma) \rightarrow (\pi/3, \pi/3, \pi/3)$ corresponding to a Brownian excursion in the cone spanned by a tiny equilateral spherical triangle.

---

[4]These results extend of course to the planes $\alpha = \gamma$ and $\alpha = \beta$ due to rotational symmetry.

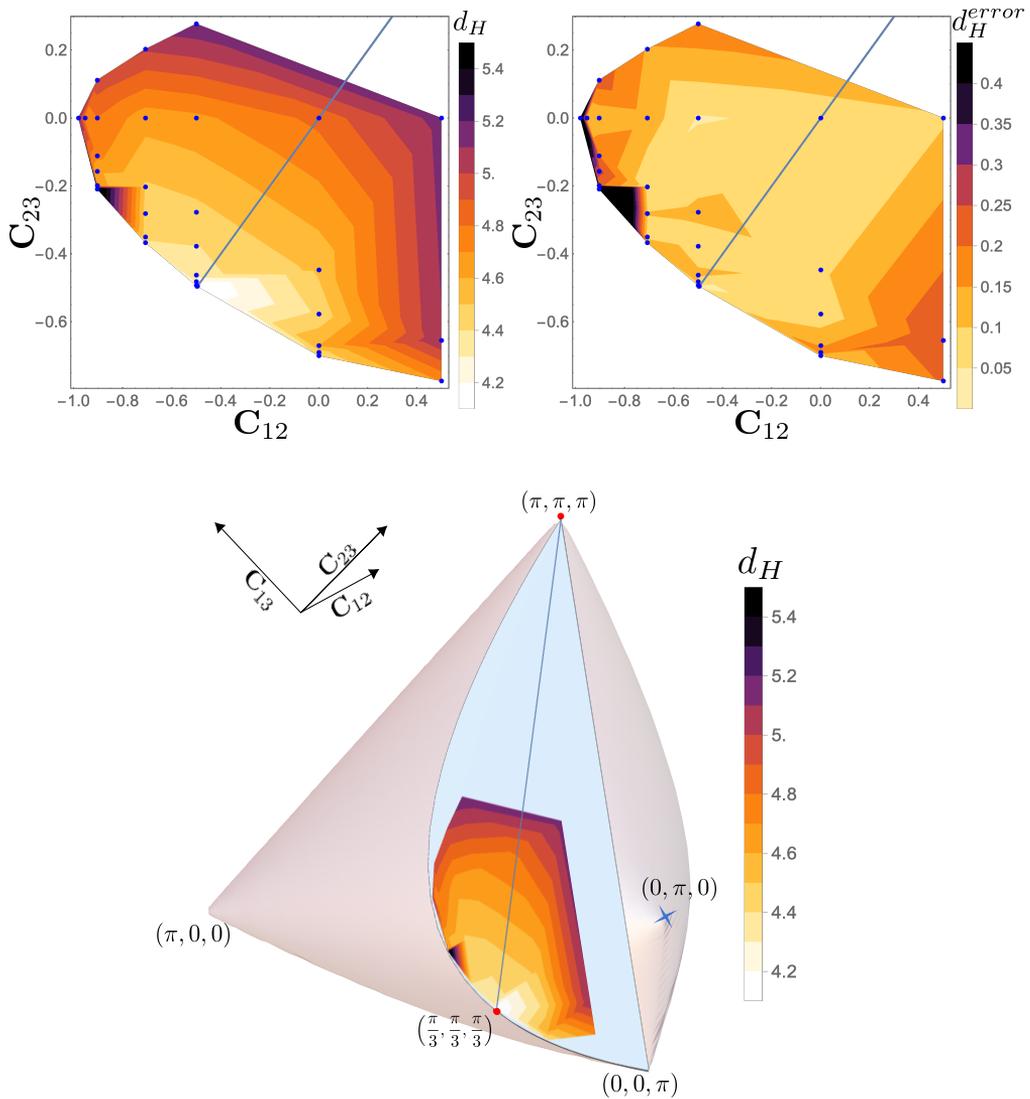

**Figure 3.17.: Top left**: Hausdorff dimension estimates from Mated-CRT graphs constructed from correlated 3D Brownian Excursions as a function of the covariances $C_{12} = -\cos(\alpha)$ and $C_{23} = C_{13} = -\cos(\beta)$. The contours are based on a linear interpolation of the simulated data points (see Table 3.2) that are shown in blue. The light blue line ($C_{12} = C_{23}$) indicates the models corresponding to equilateral spherical triangles. **Top right**: The corresponding errors in the estimates (including both systematic and statistical contributions). **Bottom**: The same contours as the top left plot shown in the full three-dimensional phase diagram.

| $(\alpha, \beta)$ | $d_H$ | $(\alpha, \beta)$ | $d_H$ |
|---|---|---|---|
| $\left(\frac{\pi}{16}, \frac{\pi}{2}\right)$ | $5.12 \pm 0.55$ | $\left(\frac{\pi}{3}, \frac{\pi}{3} + 0.04\right)$ | $4.36 \pm 0.11$ |
| $\left(\frac{25\pi}{256}, \frac{\pi}{2}\right)$ | $4.72 \pm 0.09$ | $\left(\frac{\pi}{3}, \frac{\pi}{3} + 0.80\right)$ | $5.18 \pm 0.19$ |
| $\left(\frac{9\pi}{64}, 1.14\right)$ | $5.61 \pm 1.16$ | $\left(\frac{\pi}{3}, 1.18\right)$ | $4.39 \pm 0.08$ |
| $\left(\frac{9\pi}{64}, 1.37\right)$ | $4.63 \pm 0.21$ | $\left(\frac{\pi}{3}, 1.29\right)$ | $4.47 \pm 0.11$ |
| $\left(\frac{9\pi}{64}, 1.41\right)$ | $4.61 \pm 0.26$ | $\left(\frac{\pi}{3}, \frac{\pi}{2}\right)$ | $4.67 \pm 0.04$ |
| $\left(\frac{9\pi}{64}, 1.46\right)$ | $4.73 \pm 0.20$ | $\left(\frac{\pi}{3} + 0.005, \frac{\pi}{3} + 0.005\right)$ | $4.08 \pm 0.04$ |
| $\left(\frac{9\pi}{64}, \frac{\pi}{2}\right)$ | $4.79 \pm 0.11$ | $\left(\frac{\pi}{2}, 0.79\right)$ | $4.42 \pm 0.09$ |
| $\left(\frac{9\pi}{64}, 1.68\right)$ | $4.96 \pm 0.23$ | $\left(\frac{\pi}{2}, 0.81\right)$ | $4.45 \pm 0.18$ |
| $\left(\frac{\pi}{4}, 1.19\right)$ | $4.36 \pm 0.13$ | $\left(\frac{\pi}{2}, 0.83\right)$ | $4.60 \pm 0.14$ |
| $\left(\frac{\pi}{4}, 1.21\right)$ | $4.45 \pm 0.08$ | $\left(\frac{\pi}{2}, 0.95\right)$ | $4.42 \pm 0.06$ |
| $\left(\frac{\pi}{4}, 1.28\right)$ | $4.45 \pm 0.11$ | $\left(\frac{\pi}{2}, 1.11\right)$ | $4.52 \pm 0.09$ |
| $\left(\frac{\pi}{4}, \frac{\pi}{2}\right)$ | $4.64 \pm 0.13$ | $\left(\frac{\pi}{2}, \frac{\pi}{2}\right)$ | $4.83 \pm 0.06$ |
| $\left(\frac{\pi}{4}, 1.77\right)$ | $5.01 \pm 0.14$ | $\left(\frac{2\pi}{3}, 0.68\right)$ | $4.71 \pm 0.21$ |
| $\left(\frac{\pi}{3}, \frac{\pi}{3} + 0.01\right)$ | $4.12 \pm 0.04$ | $\left(\frac{2\pi}{3}, 0.86\right)$ | $5.09 \pm 0.25$ |
| $\left(\frac{\pi}{3}, \frac{\pi}{3} + 0.02\right)$ | $4.35 \pm 0.11$ | $\left(\frac{2\pi}{3}, \frac{\pi}{2}\right)$ | $5.15 \pm 0.09$ |

**Table 3.2.:** Hausdorff dimension measurements with error bars of the Mated-CRT graph $\hat{G}_n^C$ obtained from correlated 3D Brownian Excursions. The angles $(\alpha, \beta)$ correspond to the spherical angles of the isosceles spherical triangles $T_{\alpha, \beta}$.

## 3.4. Discussion

In this study, we have proposed a sequence of discrete metric spaces $G_n^C$, Mated-CRT graphs, associated to a correlated Brownian excursion in $d$ dimensions, generalizing the Mated-CRT maps in $d = 2$. We hypothesize that upon normalization of distances these metric spaces approach a non-trivial continuous random metric as $n \rightarrow \infty$ that inherits its scaling properties from the Brownian excursion. In $d = 2$ this has largely been demonstrated as part of the mating of trees approach to Liouville Quantum Gravity, and the result is (depending on the correlation) either known or strongly suspected to yield a scale-invariant random metric with the topology of the 2-sphere. In $d = 3$ the situation is, of course, much less clear, but our numerical study indicates that for the examined correlation matrices the distance profiles of $G_n^{\mathrm{C}}$ display accurate scaling with $n$. Assuming this scaling persists to the full metric space and the Gromov–Hausdorff convergence of $n^{-1/d_h^C} G_n^{\mathrm{C}}$ as $n \longrightarrow \infty$ holds, this would establish a family of new universality classes of random geometries constructed from triples of correlated CRTs. While the characteristics of the random geometries are yet to be studied in detail, two critical exponents of these prospective universality classes can be calculated or estimated from our data: the string susceptibility and the Hausdorff dimension.

Our measurements pinpoint an interesting point on the boundary of parameter space where the off-diagonal elements of $\mathbf{C}$ approach $-1/2$, corresponding to a tiny equilateral spherical triangle ($\alpha = \beta = \gamma = \frac{\pi}{3}$), where the string susceptibility diverges ($\gamma_s \longrightarrow -\infty$) and the Hausdorff dimension appears to reach a minimum just above 4. This limit is analogous to the $\alpha \longrightarrow 0$ limit of mating of trees in $d = 2$, corresponding to the semi-classical limit $\gamma \longrightarrow 0$ in 2-dimensional Liouville Quantum Gravity. Note that in both cases the covariance matrix $\mathbf{C}$ degenerates, and the Brownian motion effectively becomes $(d - 1)$-dimensional, moving on the plane perpendicular to the diagonal. However, since one is forcing the curve to perform a unit-time excursion in $\mathbb{R}_{\geq 0}^d$ the limit is rather singular, so it is not entirely clear how the classical $\gamma = 0$ geometry is to be retrieved at $\alpha = 0$ in $d = 2$. If one relaxes the positivity constraint in $d = 2$, which naturally happens when consider infinite-volume limits, and considers Brownian motion that is nearly supported on the anti-diagonal in $\mathbb{R}^2$, the $\alpha \longrightarrow 0$ limit leads to identifications of points at equal height in a single CRT, resembling the foliated structure of two-dimensional Causal Dynamical Triangulation [9] (see the final remarks of [50]). The analogous interpretation in the case $d = 3$ amounts to the following. If we consider a two-dimensional Brownian motion on the $x + y + z = 0$-plane in $\mathbb{R}^3$, then the first two components have covariance $\mathbf{C}_{12} = -1/2$. Mating these two infinite correlated CRTs results in a random measure on $\mathbb{R}^2$ that is an infinite analogue of the unit-area $\gamma$-quantum sphere with $\gamma = \sqrt{4/3}$. The third tree then leads to identification of certain pairs of points of $\mathbb{R}^2$ that have equal sum of heights within the two embedded trees. It is a natural question to ask whether the discrete mating of trees bijection for bipolar-oriented triangulations, which lives in the $\gamma = \sqrt{4/3}$ universality class [98], can incorporate such identifications to describe three-dimensional discrete geometries.

This picture extends to other points on the two-dimensional boundary of parameter space where $\alpha + \beta + \gamma = \pi$, but with the random metric on the plane replaced by that of Liouville Quantum Gravity with $\gamma = 2\sqrt{\alpha/\pi}$ and the identification performed by equal linear combination of the two heights (with coefficients $\sin \gamma$ and $\sin \beta$ respectively). Here, as well as in the interior of the parameter space where $\det \mathbf{C} > 0$, one may ask the same question of whether discrete mating of trees bijections, like the ones in Section 3.1.1 and 3.1.2, have a combinatorial interpretation at the level of discrete 3-manifolds. It would be preferable to take the opposite route, in which one starts with a combinatorial model of discrete 3-manifolds, like three-dimensional Dynamical Triangulations [11], perhaps dressed with some matter statistical system and one would identify a bijective encoding into a triple of trees. However, the combinatorics of discrete 3-manifolds is still poorly understood, making this a challenging route. The first steps towards encoding 3-sphere triangulations in trees have been taken in [42] by greatly restricting the type of triangulations considered.

In regard to Hausdorff dimension estimates using the numerical implementation of the Mated-CRT maps in the 2-dimensional case, our estimates for $\gamma = 1, \sqrt{4/3}, \sqrt{2}, \sqrt{8/3}$ are statistically compatible with previous numerical results [21] and rigorous bounds [60, 84, 13]. Moreover, this numerical toolbox proved to be reliable in sampling random geometries in the region $\gamma < 1$ which has been inaccessible with other methods. We measured $d_H$ with good accuracy for $\gamma = 3/8, 1/2, 5/8, 3/4$. These results are compatible with a guessed formula of Ding and Gwynne, based on rigorous bounds [60], and contradict Watabiki's formula, based on a heuristic heat kernel analysis in Liouville Quantum Gravity.

In the case $d = 3$, a technical problems is finding sufficiently accurate numerical solutions to the harmonic equation (3.21) in general cones, which could be further improved with the methods of [51]. However, the main challenge in extending the results further out in the parameter space (and even to higher dimensions) is due to the large system sizes required, because of the following two reasons. As $1 - \gamma_s = \alpha_1$ becomes smaller, the distribution of the time extents of the Brownian motions $X^1(t)$ and $X^2(t)$ out of which we construct the excursion $\hat{X}(t)$ becomes increasingly heavy tailed (see Section 3.2.3), making it harder to produce unbiased samples. Secondly, the occurrence of higher Hausdorff dimensions means that a larger number of vertices is necessary to reach metric spaces of the same diameter, and this number is limited by the computing power available.

Perhaps the most important question that we leave open in this work is whether the new family of scale-invariant random geometries, if it exists, describes anything resembling spacetime geometry, in particular whether it has manifold topology. Deciding whether this is the case is considerably more difficult in $d = 3$ compared to $d = 2$. One of the reasons is the lack of a natural interpretation of the Mated-CRT graphs as a discrete geometry of deterministic topology. The other reason is that even if one has such a topology at the discrete level, there are many ways in which it can degenerate in the scaling limit. In the two-dimensional case, there exist practical sufficient criteria that ensure the limit has 2-sphere topology (see [115] for a discussion and application to the Brownian sphere), while the situation in $d = 3$ is less clear.

Short of answering these questions, having a catalogue of potential scale-invariant random geometries available is of value to research in Quantum Gravity. It opens up the possibility of comparing characteristics of the UV fixed point in asymptotically safe gravity, established through other approaches, to the concrete list of models arising from mating of trees. Drawing a bridge at the level of the dynamics is difficult, but a natural starting point is to compare critical exponents in various approaches. Hausdorff dimensions are often difficult to assess, since in approaches where the quantum geometry of spacetime is approximated with differentiable metrics, they tend to come out identical to the topological dimension. On the other hand, the string susceptibility, seen as the scaling behaviour of the partition function or as the distribution of sizes of minbus, should be easier to compare. Finally, the best studied critical exponent in Quantum Gravity appears to be the spectral dimension [129, 45, 89, 65], which characterizes diffusion processes in the geometry. It is consistently found to decrease below the topological dimension in the UV. In the case of 2-dimensional mated-CRT maps, the spectral dimension is exactly equal to 2 for all $\gamma \in (0, 2)$ [82]. A numerical estimation of the spectral dimension of the Mated-CRT graphs in $d = 3$ would be a logical follow-up for the numerical methods developed in this work.

# Hyperbolic random geometries and JT Gravity

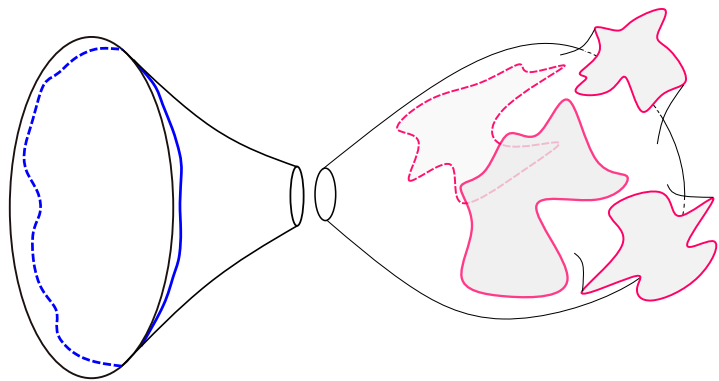

**Figure 4.1.:** Example of a configuration of critical JT gravity. The blue line represents the nearly AdS boundary and in red are the geodesics boundaries. In the non-generic critical regime, these boundaries become macroscopic.

As we have seen in Chapter 1, two-dimensional Euclidean quantum gravity is a useful toy model from both the physical and mathematical perspectives. In particular, the 2D Einstein-Hilbert action is topological, so the Euclidean path integral reduces to counting geometries with a weight given by $e^{-S_0\chi}$. When we restrict our attention to geometries with a constant negative curvature, one lands in the realm of JT gravity [91, 139], a model of Euclidean 2-dimensional quantum gravity on the hyperbolic plane. JT gravity has raised interest due to its simplicity and rich physical content [114]. One of the fascinating features of JT gravity is its connection to random geometries [152]. In particular, it has been demonstrated that the partition function of JT gravity can be split into two parts: one is proportional to volumes of moduli spaces of hyperbolic surfaces, while the other only involves information about the asymptotic behavior of AdS$_2$. The latter component can be interpreted in terms of a holographic correspondence, as it is the partition function of a quantum mechanical system given by the Schwarzian derivative effective action near the asymptotic boundary [108, 99]. Furthermore, the Weil-Petersson volumes carry the geometrical information of the negatively-curved surface. In addition, studying the higher genus contributions to this partition function revealed its relation with yet another random geometry field, matrix models [132, 137]. This allowed for significant progress in the field, since the formal machinery of matrix models [66] can be applied to JT gravity [93].

In this chapter (based on [46]), we apply novel techniques from random geometry [41] to JT gravity. We begin by examining the generating function of Weil-Petersson volumes with non-trivial weights on its geodesic boundaries. This allows us to investigate the behavior of the JT gravity partition function, in particular, the regime where there is a non-analytical behavior of the disk function and density of states, which are indicative of non-generic criticality. Such critical regimes are reminiscent of the planar map case where there is a transition from a phase of surfaces with microscopic holes to a phase with macroscopic ones (see Section 2.3.2). This translates, in the hyperbolic geometry case, to the proliferation of large geodesic boundaries. This phase transition is driven by a fine-tuning of the geodesic weights and has a significant impact on the density of states of the theory in the boundary. Our approach allows us to identify a family of models that interpolates between systems with $\rho_0(E) \sim \sqrt{E}$ and $\rho_0(E) \sim E^{3/2}$, which are commonly found in JT gravity coupled to dynamical end-of-the-world (EOW) [77] and FZZT branes [124].

## 4.1. Hyperbolic surfaces

In Chapter 2, we introduced the mathematical concept of a random map, which represents a discrete random geometry. Then, in Chapter 3, we briefly discussed an instance of a continuous random geometry, as described by Liouville Quantum Gravity. In this chapter, we will introduce yet another category of random geometry, specifically, hyperbolic random geometry. To begin our exploration, we will first outline the approach to studying Riemann surfaces.

A Riemann surface is a surface equipped with a complex structure. The space of all orientable compact Riemann surfaces of genus $g$ and $n$ boundaries up to conformal mappings is called *moduli space of Riemann surfaces*, $\mathcal{M}_{g,n}$. One can think of this space as a continuum version of the space of all possible genus-$g$ discrete maps with $n$ faces. The real dimension of this space is $d_{g,n} = 6g - 6 + 2n$. Each point in this space represents a conformal equivalence class of Riemann surfaces.

Within this space, Riemann surfaces can be characterized by the size of their automorphism group. *Stable surfaces* are those that have a finite set of automorphisms, this is equivalent to having $\chi < 0$, with $\chi = 2 - 2g - n$ its Euler characteristic. On the other side, *unstable surfaces* have an infinite set of automorphisms and have $\chi \geq 0$. The latter set is very small, it includes the sphere, the disc, the cylinder and the torus. The rest of the Riemann surfaces are stable.

Given a conformal equivalence class of stable surfaces with $n$ boundaries, there exists a Riemannian metric of constant negative curvature $-1$ such that the boundaries are geodesic. As these surfaces are locally isometric to the hyperbolic plane[1], they are called *hyperbolic surfaces*.

Given a hyperbolic surface, it can be decomposed into building blocks called pairs of pants, which is a hyperbolic surface of genus 0 and three geodesic boundaries. A pair of pants has $\chi = -1$, therefore, to build a genus-$g$ surface with $n$ boundaries, we need to glue $2g-2+n$ pairs of pants along $3g-3+n$ of their geodesic boundaries making sure their lengths match. However, there is still a degree of freedom left, we can rotate each boundary by an angle before gluing it. This implies that any hyperbolic surface of genus $g$ and $n$ boundaries is characterized by $3g-3+n$ lengths $\{l_i\}$ and $3g-3+n$ twists $\{\theta_i\}$, i.e. $6g-6+2n$ real parameters. Thus, the pants decomposition provides a set of local coordinates for $\mathcal{M}_{g,n}$.

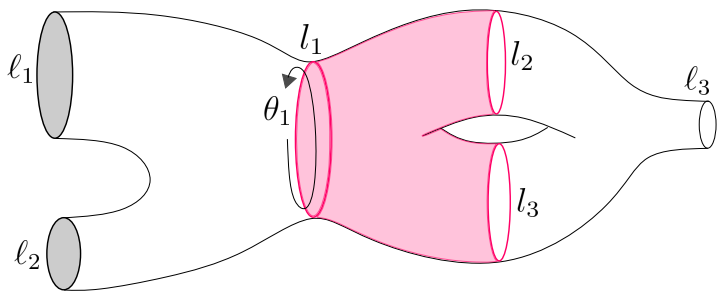

**Figure 4.2.:** Example of a hyperbolic surface decomposed in three pairs of pants.

---

[1]More precisely, they admit a subset of the hyperbolic plane as universal cover.

### 4.1.1. Weil-Petersson volumes

Now that we have a (local) parametrization of the moduli space of genus-$g$ hyperbolic surfaces with $n$ boundaries, one can define the Weil-Petersson symplectic structure on it [153],

$$\omega = \sum_{i=1}^{3g-3+n} dl_i \wedge d\theta_i. \tag{4.1}$$

This symplectic structure gives rise to a natural volume form $\omega^{3g-3+n}/(3g-3+n)!$ of $\mathcal{M}_{g,n}(\ell_1, \ldots \ell_n)$. Therefore, a natural volume of this moduli space is

$$V_{g,n}^{WP}(\ell_1, \ldots, \ell_n) = \int_{\mathcal{M}_{g,n}(\bar{\ell})} \prod_{i=1}^{3g-3+n} dl_i \wedge d\theta_i, \tag{4.2}$$

this is called the *Weil-Petersson volume*.

Let us now turn to a related space, first, we note that one can always convert a Riemann surface with $n$ boundary components into a compact Riemann surface with $n$ marked points by gluing a disc to each boundary (See Figure 4.3). Consider a closed Riemann surface of genus $g$ with $n$ marked points $p_i$ and let $\mathcal{L}_i$ be the cotangent space at the point $p_i$. We can compute algebraic invariants associated with these spaces that capture information about their topological or geometric properties, called *Chern-classes*, $c(\mathcal{L}_i)$.

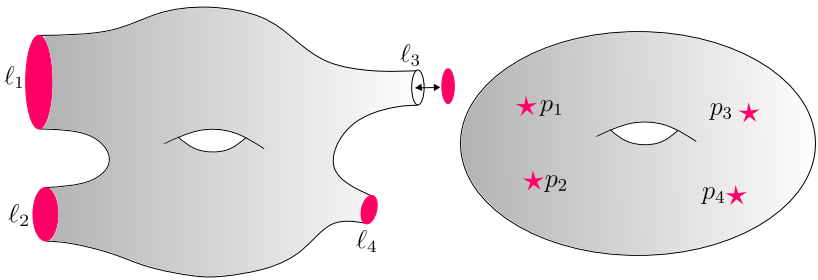

**Figure 4.3.:** A closed Riemann surface of genus $g$ with $n$ marked points (Right) can be obtained by gluing a disc to each boundary of a genus-$g$ Riemann surface with $n$ boundary components (Left).

These are cohomology classes of 2-forms, therefore some combinations of powers of these are top-forms that can be integrated over the moduli space to give scalars. These scalars are called intersection numbers and can be thought of as generalizations of the Euler characteristic. They correspond to

$$\langle \tau_{d_1} \dots \tau_{d_n} \rangle_{g,n} := \int_{\bar{\mathcal{M}}_{g,n}} c(\mathcal{L}_1)^{d_1} \dots c(\mathcal{L}_n)^{d_n}, \qquad (4.3)$$

with $d_1, \dots, d_n \in \mathbb{Z}_{\geq 0}$ and $d_{g,n} = 2(d_1 + \dots + d_n)$. $\bar{\mathcal{M}}_{g,n}$ is the Deligne-Mumford compactification of $\mathcal{M}_{g,n}$, which includes degenerate surfaces [54]. Another example of a topological class is the 2-form (4.1) which helped us define the Weil-Petersson volume, this is called the $\kappa_1$ Mumford-Morita-Miller class. Therefore, it is no surprise that the Weil-Petersson volume can be written in terms of these topological invariants [120]

$$V_{g,n}^{WP}(\ell_1, \dots, \ell_n) = 2^{3-3g-n} \sum_{d_0 + d_1 + \dots + d_n = d_{g,n}/2} \frac{(2\pi)^{2d_0}}{d_0!} \prod_{i=1}^{n} \frac{\ell_i^{2d_i}}{d_i!} \langle \kappa_1^{d_0} \tau_{d_1} \dots \tau_{d_n} \rangle_{g,n}. \qquad (4.4)$$

We will come back to this in Chapter 5, for now let us use the Weil-Petersson volume to define a random hyperbolic surface.

## 4.2. Random hyperbolic surfaces

Along the lines of random maps, we want to illustrate how to consistently define a random hyperbolic surface by constructing a probability measure in the space of hyperbolic surfaces of genus $g$ and $k$ geodesic boundaries, $\mathcal{M}_{g,k}^{WP}(b_1, \dots, b_k)$. The starting point is to write the 'cardinality' of this space, this is given by the Weil-Petersson volume $V_{g,k}^{WP}(b_1, \dots, b_k)$ [120]. The second step is to write a generating function of Weil-Petersson volumes of genus $g$ [97]

$$F_g^{WP}[\mu] = \sum_{k=0}^{\infty} \frac{2^{2-2g-k}}{k!} \int_0^{\infty} d\alpha_1 \mu(\alpha_1) \cdots \int_0^{\infty} d\alpha_k \mu(\alpha_k) V_{g,k}^{WP}(\alpha_1 \dots, \alpha_k), \qquad (4.5)$$

Note that, similarly to the Boltzmann map case, $\mu(\ell_k)$ can be interpreted as the likelihood of having geodesic boundaries of length $\ell_k$, i.e. it is a weight[2]. In Section 4.3 we will explain the role that weights play in JT gravity with defects.

---

[2]Another way to see this is to consider a weight $\mu(\ell) = x\tilde{\mu}(\ell)$ to make a clear distinction between the formal variable of the generating function, $x$, and the statistical weight $\tilde{\mu}$ in the sense of a non-negative valued label on the geodesic boundaries of the hyperbolic surface.

The generating function of Weil-Petersson volumes can be expressed in terms of the string equation

$$x(u) = u - \sum_{k=0}^{\infty} \frac{t_k + \gamma_k}{k!} u^k, \tag{4.6}$$

where [97, 120]

$$\gamma_k = \frac{(-1)^k \pi^{2k-2}}{(k-1)!} 1_{k \geq 2} \tag{4.7}$$

are called the times of pure JT gravity in the physics literature [93, 122], and

$$t_k = \frac{4^{-k}}{k!} \int_0^{\infty} d\ell \, \mu(\ell) \ell^{2k} \tag{4.8}$$

are the times associated with geodesic boundaries with labels $\mu(\ell)$. Equivalently, the string equation (4.6) can be written in the following form

$$x(u) = \frac{\sqrt{u}}{\pi} J_1 \left( 2\pi \sqrt{u} \right) - \int_0^{\infty} d\ell \, \mu(\ell) I_0 \left( \ell \sqrt{u} \right). \tag{4.9}$$

The first term is well-known in the physics literature as *pure JT gravity*, up to a change of sign $u \rightarrow -u$. The second term accounts for the presence of geodesic boundaries of weight $\mu(\ell)$ and it has been used in the physics literature to couple JT gravity to conical defects and branes [113, 123, 77].

The genus-0 generating function of Weil-Petersson volumes can be expressed in terms of the string equation, in a similar way to Boltzmann planar maps [36],

$$F_0^{WP}[\mu] = \frac{1}{4} \int_0^{E_\mu} du \, x(u)^2, \tag{4.10}$$

where $E_\mu > 0$ is the smallest solution to $x(E_\mu) = 0$, and, in the physics literature, it is called the threshold energy $E_\mu = -E_0$.

For $g > 0$, we need other constituents that also depend on the string equation, the *moments*

$$M_0^{WP} = \left( \frac{\delta E_\mu}{\delta \mu(0)} \right)^{-1}, \qquad M_p^{WP} = M_0^{WP} \frac{\delta M_{p-1}^{WP}}{\delta \mu(0)}, \tag{4.11}$$

or, alternatively

$$M_p^{WP} = \frac{\partial^{p+1} x}{\partial u^{p+1}} (E_\mu). \tag{4.12}$$

The genus-$g$ generating functions for $g > 0$ are proven to be [36]

$$F_1^{WP}[\mu] = -\frac{1}{24} \log M_0^{WP}, \tag{4.13}$$

$$F_g^{WP}[\mu] = \left(\frac{2}{(M_0^{WP})^2}\right)^{g-1} \mathcal{P}_g\left(\frac{M_1^{WP}}{M_0^{WP}}, \ldots, \frac{M_{3g-3}^{WP}}{M_0^{WP}}\right) \quad \text{for } g \geq 2. \tag{4.14}$$

The polynomials $\mathcal{P}_g$ are given by

$$\mathcal{P}_g(m_1, \ldots, m_{3g-3}) = \sum_{\sum_{k \geq 2}(i-1)d_i = d_{g,n}/2} \prod_{k \geq 2} \frac{(-m_{i-1})^{d_i}}{d_i!} \langle \tau_2^{d_2} \tau_3^{d_3} \ldots \rangle_{g,n}, \tag{4.15}$$

where $d_2, d_3, \cdots \geq 0$. This allows us to define a probability density measure on the space of hyperbolic genus-$g$ surfaces

$$p^{WP}(\ell_1, \ldots, \ell_k) = \frac{1}{k!} \frac{\prod_{i=1}^k \mu(\ell_i)}{F_g^{WP}[\mu]} 2^{2-2g-k} \frac{\omega^{3g-3+k}}{(3g-3+k)!}. \tag{4.16}$$

In this way, given an admissible weight $\mu$ such that $F_g^{WP}[\mu] < \infty$, we can call ***random genus-$g$ hyperbolic surface with $k$-boundaries of random lengths*** to a surface sampled with probability (4.16) given a fixed $k$.

In the rest of this section, we will restrict our study to the case $g = 0$ for convenience, but we will come back to the $g > 0$ case in Chapter 5. Once again, it is useful to introduce the disk function, which is the generating function of hyperbolic surfaces with one marked geodesic boundary of length $b$

$$W_\mu(b) = \frac{\delta F_0^{WP}}{\delta \mu(b)} = \sum_{k=2}^{\infty} \frac{2^{2-(k+1)}}{k!} \int_0^\infty \mathrm{d}\ell_1 \mu(\ell_1) \ldots \int_0^\infty \mathrm{d}\ell_k \mu(\ell_k) V_{0,k+1}^{WP}(b, \ell_1, \ldots, \ell_k) \tag{4.17}$$

$$= -\frac{1}{2} \int_0^{E_\mu} \mathrm{d}u \, x(u) \, I_0(\ell \sqrt{u}). \tag{4.18}$$

## 4.2.1. Critical hyperbolic surfaces and the hyperbolic $O(n)$ loop model

A natural question that arises is if a critical behavior, analogous to that observed in planar maps in Section 2.3, happens for hyperbolic surfaces too. Specifically, whether there exist weights $\mu$ such that the string equation (4.9) behaves in a similar manner to the ones presented in Table 2.1. The key difference between the Boltzmann maps and hyperbolic surfaces string equations is the first term of (4.9), which does not depend on the weights. This implies that, in order to reach a non-generic critical regime in the hyperbolic case, the weight $\mu$ needs to be fine-tuned to dominate the Bessel function in the first term of (4.9), instead of the simpler linear term in the Boltzmann map case (2.9). Therefore, it appears to be a significantly larger challenge to find non-generic critical weights for hyperbolic surfaces. The existence of such regimes for hyperbolic surfaces is formally proved in [41]. In this subsection, we will present an alternative argument for their existence, drawing upon an analogy with the gasket decomposition of the hyperbolic $O(n)$ loop model. Consider the string equation (4.9), which, for convenience, we write again

$$x(u) = \frac{\sqrt{u}}{\pi} J_1 \left( 2\pi \sqrt{u} \right) - \int_0^\infty d\ell \mu(\ell) I_0 \left( \ell \sqrt{u} \right),$$

and a weight $\mu(\ell) = \kappa \, e^{-\lambda \ell} \hat{\mu}(\ell)$ with $\kappa, \lambda > 0$ and $\hat{\mu}(\ell)$ a bounded, non-negative function on $[0, \infty)$ with asymptotics $\hat{\mu}(\ell) \sim C \, \ell^{-\alpha}$ as $\ell \longrightarrow \infty$, e.g. $\hat{\mu}(\ell) = (\ell + 1)^{-\alpha}$ or $\hat{\mu}(\ell) = \ell^{-\alpha} \Theta(\ell - 1)$, where $\Theta$ is the Heaviside step function. We will demonstrate how we can choose constants $\kappa, \lambda > 0$ to reach non-generic criticality, i.e. $x(E_\mu) = 0$, $x'(E_\mu) = 0$ and $x(u) \sim c (E_\mu - u)^{\alpha - 1/2}$ as $u \longrightarrow E_\mu$, with $\alpha \in (3/2, 5/2)$.

To see this, we note that if we have $\mu$ such that $x'(E_\mu) = 0$, then $x''(E_\mu) = \infty$. This last expression requires that $\lambda = \sqrt{E_\mu}$ because $I_0(\ell \sqrt{E_\mu})'' \sim \ell^{3/2} e^{\ell \sqrt{E_\mu}}$ as $\ell \longrightarrow \infty$. Then, $x(E_\mu) = 0$ allows us to fix $\kappa$ to

$$\kappa = \frac{\frac{\sqrt{E_\mu}}{\pi} J_1(2\pi \sqrt{E_\mu})}{\int_0^\infty d\ell \hat{\mu}(\ell) e^{-\ell \sqrt{E_\mu}} I_0(\ell \sqrt{E_\mu})}. \tag{4.19}$$

If we take $E_\mu$ to be the first zero of

$$x'(E_\mu) = J_0(2\pi \sqrt{E_\mu}) - \frac{\sqrt{E_\mu}}{\pi} J_1(2\pi \sqrt{E_\mu}) \frac{\int_0^\infty d\ell \hat{\mu}(\ell) e^{-\ell \sqrt{E_\mu}} \frac{\ell I_1(\ell \sqrt{E_\mu})}{\sqrt{E_\mu}}}{\int_0^\infty d\ell \hat{\mu}(\ell) e^{-\ell \sqrt{E_\mu}} I_0(\ell \sqrt{E_\mu})}, \tag{4.20}$$

which exists since $x'(0) = 1$ and it is negative for $\sqrt{2} E_\mu = c_0^2/(8\pi^2) = 0.0732 \ldots$, where $c_0$ is the first zero of the Bessel function $J_0$. Accordingly,

$$x''(u) = \left( \frac{\sqrt{u}}{\pi} J_1\left(2\pi\sqrt{u}\right) \right)'' - \int_0^\infty d\ell\, \mu(\ell) \frac{\ell^2}{u} I_2(\ell\sqrt{u}) \tag{4.21}$$

$$= (\cdots)'' - \int_0^\infty d\ell\, \kappa\, C\, e^{-\sqrt{E_\mu}\ell}\, \ell^{-\alpha} \frac{\ell^2}{u} \frac{e^{\sqrt{u}\ell}}{\sqrt{2\pi\sqrt{u}\ell}} + o(\ell^{-\alpha+5/2}) \tag{4.22}$$

$$= (\cdots)'' - \frac{\kappa\, C}{u\sqrt{2\pi\sqrt{u}}} \Gamma\left(\frac{5}{2}-\alpha\right) (\sqrt{E_\mu} - \sqrt{u})^{\alpha-5/2} + o((\sqrt{E_\mu} - \sqrt{u})^{\alpha-5/2}) \tag{4.23}$$

$$= -\frac{\kappa\, C}{E_\mu\sqrt{2\pi\sqrt{E_\mu}}} \Gamma\left(\frac{5}{2}-\alpha\right) (2\sqrt{E_\mu})^{5/2-\alpha}(E_\mu - u)^{\alpha-5/2} + o((E_\mu - u)^{\alpha-5/2}), \tag{4.24}$$

where we use Karamata's Tauberian Theorem (see e.g. [70], Section XIII.5, Theorem 2) in the last line. Integrating twice and using that $x(E_\mu) = x'(E_\mu) = 0$ gives

$$x(u) = -\frac{\kappa\, C}{\sqrt{2\pi}} \Gamma\left(\frac{1}{2}-\alpha\right) 2^{5/2-\alpha} E_\mu^{-\alpha/2}(E_\mu - u)^{\alpha-1/2} + o((E_\mu - u)^{\alpha-1/2}) \tag{4.25}$$

$$= -\frac{\kappa\, C}{\sqrt{2\pi\sqrt{E_\mu}}} \Gamma\left(\frac{1}{2}-\alpha\right) 2^2 (\sqrt{E_\mu} - \sqrt{u})^{\alpha-1/2} + o((E_\mu - u)^{\alpha-1/2}). \tag{4.26}$$

This ends the demonstration. A similar reasoning follows for the disk function (4.18),

$$W_\mu(\ell) = -\frac{1}{2} \int_0^{E_\mu} du\, x(u) I_0(\ell\sqrt{u}) \tag{4.27}$$

$$= \frac{1}{2} \frac{\kappa\, C}{\sqrt{2\pi\sqrt{E_\mu}}} \Gamma\left(\frac{1}{2}-\alpha\right) 2^2 \int_0^{E_\mu} (\sqrt{E_\mu} - \sqrt{u})^{\alpha-1/2} \frac{e^{\ell\sqrt{u}}}{\sqrt{2\pi\ell\sqrt{u}}} du + \cdots \tag{4.28}$$

$$= \frac{\kappa\, C}{\sqrt{2\pi\sqrt{E_\mu}}} \Gamma\left(\frac{1}{2}-\alpha\right) 2^2 \int_0^{\sqrt{E_\mu}} y^{\alpha-1/2} \frac{e^{-\ell y+\ell\sqrt{E_\mu}}}{\sqrt{2\pi\ell}} E_\mu^{1/4} dy + \cdots \tag{4.29}$$

$$= \frac{2}{\pi}\kappa\, C\, \Gamma\left(\frac{1}{2}-\alpha\right) \Gamma(\alpha+1/2)\ell^{-\alpha-1} e^{\ell\sqrt{E_\mu}} + o(\ell^{-\alpha-1}) \tag{4.30}$$

$$= \frac{2}{\sin\left(\pi\left(\alpha-\frac{1}{2}\right)\right)} \kappa\, C\, \ell^{-\alpha-1} e^{\ell\sqrt{E_\mu}} + o(\ell^{-\alpha-1}) \tag{4.31}$$

where we used Euler's reflection formula for the gamma function to write the second line.

Now, we would like to show how this fits into the hyperbolic $O(n)$ loop model in a similar way as for Boltzmann planar maps [30]. The hyperbolic $O(n)$ loop model is defined by the generating function

$$F_0^{(n)}[\mu] = \sum_{k=3}^{\infty} \frac{1}{k!} \int_0^{\infty} d\ell_1 \mu(\ell_1) \cdots \int_0^{\infty} d\ell_k \mu(\ell_k) \sum_{\Gamma} \int_{\mathcal{M}_{0,k}(\ell_1,\ldots,\ell_k)} d_{WP} \prod_{\gamma \in \Gamma} n \ e^{-xL_\gamma} \tag{4.32}$$

where $n \in (0, 2)$, $\Gamma$ is a collection of non-intersecting simple geodesic loops $\gamma$ of length $L_\gamma$ and $d_{WP}$ is the Weil-Petersson measure of the hyperbolic surface. The coupling constant $n$ introduces a weight $n \ e^{-xL_\gamma}$ to each loop drawn on top of a hyperbolic surface (see Figure 4.4).

What we aim to investigate is if there is a consistent weight substitution $\mu \longrightarrow \hat{\mu}$ that relates the disk functions of the hyperbolic $O(n)$ loop model and the hyperbolic non-generic critical surfaces one, $W_\mu^{(n)}(\ell) = W_{\hat{\mu}}(\ell)$. From (4.31), we can observe that our non-generic critical weight behaves asymptotically as

$$\hat{\mu}(\ell) = 2 \ \sin\left(\pi \left(\alpha - \frac{1}{2}\right)\right) \ e^{-2\sqrt{E_\mu}\ell} \ 2^{-2} \ W_{\hat{\mu}}(\ell) + o(\ell^{-\alpha}) \tag{4.33}$$

Therefore, we have shown that a weight substitution

$$\hat{\mu}(L_\gamma) = \mu(L_\gamma) + n \ e^{-L_\gamma x} L_\gamma \ 2^{-2} \ W_{\hat{\mu}}(L_\gamma), \tag{4.34}$$

for[3] $x = 2\sqrt{E_\mu}$ and $n = 2\sin\left(\pi\left(\alpha - \frac{1}{2}\right)\right)$ consistent with the disk function substitution and with asymptotics $\hat{\mu}(L_\gamma) \sim C' e^{\sqrt{E_\mu}L_\gamma} L_\gamma^{-\alpha}$ exists. In geometrical terms, this means that one can substitute each part of the hyperbolic surface surrounded by a geodesic loop $\gamma$ by a geodesic boundary with weight (4.34) (see Figure 4.4).

To conclude this section, we present the qualitative phase diagram of the $O(n)$ loop model (Figure 4.5) as studied in [30, 31], which we expect to generalise for the hyperbolic $O(n)$ loop model. This will be relevant when we identify non-generic criticality in JT gravity (Section 4.4) and draw similarities with hyperbolic geometries.

---

[3] The factor of $2^{-2}$ is necessary for consistency with the prefactor in the generating function of Weil-Petersson volumes and in (4.18). Thus, it does not affect the relation between $n$ and $\alpha$.

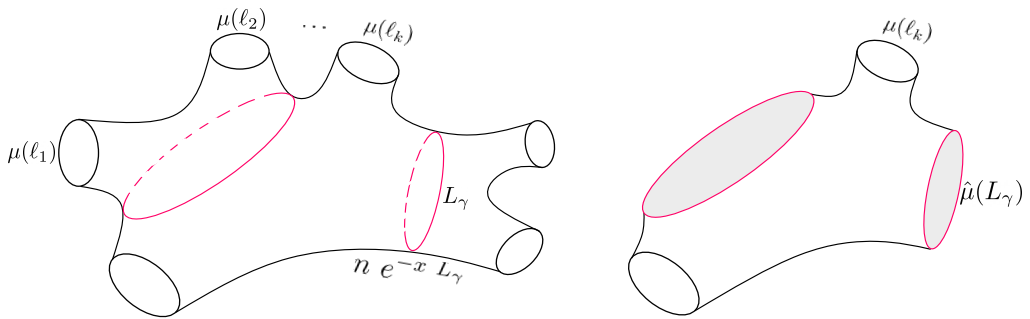

**Figure 4.4.: Left**: Example of a hyperbolic $O(n)$ loop model with two non-intersecting geodesic loops in red. **Right**: The corresponding gasket where the marked geodesic boundaries are colored in gray.

As we can see in Figure, 4.5, within the non-generic critical family of weights, there are two cases, the ***dense*** and ***dilute*** phases. In the dense phase ($\alpha > 2$), the length of the loops is big and they can touch tangentially. In the dilute phase ($\alpha < 2$), the loops tend to avoid each other and also themselves.

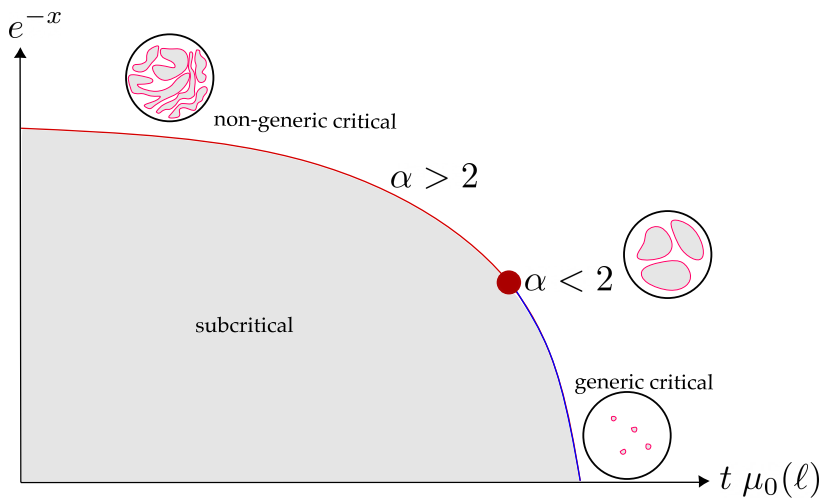

**Figure 4.5.:** Qualitative phase diagram of the $O(n)$ model [30, 31]. The line $x = 2\sqrt{E_\mu}$ corresponds to non-generic criticality, $t > 0$ and $\mu_0(\ell)$ is a reference weight.

## 4.3. JT Gravity

Let us turn now to the standard formulation of JT gravity before reintroducing it in terms of random geometry. We revisit the action of JT Gravity introduced in Section 1.2.4,

$$I_{JT} = -\frac{S_0}{2\pi} \left( \frac{1}{2} \int_{\mathcal{M}} \sqrt{g}R + \int_{\partial\mathcal{M}} \sqrt{h}K \right) - \left( \frac{1}{2} \int_{\mathcal{M}} \sqrt{g}\phi(R+2) + \int_{\partial\mathcal{M}} \sqrt{h}\phi(K-1) \right),$$

where $S_0 = \frac{1}{4G}$. From now on we will focus on the case $n = 1$, i.e. one asymptotic boundary. In this case, the one boundary JT partition function is

$$\mathcal{Z}(\beta) = \sum_{g=0}^{\infty} e^{-S_0(2g-1)} \tilde{\mathcal{Z}}_g(\beta) = e^{S_0} \tilde{\mathcal{Z}}_0(\beta) + \sum_{g=1}^{\infty} e^{-S_0(2g-1)} \int_0^{\infty} b\,\mathrm{d}b\, V_{g,1}^{WP}(b) \mathcal{Z}_{tr}(\beta, b). \tag{4.35}$$

We separate the contribution for $g = 0$ because $V_{0,1}$ vanishes. However, this geometry corresponds to a hyperbolic disk for which the path integral (1.23) is not vanishing, this is what we call $\tilde{\mathcal{Z}}_0(\beta)$ in this case, and it is given by [132]

$$\tilde{\mathcal{Z}}_0(\beta) = \frac{1}{\sqrt{16\pi\beta^3}} e^{\frac{\pi^2}{\beta}}. \tag{4.36}$$

### 4.3.1. JT Gravity with defects

In a similar fashion, one can consider the case when the hyperbolic surface has geodesic boundaries of length $\ell$ that are not attached to an asymptotic boundary by a trumpet, these are called *defects* [113] and are assigned an effective weight $\mu(\ell)$ that compile the physical system they represent e.g. conical defects or branes (See Fig. 4.6). Then, the partition function can be expanded in

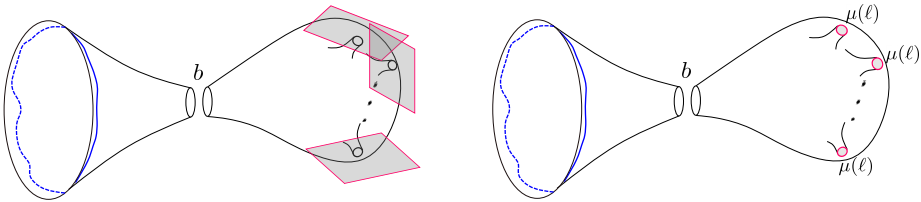

**Figure 4.6.:** Left: JT spacetime ending on $n$ branes. Right: JT spacetime with $n$ geodesic boundaries with weight $\mu$.

a sum over genus and a sum over the number of defects

$$\mathcal{Z}(\beta) = \sum_{g=0}^{\infty} e^{-S_0(2g-1)} \sum_{k=0}^{\infty} e^{-S_0 k} \mathcal{Z}_{g,k+1}(\beta), \tag{4.37}$$

where for $g > 0$ and $k \geq 0$,

$$\mathcal{Z}_{g,k+1}(\beta) = \frac{1}{k!} \int_0^{\infty} d\ell_1 \mu(\ell_1)... \int_0^{\infty} d\ell_k \mu(\ell_k) \int_0^{\infty} b \, db \, V_{0,k+1}^{WP}(b, \ell_1, ..., \ell_k) \mathcal{Z}_{tr}(\beta, b). \tag{4.38}$$

In particular, for $g = 0$ we know that the Weil-Petersson volumes vanish for the cases $k = 0, 1$, but these partition functions can be explicitly computed using the Schwarzian action and they correspond to the disk ($\tilde{\mathcal{Z}}_0$) and the trumpet ending on a defect (sometimes called *half-wormhole*). Therefore, the genus-0 partition function can be explicitly written as

$$\mathcal{Z}_0(\beta) = \frac{1}{\sqrt{16\pi\beta^3}} e^{\frac{\pi^2}{\beta}} + e^{-S_0} \int_0^{\infty} d b \mu(b) \frac{e^{-b^2/4\beta}}{2\sqrt{\pi\beta}} \tag{4.39}$$

$$+ \sum_{k=2}^{\infty} \frac{e^{-S_0 k}}{k!} \int_0^{\infty} d\ell_1 \mu(\ell_1)... \int_0^{\infty} d\ell_k \mu(\ell_k) \int_0^{\infty} b \, db \frac{e^{-b^2/4\beta}}{2\sqrt{\pi\beta}} V_{0,k+1}^{WP}(b, \ell_1, ..., \ell_k). \tag{4.40}$$

More examples of the geometries that contribute to (4.37) are shown in Figure 4.7.

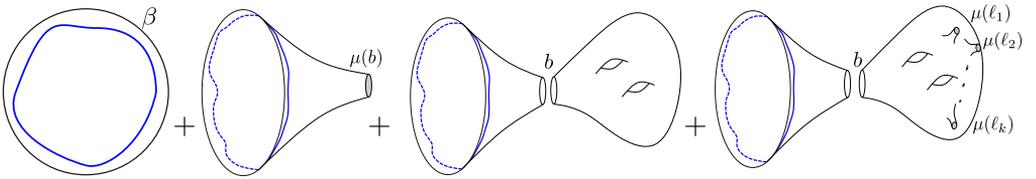

**Figure 4.7.:** Examples of geometries that contribute to the partition function (4.37). From left to right: ($g = 0$, $n = 1$, $k = 0$), ($g = 0$, $n = 1$, $k = 1$), ($g = 2$, $n = 1$, $k = 0$) and ($g = 2$, $n = 1$, $k > 0$).

So far, we have introduced JT gravity as it is mainly done in the physics literature, yet, one of our main goals is to reintroduce techniques from random geometry that we will prove are highly useful for JT gravity (Section 4.4). In particular, the power of the string equation formulation has been used in JT gravity [140, 73][4], but, in the physics literature, one usually computes the partition function and density of states using a holographic quantum mechanical system in the asymptotic boundary and then 'guess' the form of the string equation. Here, we take the opposite point of view and start with the string equation.

Using this, we can re-express (4.40) in a more convenient way which explicitly makes use of the string equation [93, 122, 113]

$$\mathcal{Z}(\beta)_{g=0} = \frac{1}{2\sqrt{\pi\beta}} \int_{-\infty}^{0} \mathrm{d}x \, e^{u(x)\beta} = \frac{1}{2\sqrt{\pi\beta}} \int_{-\infty}^{E_\mu} \mathrm{d}u \, e^{u\beta} \frac{\partial x(u)}{\partial u} \tag{4.41}$$

$$= -\frac{1}{2} \int_{-\infty}^{0} \mathrm{d}u \, e^{u\beta} x(u) \sqrt{\frac{\beta}{\pi}} - \frac{1}{2} \int_{0}^{E_\mu} \mathrm{d}u \, e^{u\beta} x(u) \sqrt{\frac{\beta}{\pi}} \tag{4.42}$$

$$= \frac{1}{\sqrt{16\pi\beta^3}} e^{\frac{\pi^2}{\beta}} + \int_{0}^{\infty} \mathrm{d}b \mu(b) \frac{e^{-b^2/4\beta}}{2\sqrt{\pi\beta}} + \int_{0}^{\infty} b \mathrm{d}b \frac{e^{-b^2/4\beta}}{2\sqrt{\pi\beta}} W(b), \tag{4.43}$$

where we set $e^{S_0} = 1$ for convenience and in the last line we used the identity

$$e^{-bz} = 2z \int_{0}^{\infty} \mathrm{d}\beta e^{-\beta z^2} \frac{e^{-b^2/4\beta}}{2\sqrt{\pi\beta}}. \tag{4.44}$$

The density of states is obtained by Laplace transforming the partition function, and is given by

$$\rho_0(E) = \frac{1}{2\pi} \int_{-E}^{-E_0} \mathrm{d}u \frac{1}{\sqrt{E+u}} \frac{\partial x}{\partial u}. \tag{4.45}$$

In order to draw a parallel between this language and some well-known results in the JT gravity literature, we present two examples: FZZT and end-of-the-world branes.

## 4.3.2. Example: FZZT Branes

Dynamical FZZT Branes were introduced in the context of Liouville Field Theory in [69] as a family of conformally invariant (Dirichlet) boundary conditions parametrized by the so-called *boundary cosmological constant*, which controls

---

[4] It is worth noting that what we call the string equation is known in the physics literature as the genus $g = 0$ string equation in the limit $\hbar \to 0$

the length of the boundary. Beyond the Physics aspects, the convenience of working with these branes comes from the fact that they have well-defined analytic expressions in Matrix Models, they are determinant insertions [1]. In the JT gravity literature, FZZT branes have been studied in this language [132, 28], and it was shown that these branes can be effectively encoded by geodesic boundaries of length $\ell$ with weight [123] (see Figure 4.6)

$$\mu_{FZZT}(\ell) = -e^{-z\ell}.$$ (4.46)

where the constant $z$ plays the role of the boundary cosmological constant. Using (4.8), we can compute the times

$$t_k^{FZZT} = -\frac{4^{-k}(2k)!}{k!}z^{-2k-1},$$ (4.47)

which are consistent with [123]. Then, the string equation (4.9) takes the well-known form

$$x(u) = \frac{\sqrt{u}}{\pi}J_1\left(2\pi\sqrt{u}\right) + \frac{1}{\sqrt{z^2 - u}}.$$ (4.48)

It is worth noting that the term 'dynamical' in this context means that all physical quantities, e.g. (4.41) and (4.45), are computed using the shifted weights $t_m + \gamma_m$ instead of the pure JT gravity ones, $\gamma_m$. In these terms is that our computations are also done in the dynamical framework.

### 4.3.3. Example: EOW branes

In JT gravity, End-of-the-world branes are boundaries of space where Neumann boundary conditions are usually imposed. These have been mainly used to investigate black hole evaporation and the Page curve[126], and correspond to introducing vectors in the JT gravity Matrix Model. However, for our purposes, EOW branes are geodesic boundaries with weights [77]

$$\mu_{EOW}(\ell) = \frac{e^{-m\ell}}{2\sinh\left(\ell/2\right)}$$ (4.49)

and, equivalently, times

$$t_k^{EOW} = \frac{4^{-k}(2k)!\zeta\left(2k+1, m+\frac{1}{2}\right)}{k!}1_{k\geq1},$$ (4.50)

where $m$ is the value of the normal derivative of the dilaton at the EOW boundary.

## 4.4. Critical JT Gravity

Our objective is to investigate the applicability of the criticality results observed in Boltzmann planar maps and more recently in hyperbolic surfaces [41] to JT gravity. In this last work, it is proven that the generating function of Weil-Petersson volumes experiences a similar critical behavior to Boltzmann planar maps (see Table 2.1), a result that we also showed in Subsection 4.2.1 at the level of the string equation and disk function. Furthermore, in the non-generic critical regime ($M_0^{WP} = 0$), the hyperbolic surfaces also develop large faces, which correspond to geodesic boundaries. We are especially interested in how the non-generic criticality in the bulk translates into the holographic quantum system on the asymptotic boundary, and, in particular, how it affects its partition function and density of states. More precisely, we will show that this family of non-generic critical models consistently interpolates between two known cases in the JT literature, $\rho(E) \sim \sqrt{E}$ and $\rho(E) \sim E^{3/2}$, which correspond to the $k = 1$ and $k = 2$ minimal models respectively.

For simplicity, we will assume a general form of the non-generic critical weight $\mu_\alpha$ that satisfies the conditions deduced in Section 4.2.1 such that the string equation is

$$x_{crit}(u) = -t_{crit} \left(E_\mu - u\right)^{\alpha - \frac{1}{2}} + o((E_\mu - u)^{\alpha - \frac{1}{2}}) \tag{4.51}$$

where $3/2 < \alpha < 5/2$. The genus-0 generating function is

$$F_0^{WP}[\mu_\alpha] = \frac{E_\mu^{2\alpha}}{8\alpha} t_{crit} + o(E_\mu^{2\alpha}), \tag{4.52}$$

and the disk function

$$W(b) = \frac{E_\mu^{\alpha + \frac{1}{2}} {}_0F_1\left(\alpha + \frac{3}{2}; \frac{b^2 E_\mu}{4}\right)}{2\alpha + 1} t_{crit} + o(b^{-\alpha - 1}). \tag{4.53}$$

### 4.4.1. Density of states

As emphasized in Section 4.3, the string equation provides sufficient information for determining the leading-order genus zero partition function and density of states. So, using (4.51), (4.41) and (4.45), and neglecting terms of higher

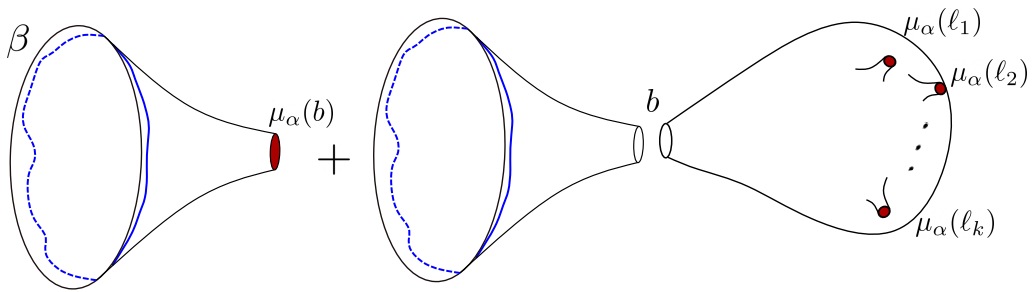

**Figure 4.8.:** Geometries that contribute to the partition function of critical JT gravity with defects of weight $\mu_\alpha$.

order, the partition function of critical JT gravity is

$$\mathcal{Z}_0(\beta) = \frac{\Gamma\left(\alpha + \frac{1}{2}\right) e^{-\beta E_0} t_{crit}}{2\sqrt{\pi}} \beta^{-\alpha}, \tag{4.54}$$

and its density of states is given by (4.45),

$$\rho_0(E) = \frac{\Gamma\left(\alpha + \frac{1}{2}\right) t_{crit}}{2\sqrt{\pi}\Gamma(\alpha)} (E - E_0)^{\alpha-1}, \tag{4.55}$$

where we substituted $E_\mu \longrightarrow -E_0$. In Figure 4.9, we show (4.55) for different values of $\alpha$. We note that this non-generic critical family of models interpolates constantly between $\sqrt{E - E_0}$ and $(E - E_0)^{3/2}$.

## 4.4.2. Phase transitions

We dedicate this subsection to clarify the physical and geometrical meaning of the JT gravity non-generic critical family of models described by the density of states (4.55). First, let us remark that it is known that in JT gravity with FZZT [124] and EOW [77] branes some parameters can be tuned to shift $\rho_0(E)$ from $\sqrt{E - E_0}$ to $(E - E_0)^{3/2}$, and our model bridges the gap between these two cases by providing an interpolation. However, the framework that we present in this paper makes clear and precise in which sense there is a phase transition. The explanation is that by fine-tuning the first derivative of the string equation to vanish, the weight goes from a subcritical to a generic critical phase. What our model does is fine-tune the weight to be non-generic critical. In principle, this non-generic critical weight can be linked to modifications to the FZZT and EOW brane weights. Given that our model offers this interpolation, it seems plausible this non-generic critical weight can be linked to deformations of the FZZT and EOW brane weights and its precise form could yield interesting results. However, we will not explore this direction in the present work, but rather the general properties of these deformations.

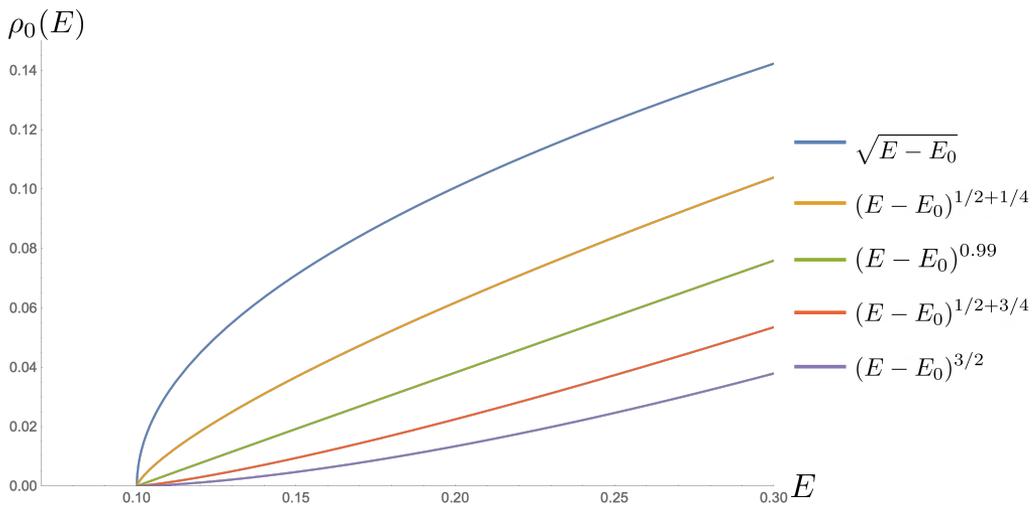

**Figure 4.9.:** Genus zero density of states (4.55) for different values of $\alpha$ and $t_{crit} = 1$. This family of non-generic critical models continuously interpolates between $\sqrt{E - E_0}$ and $(E - E_0)^{3/2}$. Our simplified forms of the weight and string equation do not fix a value for $E_0$. For convenience, we consider the threshold energy as constant ($E_0 = 0.1$) for all $\alpha$, but, in general, it will also be a function of $\alpha$.

The aspect we highlight now is the geometrical features of this family of non-generic critical models. Within this critical regime, the system undergoes significant transformations. Notably, the analysis in [41] reveals formal divergences in the expected number of boundaries and their corresponding total length, indicating substantial geometrical similarities with the Boltzmann planar map case as the model approaches criticality. We can see this by using the probability measure (4.16) to compute the expected number of geodesic boundaries

$$\mathbb{E}(n) = \sum_{n=3}^{\infty} n \cdot \frac{2^{2-n}}{n!} \frac{\prod_{i=1}^{n} \int_0^{\infty} d\ell_i \mu(\ell_i) V_{0,n}^{WP}(\ell_1, ..., \ell_n)}{F_0^{WP}}, \qquad (4.56)$$

The main argument is that when two of the defects have length zero, we can easily see that a special case of the last expression behaves as

$$\frac{\delta E_\mu}{\delta \mu(\ell)} = \sum_{n=3}^{\infty} \frac{2^{3-n}}{(n-3)!} \left( \prod_{k=4}^{n} \int_0^{\infty} d\ell_i \, \mu(\ell_i) \right) V_{0,n}^{WP}(0, 0, \ell, \ell_4, ..., \ell_n) \longrightarrow \infty \quad (4.57)$$

since, from the string equation, we have that $\frac{\delta E_\mu}{\delta \mu(\ell)} = \frac{\delta u / \delta \mu(E_\mu)}{M_0^{WP}}$, and $M_0^{WP} = 0$ at criticality. Therefore, $\mathbb{E}(n) \longrightarrow \infty$. Additionally, we can analyse the expected number of boundaries in a disk of diameter $b$ in the non-generic critical regime,

$$\mathbb{E}_{W_\mu(b)}(n) = \int_0^{\infty} d\ell \frac{\delta W_\mu(b)}{\delta \mu(\ell)} \mu(\ell) \propto \int_0^{\infty} d\ell \, \ell^{-1/2} e^{\ell \sqrt{E_\mu}} \, \ell^{-\alpha} e^{-\ell \sqrt{E_\mu}} + o(\ell^{-\alpha-1/2}).$$
$$(4.58)$$

The expression inside the integral is the large $\ell$ probability distribution of the number of geodesic boundaries of size $\ell$ in a disk of diameter $b$. Then, we can see that the variance of this distribution is $\mathrm{Var}(\ell^{-\alpha-1/2}) \longrightarrow \infty$ for $\alpha > 3/2$. This implies that in the non-generic critical regime of JT gravity, one can expect a very large number of defects, and if $b \longrightarrow \infty$ there will always be boundaries of comparable size. Even though we are not explicitly proving the existence of a dilute and dense phase, we anticipate a similar behavior to Boltzmann maps due to the relationship with the hyperbolic $O(n)$ loop model discussed in Section 2.3. A schematic illustration of this behavior is provided in Figure 4.10.

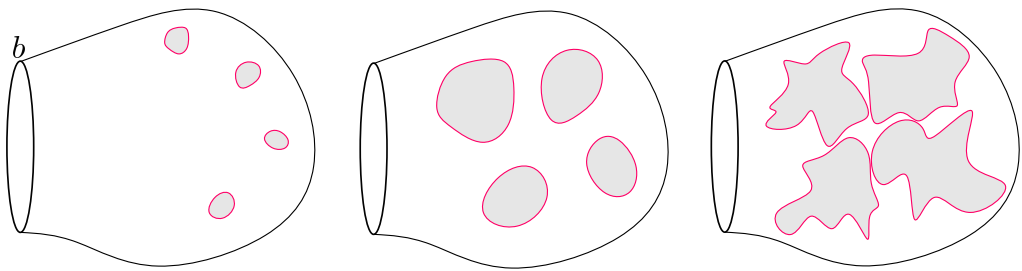

**Figure 4.10.:** Schematic illustration of a hyperbolic surface with: subcritical, dilute, and dense non-generic critical weights.

## 4.5. Critical Dilaton Gravity

In the field of JT gravity with defects, such as conical defects and branes, it has been proposed that these can be integrated out, leading to an effective dilaton action [150, 113, 77], which is a deformation of pure JT gravity. While the formality of this approach may remain subject to an ongoing debate, we adopt this particular conjectural view as a means to analyze non-generic critical models within the field of thermodynamic solutions in dilaton gravity.

Dilaton gravity is a 2-dimensional theory defined by the action

$$I_{dil} = \frac{1}{2} \int_{\mathcal{M}} \sqrt{g}(\Phi R + W(\Phi)). \tag{4.59}$$

where $\Phi$ is the dilaton and the potential is given by[5]

$$W(\Phi) = 2\Phi + e^{-2\pi\Phi} \int_0^\infty d\ell \, \mu(\ell) \cos(\ell\Phi). \tag{4.60}$$

The first term corresponds to the pure JT gravity case and the second term is determined by the weight per geodesic boundary. This potential satisfies the asymptotically $AdS_2$ condition, i.e. $W(\Phi) \sim 2\Phi$ as $\Phi \longrightarrow +\infty$. Yet, its closed form

---

[5][123] highlights the caveats of constructing the dilaton potential for a time modification different from the conical defect.

is only known for a few cases e.g. FZZT branes.

We can also write it as a power series

$$W(\Phi) = 2\Phi + e^{-2\pi\Phi} \sum_{m=0}^{\infty} \frac{(-1)^m}{(2m)!} \Phi^{2m} \int_0^\infty d\ell \, \mu(\ell) \, \ell^{2m}, \qquad (4.61)$$

which makes it clear that the weight determines the coefficients of the expansion. In particular, for critical JT gravity, we compute these coefficients using the form of the non-generic critical weight we derived in Subsection 4.2.1, $\mu_\alpha(\ell) = t_{crit} \, e^{-\ell \sqrt{E_\mu}} \ell^{-\alpha} \Theta(\ell - \epsilon)$ with $\epsilon > 0$ a cutoff we use to ensure that $\mu_\alpha(\ell) < \infty$ for all $\ell$. We get that

$$\int_0^\infty d\ell \, \mu_\alpha(\ell) \, \ell^{2m} = t_{crit} \, \epsilon^{-\alpha+2m+1} E_{\alpha-2m} \left( \sqrt{E_\mu} \epsilon \right), \qquad (4.62)$$

where $E_\gamma(x)$ is the generalized exponential integral function. Since the second term of the dilaton potential (4.61) rapidly becomes subdominant for $\Phi > 1$, due to the exponential suppression, the smallest values of $m$ determine the general features of the potential. Figure 4.11 (left) shows the first coefficient ($m = 0$) of this potential for different values of $\epsilon$. In particular, when the cutoff is removed ($\epsilon = 0$), this term becomes $\left( \sqrt{E_\mu} \right)^{\alpha-1} \Gamma(1 - \alpha)$, which diverges at $\alpha = 2$. This is intriguing since we hypothesize that $\alpha = 2$ corresponds to a phase transition of the hyperbolic $O(n)$ model. The plot of the first three coefficients of the dilaton potential in this scenario (Figure 4.11, right) reveals that only the term for $m = 0$ exhibits discontinuity at $\alpha = 2$. Upon further examination of (4.62) with $\epsilon = 0$, it can be proven that all coefficients $m > 0$ remain continuous. We will further analyze the case $\epsilon = 0$, in this case, the dilaton potential is

$$W(\Phi) = 2\Phi + t_{crit} \, e^{-2\pi\Phi} \sum_{m=0}^{\infty} \frac{(-1)^m}{(2m)!} \Phi^{2m} \left( \sqrt{E_\mu} \right)^{\alpha-2m-1} \Gamma(2m + 1 - \alpha), \qquad (4.63)$$

and it is shown in Figure 4.12 for different values of $\alpha$ and in Figure 4.13 for different values of $t_{crit}$ for the dilute ($\alpha < 2$) and dense ($\alpha > 2$) phases.

It should be emphasized that in the case $\epsilon = 0$ the weight $\mu_\alpha(\ell) = e^{-\ell \sqrt{E_\mu}} \ell^{-\alpha}$ is not finite at $\ell = 0$, so it is not an admissible weight in the probabilistic sense. But, as we will show in the following subsection, this is an interesting thermodynamical case that further relates this non-generic critical family of models with FZZT and EOW branes, in the latter case, the weight (4.49) also diverges at $\ell = 0$.

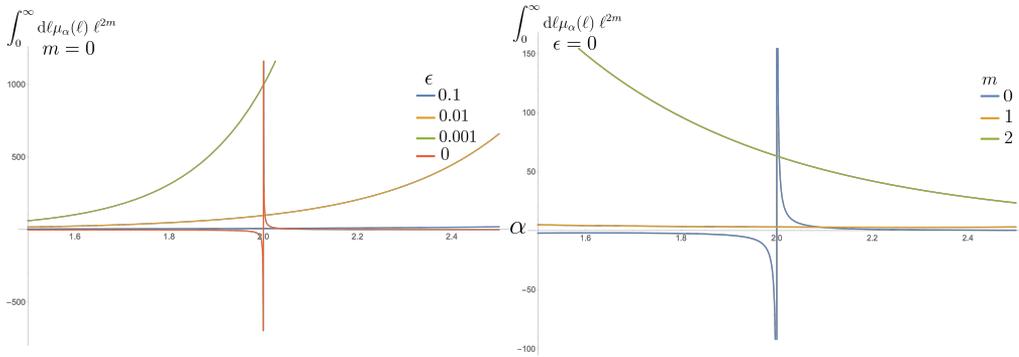

**Figure 4.11.:** Left: The $m = 0$ coefficient (4.62) for decreasing values of $\epsilon$. Right: First three coefficients of the potential (4.61) for $\epsilon = 0$ as functions of $\alpha$. In both cases, $E_\mu = 0.1$ and $t_{crit} = 1$.

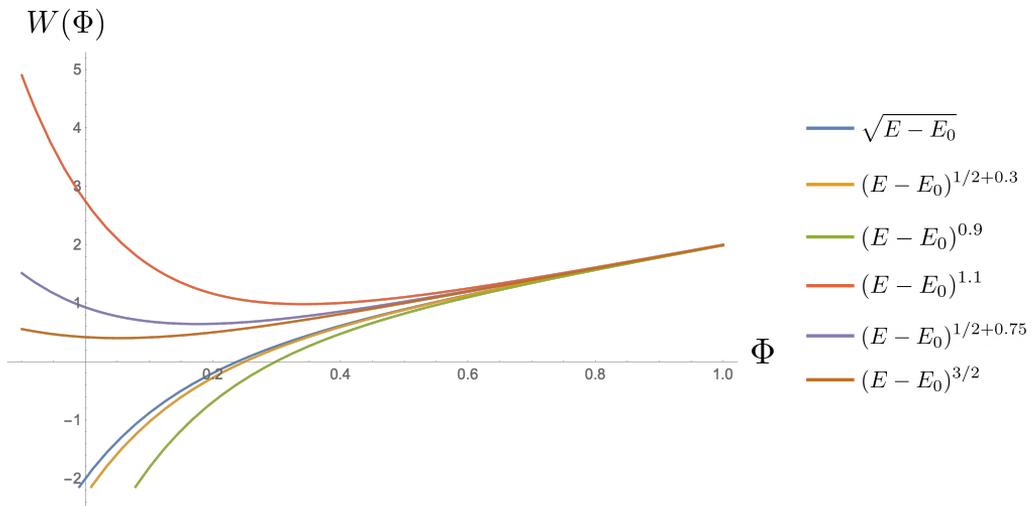

**Figure 4.12.:** Dilaton potential (4.63) for different values of $\alpha$, $t_{crit} = 1$ and $E_\mu = 0.1$. We label the graphs using the corresponding energy density they correspond to.

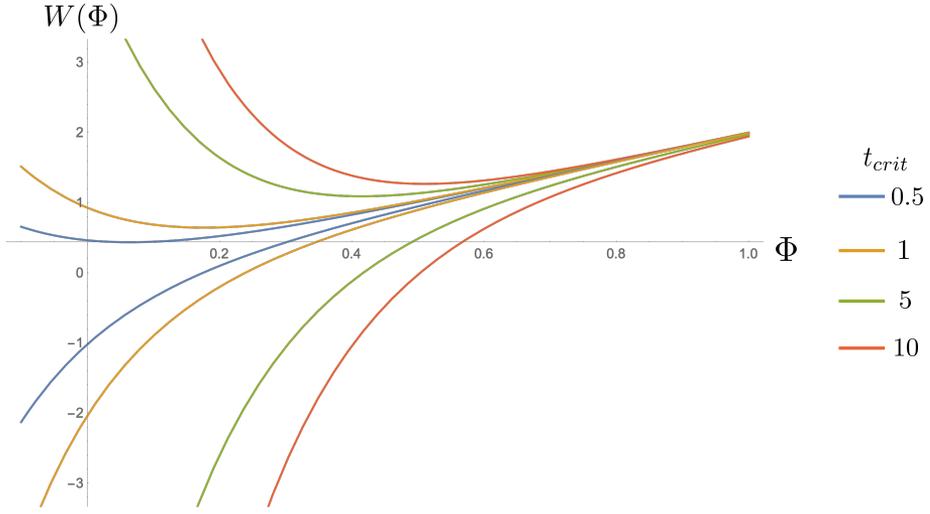

**Figure 4.13.:** Dilaton potential (4.63) for different values of $t_{crit}$ and $E_\mu = 0.1$. The top four graphs correspond to $\alpha = 2.25$ $((E - E_0)^{5/4})$ and the bottom four correspond to $\alpha = 1.75$ $((E - E_0)^{3/4})$.

## 4.5.1. Thermodynamics

Following the steps of [149], we can write the classical euclidean black hole solution to the equations of motion of (4.59) in the gauge $\Phi(r) = r$,

$$ds^2 = A(r)^2 dt^2 + \frac{1}{A(t)^2} dr^2, \qquad A(r) = \int_{r_h}^r dx\, W(x) \qquad (4.64)$$

where $r = r_h$ is the position of the euclidean horizon. To have a well-defined exterior solution, we need this function to be positive $A(r) > 0$ for $r > r_h$ and vanish at the horizon. Expanding around the horizon one gets $A(r) = W(\Phi_h)(r - r_h) + o((r - r_h)^2)$, therefore the latter condition translates into $W(\Phi_h) > 0$. Substituting this expansion in the metric and demanding smoothness at $r = r_h$, one obtains the euclidean temperature of the black hole

$$T(\Phi_h) = \frac{W(\Phi_h)}{4\pi}. \qquad (4.65)$$

Additionally, it can be observed that the energy difference between two black holes with horizon radii $\Phi(r_1) = \Phi_1$ and $\Phi(r_2) = \Phi_2$ is

$$\Delta E = \frac{1}{2} \int_{\Phi_1}^{\Phi_2} d\Phi\, W(\Phi). \qquad (4.66)$$

Thermodynamically speaking, a system is stable if its heat capacity is positive, i.e. $\frac{dE}{dT} > 0$, which implies that the stability of the euclidean black hole solution with a horizon at $\Phi_h$ in the canonical ensemble ($T$ fixed) requires that $W'(\Phi_h) > 0$. In the case of our potential (4.63), it is monotonically increasing in the dilute phase ($3/2 < \alpha < 2$) and convex in the dense phase ($2 < \alpha < 5/2$), as shown in Figure 4.12. Qualitatively similar behavior has been observed in [124], but in that case, one regime corresponds to FZZT branes and the other to anti-FZZT branes and the density of states changes abruptly from $\sqrt{E - E_0}$ to $(E - E_0)^{3/2}$. On the other side, in [77], where they observe the same change in $\rho_0(E)$ ($\sqrt{E - E_0}$ to $(E - E_0)^{3/2}$) for EOW branes, the dilaton potential is always convex. In our case, this family of non-generic critical models continuously interpolates between these two regimes, and this change is directly linked to the varying scaling of $\rho_0$.

To exemplify the thermodynamics of this family of black holes, we showcase four instances of the potential (4.63) for different values of $\alpha$ in Figure 4.14. We start by fixing a temperature $T$. The first observation is that in the dilute phase, there is only one black hole solution and it is stable. On the other side, in the dense phase, there are either no black hole solutions (yellow graph), one thermodynamically unstable solution (green), or two black holes (red), one unstable ($\Phi_1$) and one stable ($\Phi_3$). Therefore, in this case, the only two stable black holes at temperature $T$ are $\Phi_3$ and $\Phi_4$. As it was noted in [149], the stable solution that minimizes the free energy of the system is the one with the largest horizon, which in this case corresponds to $\Phi_4$. In other words, given a fixed temperature and a fixed value for $t_{crit}$, there is a thermodynamically preferred value of $\alpha$ and, therefore, in the brane picture, a preferred geometry, either dense or dilute (see Figure 4.10), the latter is the case for this example.

Another remark we want to make is that the two thermodynamical conditions of the potential, $W(\Phi_h) > 0$ and $W'(\Phi_h) > 0$, define the physical consistency requirements for any non-generic critical potential we can construct, which are additional constraints for the form of the weights $\mu_\alpha(\ell)$. So, if we start with a general potential (4.60) without assuming the exact form of $\mu_\alpha$ and want to construct thermodynamically stable black hole solutions, it implies that

$$0 < \exp^{-2\pi\Phi_h} \int_0^\infty d\ell \; \ell\mu_\alpha(\ell) \sin(\ell\Phi_h) < 2 + 4\pi\Phi_h \qquad (4.67)$$

needs to be satisfied. It is to be remarked that this inequality admits weights that diverge at $\ell = 0$, even though these are not allowed if we want a proba-bilistic interpretation of non-generic critical JT gravity. Furthermore, in terms of the physical interpretation of this family of non-generic critical models, they can be interpreted as a family of thermodynamic stable brane solutions with finite entropy, to which evaporating black holes can decay. However, it is important to note that the rigorous derivation of these branes from String Theory is still an open question.

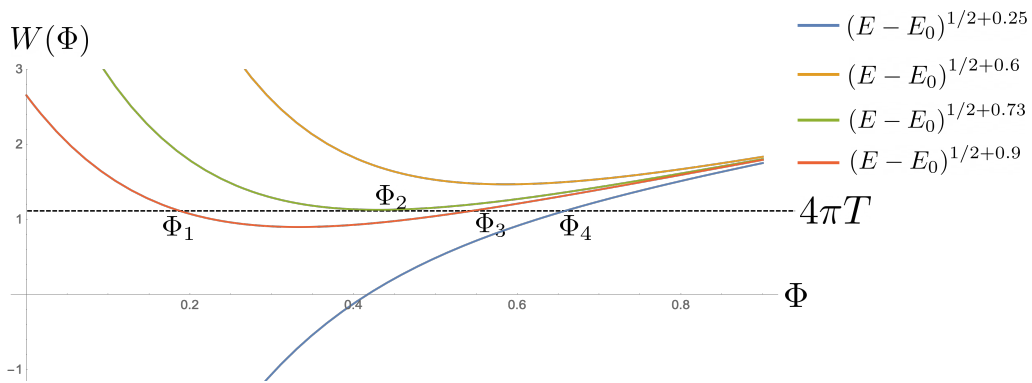

**Figure 4.14.:** The potential (4.63) for $\alpha = 1.75$ (blue), $\alpha = 2.1$ (yellow), $\alpha = 2.23$ (green) and $\alpha = 2.4$ (red), and fixed values $t_{crit} = 5$ and $E_\mu = 0.1$. A fixed temperature $T$ is shown as a dotted line. The black hole solutions at temperature $T$ are denoted by $\Phi_1$, $\Phi_2$, $\Phi_3$ and $\Phi_4$ labelled by increasing size.

## 4.6. Spectral form factor

The spectral form factor in JT gravity is given by the time-ordered two-point correlation function, which in geometrical terms corresponds to the double trumpet for genus zero and no defects. In matrix models, it displays distinctive and universal characteristics, notably the presence of a ramp followed by a plateau structure. Interestingly, in JT gravity the spectral form factor exhibits a ramp-like behavior without a well-defined plateau. The absence of a plateau is a consequence of the geometric structure of the double trumpet, whose dominant contribution leads to a continuous growth of the spectral form factor without reaching a plateau. The absence of a plateau in the spectral form factor indicates that the dual CFT does not possess a well-defined time scale for thermalization.

In earlier studies (e.g. [123]), it has been investigated if a plateau can be recovered with the inclusion of branes. Hence, it arises as a natural question to investigate it using our configuration of branes. In our case, the main contribution to the double trumpet corresponds to two trumpets (or half-wormholes) glued at geodesic boundaries, each with non-generic critical weight $\mu_\alpha(\ell) = e^{-\ell \sqrt{E_\mu}} \ell^{-\alpha}$ (see Figure 4.15). This corresponds to

$$\mathcal{Z}_{0,2}(\beta_1, \beta_2) = \int_0^\infty \mathrm{d}b \left( \mu_\alpha(b) \frac{e^{-b^2/4\beta_1}}{2\sqrt{\pi\beta_1}} \right) \left( \mu_\alpha(b) \frac{e^{-b^2/4\beta_2}}{2\sqrt{\pi\beta_2}} \right) \tag{4.68}$$

$$= \frac{t_{crit}}{4\pi} \frac{\Gamma(1-\alpha) \left( \frac{\beta_1\beta_2}{\beta_1+\beta_2} \right)^{\frac{1-\alpha}{2}} U\left( \frac{1}{2} - \frac{\alpha}{2}, \frac{1}{2}, E_\mu \frac{\beta_1\beta_2}{\beta_1+\beta_2} \right)}{\sqrt{\beta_1\beta_2}}, \tag{4.69}$$

where $U$ is the confluent hypergeometric function. Therefore, the spectral form factor is

$$\mathcal{Z}_{0,2}(\beta + it, \beta - it) = \frac{t_{crit}}{\pi} \frac{2^{\alpha-\frac{5}{2}}\Gamma(1-2\alpha) \left( \beta + \frac{t^2}{\beta} \right)^{\frac{1}{2}-\alpha} U\left( \frac{1}{2} - \alpha, \frac{1}{2}, \frac{2E_\mu(t^2+\beta^2)}{\beta} \right)}{\sqrt{\beta^2 + t^2}}. \tag{4.70}$$

This is shown in Figure 4.16, where we can observe that the spectral form fac-tor grows linearly without reaching a plateau. Regardless, this is a computation at $g = 0$ and $\hbar \to 0$, so the possibility that these and more non-perturbative corrections change the late time behavior or (4.70) cannot be ruled out. It is also worth noting that the low temperature and late time ($\beta \gg 1$ and $t \gg 0$) behavior of the spectral form factor does not depend on the values of $\alpha$.

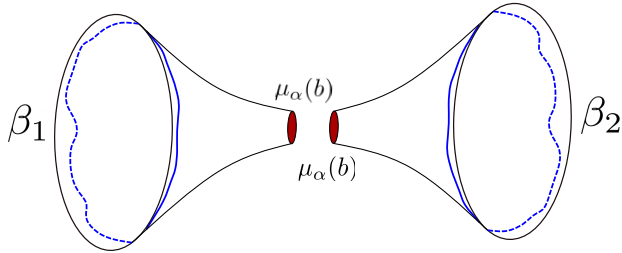

**Figure 4.15.:** Double trumpet contribution arising from gluing two half-wormholes ending each on a brane with non-generic critical weight.

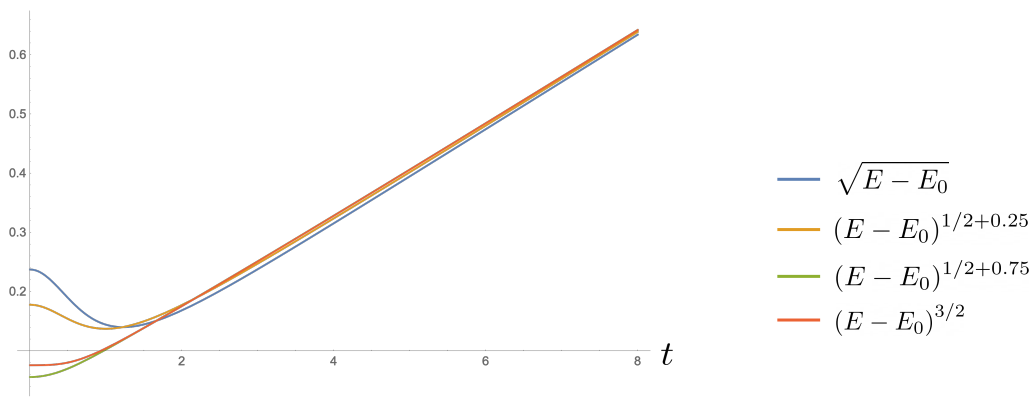

**Figure 4.16.:** Spectral form factor (4.70) for different values of $\alpha$, $t_{crit} = 1$, $\beta = 50$ and $E_\mu = 0.1$.

## 4.7. Discussion

Probably the most important result of this chapter is to show how JT gravity and random geometry are deeply connected and how tools from both fields are useful for computational goals but also for our conceptual understanding. In this paper, we explored the random geometry aspects of JT gravity, offering insights into its critical regimes. Furthermore, we have remarked once again 'the power of the string equation', already highlighted in multiple works [93, 94, 140], noting its capacity not only to compute partition functions and density of states at the leading order but as a translation tool between random geometry and JT gravity.

By using tools coming from the study of Boltzmann planar maps, we uncovered what we call critical phases of JT gravity, which bear resemblance to phase transitions observed in planar maps, from surfaces with microscopic holes in the subcritical regime, to those with macroscopic ones in the non-generic critical regime. The occurrence of such transitions is the result of fine-tuning the weights of geodesic boundaries, which can be thought of as brane configurations. Furthermore, we made this analogy precise by determining the relation between hyperbolic surfaces with non-generic critical weights with $3/2 < \alpha < 5/2$ and the hyperbolic $O(n)$ loop model with $n = 2\sin\left(\pi\left(\alpha - \frac{1}{2}\right)\right)$.

We showed that the different values of $\alpha$ significantly impact the density of states of the system. Through our analysis, we have identified a family of models that smoothly interpolate between the density of states behaviors of $\rho_0(E) \sim \sqrt{E - E_0}$ and $\rho_0(E) \sim (E - E_0)^{3/2}$, thereby establishing connections to JT gravity coupled to dynamical EOW branes [77] and FZZT branes [124]. Moreover, we showed in which strict sense this transition from $\rho_0(E) \sim \sqrt{E - E_0}$ to $\rho_0(E) \sim (E - E_0)^{3/2}$ in the density of states is a proper phase transition in the physical sense and in which sense it is different compared to other transitions that have been identified previously [96]. This is when the expected number of geodesic boundaries, or defects, diverge. Additionally, by analysing the effective dilaton potential for these models, one can distinguish between two subsets of models, $\alpha > 2$ and $\alpha < 2$ which suggests another phase transition at $\alpha = 2$.

In light of our results, there are several interesting paths for future research to explore. Firstly, it remains essential to address the non-perturbative stability of the identified critical regimes and their compatibility with alternative definitions of EOW branes [94]. This is related to an important question that arises, whether there exists a matrix model whose double-scaled density of states follows the form $(E - E_0)^{\alpha-1}$. Such a matrix model was constructed in [4], but its non-perturbative study poses a big challenge due to the infinite order of its KdV differential equations [5]. We plan on addressing this and investigating the properties of such a matrix model.

Additionally, from the mathematical point of view, exploring the phase transition at $\alpha = 2$ in the hyperbolic $O(n)$ loop model is an open problem. Our conjecture is that a transition from non-touching geodesic boundaries to overlapping ones should occur similarly to the planar map case. This line of work may help to better understand the geometries in Figure 4.10 and their physical meaning either from a fluctuating or from a decorated hyperbolic geometry point of view in a similar way as for planar maps.

Furthermore, we have presented a way to couple JT gravity to any kind of (admissible) defects using the string equation. This allows us to compare different results in the JT gravity literature in a single and precise context. For example, in [26], the weights are chosen to achieve factorization in JT gravity. Such weight, $\mu(\ell_1, \ell_2) = -\frac{1}{\ell_1}\delta(\ell_1 - \ell_2)$, is bilocal and negative valued which is not compatible with a probabilistic interpretation, but allows to 'unglue' the double trumpet allowing for the factorization of the two boundary amplitude $\langle Z(\beta_1)Z(\beta_2)\rangle$. An even more interesting case for us is the weight choice used in [27] to obtain a discrete JT gravity spectrum. In this context, our non-generic critical weights are equivalent to finding a background matrix $H_0$ such that $\text{Tr}(H_0)^k \sim k^{-\alpha}$ for $k \gg 1$. In this regime, when big holes appear in the hyperbolic surface, the density of states $\rho_0$ is conjectured to become discrete. Nevertheless, our results for $\rho_0$ do not agree with it, its leading order behavior changes from a square root in a non-trivial way but the spectrum stays continuum for all $\alpha$. The incompatibility of these two results poses an interesting avenue for further research. The last case we want to comment on is [124], where the weights of anti-FZZT branes were tuned to reproduce a Page-like curve which exactly corresponds to a transition from $\rho_0(E) \sim \sqrt{E - E_0}$ to $\rho_0(E) \sim (E - E_0)^{3/2}$. Since our models interpolate between these two regimes, we think that the exploration of the non-generic critical phase raises intriguing questions regarding the implications for the Page curve and black hole evaporation.

Finally, we want to make a comment on the significance of non-generic critical models from the holographic point of view. One could forget about the geometry of the bulk and only look at the density of states of the quantum system at the holographic boundary. Then, in the subcritical regime $\rho_0(E) \sim \sqrt{E - E_0}$, which is the leading order behavior of the Schwarzian. However, in the non-generic critical regime, the density of states behaves as $\rho_0(E) \sim (E - E_0)^{\alpha-1}$, so a question that arises is if the transition from a subcritical to a non-generic critical regime can be put in the framework of deformations of the Schwarzian or if there are quantum systems that naturally present this the non-generic critical behavior, but we leave this for future work.

# Irreducible random metric maps and Topological Gravity

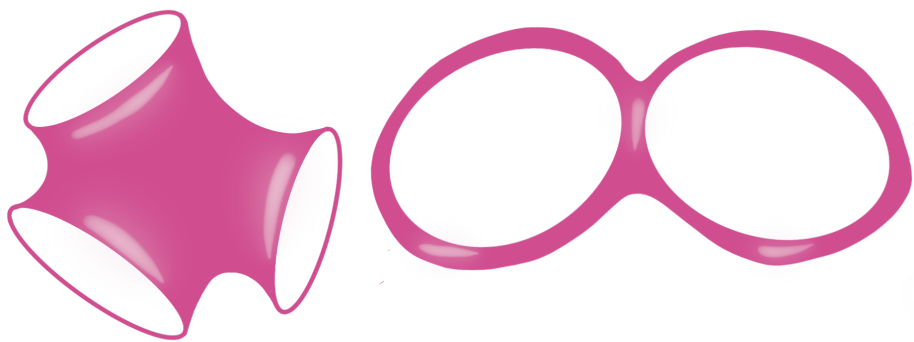

**Figure 5.1.:** A hyperbolic surface and a fatgraph.

So far in this thesis, we have explored various facets of random geometries and quantum gravity. We began by formally defining the 2D quantum gravity action as a probability measure, delving into the associated random geometry configurations inherent to Liouville Quantum Gravity. As we narrowed our focus to spacetimes with negative curvature, we entered the domain of hyperbolic surfaces and JT gravity. There, the AdS/CFT correspondence revealed that the holographic theory is precisely a random matrix theory. This observation raised the question of whether a well-defined combinatorial interpretation associated with the Feynman expansion of matrix models could extend to the realm of hyperbolic surfaces and by extension, JT gravity.

To address this question, we employ the framework of topological gravity [57]. Topological gravity is a mathematical framework that explores the inter-section theory on the moduli space of Riemann surfaces, and it plays a pivotal role in our exploration. Notably, topological gravity can be precisely expressed as a specific matrix model [101], ultimately giving rise to structures known as metric maps. Metric maps are maps with real edge lengths. The remarkable point is that JT gravity represents a specific instance within the broader framework of topological gravity.

It is well-established that a hyperbolic surface resembles a metric map when the lengths of its geodesic boundaries are very large [61]. Moreover, the integrable structure of WP volumes resembles the one in matrix models, both in the form of topological recursion [68]. However, as we have seen, random matrix models count surface discretizations and have a combinatorial explanation. These observations give rise to a central question: Can we establish a combinatorial interpretation for hyperbolic surfaces, ideally as a form of discretization of these surfaces? To address this question, we turn to the concept of irreducible metric maps.

Irreducible metric maps, which have been previously introduced in [36], possess an interesting yet not fully understood connection to hyperbolic surfaces and Weil-Peterson volumes. Remarkably, by means of a weight substitution procedure, these irreducible metric maps can be transformed into the very structures that emerge in Kontsevich's matrix model, namely, metric maps. In this final chapter, we approach the study of random metric maps and hyperbolic surfaces to bridge the gap between continuous and discrete geometry. This chapter is mainly based on work in progress [38].

## 5.1. Random metric maps

A **genus-g metric map** (mm)[101] is a genus-g map with all vertices of degree at least three and a positive real-valued label on each edge, called *length of the edge*. The length of a face is the sum of the lengths of its edges. Unlike genus-g maps, the length of the edges can be different from 1, therefore, the length of a face is in general different from its degree. Note that since vertices of degree 1 and 2 are not allowed, the 'smallest' planar metric map has at least 1 vertex and 2 edges, therefore, by Euler's formula, $|F(\mathrm{mm})| \geq 3$.

Let $\mathcal{M}^{mm}_{g,n}(\ell_1, \dots, \ell_n)$ be the set of genus-$g$ metric maps with $n$ faces of lengths $\{\ell_i\}_{1 \leq i \leq n}$. Similarly to hyperbolic surface case, the 'cardinality' of this set is given by the volumes $V^{mm}_{g,n}(\ell_1, \dots, \ell_k)$. We can define the generating function of metric map volumes of genus $g$ as

$$F_g^{mm}[\mu] = \sum_{k=1}^{\infty} \frac{1}{k!} \int_0^\infty d\ell_1 \mu(\ell_1) \cdots \int_0^\infty d\ell_k \mu(\ell_k) V_{g,n}^{mm}(\ell_1, \ldots, \ell_k), \quad (5.1)$$

where $\mu(\ell)$ are weights. By this point, it is no surprise that $F_g[\mu]$ can be written in terms of a string equation, as we will show in Section 5.1.1. For now, let us introduce the string equation

$$x(u) = u - \int_0^\infty d\ell \mu(\ell) I_0 \left( \ell \sqrt{u} \right). \quad (5.2)$$

Alternatively, this equation can be expressed as

$$x(u) = u - \sum_{k=0}^{\infty} \frac{t_k}{k!} u^k, \quad (5.3)$$

where the times are given by (4.8). The genus-$g$ generating function is proven to be [90]

$$F_0^{mm}[\mu] = \frac{1}{4} \int_0^{E_\mu} dr \ (x(u))^2, \quad (5.4)$$

$$F_1^{mm}[\mu] = -\frac{1}{24} \log M_0^{mm}, \quad (5.5)$$

$$F_g^{mm}[\mu] = \left( \frac{2}{(M_0)^2} \right)^{g-1} \mathcal{P}_g \left( \frac{M_1}{M_0}, \ldots, \frac{M_{3g-3}}{M_0} \right) \quad \text{for } g \geq 2, \quad (5.6)$$

where $E_\mu > 0$ is the smallest solution to the string equation $x(E_\mu) = 0$, $\mathcal{P}_g$ is given by (4.15) and the moments $M_p^{mm}$ are

$$M_0^{mm} = \left( \frac{\delta E_\mu}{\delta \mu(0)} \right)^{-1}, \qquad M_p^{mm} = M_0^{mm} \frac{\delta M_{p-1}^{mm}}{\delta \mu(0)} \quad (5.7)$$

or, alternatively

$$M_p^{mm} = \frac{\partial^{p+1} x}{\partial u^{p+1}}(E_\mu). \quad (5.8)$$

Another important quantity in this context is the disk function which corresponds to the generating function of genus-0 metric maps with one marked

face. This is

$$W_\mu^{mm}(\ell) = \frac{\delta F_0^{mm}[\mu]}{\delta \mu(\ell)} = -\frac{1}{2} \int_0^{E_\mu} du I_0 \left( \ell \sqrt{u} \right) x(u) \tag{5.9}$$

$$= \sum_{k=2}^\infty \frac{1}{k!} \int_0^\infty d\ell_1 \mu(\ell_1) \cdots \int_0^\infty d\ell_k \mu(\ell_k) V_{0,k+1}^{mm}(\ell, \ell_1 \ldots, \ell_k). \tag{5.10}$$

This enumeration allows us to define a probability density measure on the space of Boltzmann genus-$g$ metric maps,

$$p^{mm}(\ell_1, ..., \ell_k) = \frac{1}{k!} \frac{\prod_{i=1}^k \mu(\ell_i)}{F_g^{mm}[\mu]} \frac{\omega_{mm}^{3g-3+n}}{(3g-3+n)!}, \tag{5.11}$$

where $\omega_{mm}/(3g-3+n)!$ is the volume form. In this way, given that $F_0^{mm}[\mu] < \infty$, we can define a **random genus-$g$ metric map** with $k$-boundaries of random lengths to a surface sampled with probability (5.11).

At this point, one can observe a striking similarity in the structure of generating functions for both metric maps and hyperbolic surfaces. This resemblance becomes particularly apparent at the level of the string equation: (5.3) and (4.6) are identical, differing only by the identification $\gamma_k = 0$. This makes us wonder if there is a broader framework that encompasses both types of surfaces. The answer is yes and it is explained in the forthcoming section through the concept of intersection numbers

## 5.1.1. Intersection numbers and matrix models

As discussed in section 4.1, we can construct a compact Riemann surface with $n$ marked points by attaching a disc to each boundary of a Riemann surface with $n$ boundary components (See Figure 4.3). Within this moduli space, we briefly introduced intersection numbers (4.3). Intersection numbers, which are topological invariants, play a crucial role in understanding the geometry and topology of moduli spaces of Riemann surfaces, generalizing the Euler characteristic.

In the context of hyperbolic surfaces, we noted that Weil-Petersson volumes can be expressed in terms of intersection numbers in conjunction with the Mumford-Morita-Miller class $\kappa_1$, which is formed by local coordinates arising from the pairs of pants decomposition.

For metric maps, a similar approach can be taken, although we generally do not have the pairs of pants decomposition. Therefore, we can omit the $\kappa_1$ class and represent the volume of metric maps as follows [101]

$$V_{g,n}^{mm}(\ell_1, \ldots, \ell_n) = 2^{5-5g-2n} \sum_{d_0+d_1+\cdots+d_n=d_{g,n}/2} \prod_{i=1}^{n} \frac{\ell_i^{2d_i}}{d_i!} \langle \tau_{d_1} \ldots \tau_{d_n} \rangle_{g,n} \quad (5.12)$$

with $d_1, \ldots, d_n \geq 0$.

This means that the generating functions for both Weil-Petersson volumes and metric maps can be expressed in terms of intersection numbers. Moreover, they can be expressed in terms of a common generating function $G_g(s, t_0, t_1, \ldots)$ defined as follows

$$G_g(s, t_0, t_1, \ldots) = \sum_{n=0}^{\infty} \sum_{m+\sum_i n_i = \frac{d_{g,n}}{2}} \prod_{i=0} \frac{t_i^{n_i}}{n_i!} \frac{s^m}{m!} \langle \kappa_1^m \tau_0^{n_0} \tau_1^{n_1} \ldots \rangle_{g,n}. \quad (5.13)$$

Here $m, n_0, n_1, \cdots \geq 0$ and $\sum_i n_i = n$. It can be observed that

$$F_g^{mm}[\mu] = 2^{g-1} G_g(0, t_0, t_1, \ldots), \quad (5.14)$$

$$F_g^{WP}[\mu] = 2^{g-1} G_g(\pi^2, t_0, t_1, \ldots), \quad (5.15)$$

where the times $t_i$ are given by (4.8). Here, one might perceive this construction as tailored to accommodate both metric maps and hyperbolic surfaces. However, it is important to note that the underlying formulation and formal proofs of (5.14) and (5.15) are linked to the underlying integrable structure of $\mathcal{M}_{g,n}$. While a detailed exposition of this aspect is beyond the scope of this discussion, interested readers can check [110] for a detailed explanation and proofs. In the physics literature, this field is known as topological gravity [57].

Furthermore, as demonstrated in Kontsevich's proof of Witten's conjecture [147, 101], matrix models provide a powerful method to address the challenge of efficiently computing quantities associated with the moduli space of Riemann surfaces. The Kontsevich matrix model [101] offers a practical way to compute intersection numbers of $\mathcal{M}_{g,n}$ by evaluating matrix integrals. The key insight is that given a fatgraph of genus $g$ and $n$ with lengths real lengths on its edges $\ell_e$, a Riemann surface of genus $g$ and $n$ marked points can be constructed by gluing $n$ disks of length $L_{e'} = \sum_{e \in f} \ell_e$ along the faces of the fatgraph. This observation further extends the relation of metric maps and hyperbolic surfaces to the enumeration of maps and it gives rise to a natural question: whether we can provide a combinatorial or even geometrical interpretation for the relation between metric maps, matrix models and hyperbolic surfaces beyond the framework of the Kontsevich Matrix Model. We will address this question through the study of irreducible metric maps.

## 5.2. Irreducible metric maps

A genus-0 metric map is a ***q-irreducible metric map*** if each cycle has a length bigger or equal to $q$ and equal to $q$ only if it bounds a face of length $q$. For $g > 0$, a genus-$g$ metric map is said to be a $q$-irreducible metric map if its universal cover is $q$-irreducible.

The procedure to go from discrete irreducible maps to irreducible metric maps and from discrete maps to metric maps is basically the same. The main idea is that given an q-irreducible map, we can construct a $q$-irreducible metric map by deleting vertices of degree one, deleting $k-1$ adjacent vertices of degree two to form a single edge of total length $L_k$, and assigning a rescaled length $\ell_k = \frac{q}{2q} L_k$ to each resulting edge of the metric map (See Figure 5.2). This construction has the property that, in the large face-degree limit, the counting measure of discrete maps (upon appropriate rescaling) converges to the measure on $\mathcal{M}_{g,n}^{mm}(\ell_1, \dots, \ell_n)$ for metric maps and $\mathcal{M}_{g,n}^{(q)}(\ell_1, \dots, \ell_n)$ for irreducible metric maps. This means that the enumeration problem allows us to solve the continuum volume problem.

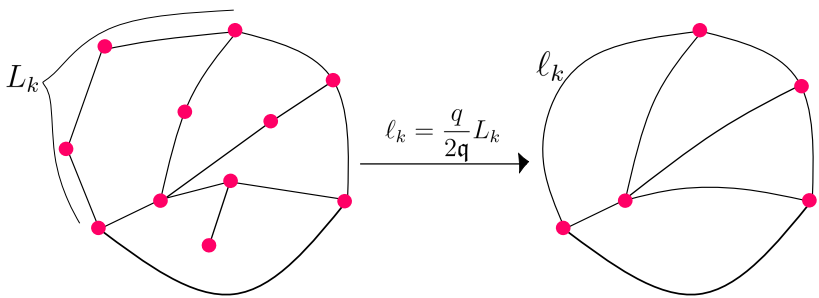

**Figure 5.2.:** Starting with a q-irreducible map (Left), one can construct a $q$-irreducible metric map (Right) by removing all vertices of degree one, and joining all edges with vertices of degree two by deleting the vertices between them. The lengths of the metric map are rescaled versions of the discrete map. The resulting metric map is 3-irreducible.

The enumeration results for $q$-irreducible metric maps, as detailed in [36], can be summarized as follows. Let $F^{(q)}[\mu]$ be the generating function of the corresponding volumes defined as

$$F_g^{(q)}[\mu] = \sum_{k=1}^{\infty} \frac{1}{k!} \int_q^{\infty} d\ell_1 \mu(\ell_1) \cdots \int_q^{\infty} d\ell_k \mu(\ell_k) V_{g,k}^{(q)}(\ell_1 \ldots, \ell_k), \qquad (5.16)$$

where $\mu : [q, \infty) \longrightarrow \mathbb{R}$ is a weight function playing the role of a generating variable for the series. This can be written in terms of the string equation

$$x^{(q)}(u) = \frac{2\sqrt{u}}{q} J_1(q\sqrt{u}) - \int_q^{\infty} d\ell \mu(\ell) I_0 \left( \sqrt{(\ell^2 - q^2)r} \right). \qquad (5.17)$$

Expanding the Bessel functions we can also write the string equation as

$$x^{(q)}(u) = u - \sum_{k=0}^{\infty} \frac{t_k^{(q)} + \gamma_k^{(q)}}{k!} u^k, \qquad (5.18)$$

with

$$\gamma_k^{(q)} = \frac{(-1)^k}{(k-1)!} \left( \frac{q}{2} \right)^{2k-2} 1_{k \geq 2} \qquad (5.19)$$

and

$$t_k^{(q)}[\mu] = \frac{4^{-k}}{k!} \int_q^{\infty} d\ell \mu(\ell) \left( \sqrt{\ell^2 - q^2} \right)^{2k}. \qquad (5.20)$$

Then, the genus-$g$ generating functions $F_g^{(q)}$ are given by

$$F_0^{(q)}[\mu] = \frac{1}{4} \int_0^{E_\mu^{(q)}} dr \left( x^{(q)}(r) \right)^2, \qquad (5.21)$$

$$F_1^{(q)}[\mu] = -\frac{1}{24} \log M_0^{(q)}, \qquad (5.22)$$

$$F_g^{(q)}[\mu] = \left( \frac{2}{(M_0^{(q)})^2} \right)^{g-1} \mathcal{P}_g \left( \frac{M_1^{(q)}}{M_0^{(q)}}, \ldots, \frac{M_{3g-3}^{(q)}}{M_0^{(q)}} \right) \quad \text{for } g \geq 2, \qquad (5.23)$$

where $E_\mu^{(q)}$ is the formal power series solution to $x^{(q)} \left( E_\mu^{(q)} \right) = 0$, $\mathcal{P}_g$ is given by (4.15) and the moments $M_p^{(q)}$ can be defined recursively via

$$M_0^{(q)} = \left( \frac{\delta E_\mu^{(q)}}{\delta \mu(0)} \right)^{-1}, \qquad M_p^{(q)} = \left( \frac{1}{p!} \left( \frac{q}{2} \right)^{2p} + \delta_{\mu(0)} M_{p-1}^{(q)} \right) M_0^{(q)}, \qquad p \geq 1, \quad (5.24)$$

or, equivalently

$$M_p^{(q)} = \sum_{k=0}^{p} \frac{1}{(p-k)!} \left(\frac{q}{2}\right)^{2(p-k)} \frac{\partial^{k+1} x^{(q)}}{\partial u^{k+1}}(E_\mu).$$ (5.25)

We observe that the string equation for $q$-irreducible metric maps (5.17) exhibits similarities to the string equation of hyperbolic surfaces (4.9) when $q = 2\pi$. Furthermore, in [36], it was observed that the volumes of $q$-irreducible metric maps are closely related to Weil-Peterson volumes in the following way

$$V_{g,n}^{(2\pi)}(\ell_1, \dots, \ell_n) = 2^{2-2g-n} V_{g,n}^{WP} \left( \sqrt{\ell_1^2 - 4\pi^2}, \dots, \sqrt{\ell_n^2 - 4\pi^2} \right),$$ (5.26)

for $g = 0$ and $n \geq 3$ or $g = 1$ and $n \geq 1$. This relation is at the level of the total volume of the moduli space and it raises the question: Is there a bijection between hyperbolic surfaces and irreducible metric maps that can explain this coincidence? Two limits of this relation are known: for $\ell_i \longrightarrow \infty$, the irreducibility constraint becomes irrelevant so irreducible metric maps become metric maps. This same limit is also observed in the context of hyperbolic surfaces. For $\ell_i = 0$, a genus-0 hyperbolic surface with cusps resembles the boundary of a convex ideal polyhedron. Nevertheless, the exact relationship between $q$-irreducible metric maps and hyperbolic surfaces for finite values of the faces/geodesic boundaries $\ell_i$ remains a question.

This question can also be approached at the level of the generating functions of Weil-Petersson volumes and metric maps through the generating function $G$ in the following way: If we can construct a weight substitution mechanism analogous to the discrete case (2.23) but directly in the continuum, this will enable us to go from $q$-irreducible metric maps to metric maps without the irreducibility constraint. Then, we can use the relation between metric maps and hyperbolic surfaces at the level of intersection numbers and this could provide a better understanding of the link (5.26) between hyperbolic surfaces and irreducible metric maps.

## 5.3. Irreducible metric maps via substitution

As described in the last section, the computation of the volumes of $q$-irreducible metric maps in [36] relies on considering limits of the enumeration of irreducible discrete maps [35]. In this section we give a new derivation of the generating function of genus-0 irreducible metric maps without reference to discrete maps, in the hope to shed light on the relation with Weil-Petersson volumes. To this end we aim to formulate a substitution approach in the continuous setting relating volumes of irreducible metric maps to those of metric maps without irreducibility constraint.

### 5.3.1. The substitution equation

Suppose $q > 0$ and $\mu$ is a weight on $[q, \infty)$. For any $\ell > q$ we denote by

$$W^{(q)}(\ell)[\mu] = \frac{\delta F_0^{(q)}}{\delta \mu(\ell)} \tag{5.27}$$

$$= \sum_{k=2}^{\infty} \frac{1}{k!} \int_q^{\infty} \mathrm{d}\ell_1 \mu(\ell_1) \cdots \int_q^{\infty} \mathrm{d}\ell_k \mu(\ell_k) V_{0,k+1}^{(q)}(\ell, \ell_1 \ldots, \ell_k) \tag{5.28}$$

the disk function for $q$-irreducible metric maps. The idea of the substitution approach is to demonstrate that one may extend $\mu$ to a weight on $[0, \infty)$, by an appropriate choice of values on the interval $[0, q)$, such that we have the identity

$$W^{(q)}(\ell)[\mu] = W^{mm}(\ell)[\mu] \tag{5.29}$$

for $\ell > q$, where $W^{mm}(\ell)[\mu] = W^{(0)}(\ell)[\mu]$ is the disk function of metric maps without irreducibility constraint (5.10). Since $q$-irreducible metric maps form a strict subset of all metric maps, we have $V_{0,k+1}^{(q)}(\ell, \ell_1 \ldots, \ell_k) < V_{0,k+1}^{mm}(\ell, \ell_1 \ldots, \ell_k)$, and, therefore, (5.29) can only be satisfied if $\mu$ is allowed to become negative on $[0, q)$. More precisely, we need to ensure that the contributions of $q$-reducible (meaning not $q$-irreducible) metric maps to the volumes in $W^{mm}(\ell)[\mu]$ exactly cancel.

Suppose $\mathfrak{m}$ is a metric map with boundary length $\ell > q$. Let $\gamma$ be the shortest simple closed cycle in $\mathfrak{m}$, which generically is unique, and denote its length by $q' < \ell$. Then $\mathfrak{m}$ is $q$-irreducible if $q' > q$ and $q$-reducible if $q' < q$, while for $q = q'$ it depends whether $\gamma$ is the contour of a face of length $q$, but maps of the latter kind do not contribute to the volumes so we disregard them. If $\mathfrak{m}$ is $q$-reducible, we would like to cancel its contribution to $W^{mm}(\ell)[\mu]$. To this end, let us consider the set of all metric maps that are identical to $\mathfrak{m}$ on the exterior of $\gamma$, meaning all the edges on $\gamma$ or not surrounded by $\gamma$, but may differ from $\mathfrak{m}$ in the region surrounded by $\gamma$, but such that $\gamma$ is still the shortest closed cycle. One such map has $\gamma$ as the contour of a face of length $q'$, while for any other map the interior of $\gamma$ corresponds to a $q'$-irreducible metric map of boundary length $q'$. Vice versa, any $q'$-irreducible metric map of boundary length $q'$ can be glued inside $\gamma$ to obtain an element in the set (see Figure 5.3). This means that the contribution of this set of maps to $W^{(0)}(\ell)[\mu]$ is proportional to $\mu(q') + q' W^{(q')}(q')[\mu]$, where the factor of $q'$ accounts for the twisting freedom in the gluing operation. Hence, having this quantity vanish would be sufficient to ensure the identity (5.29). The question thus becomes whether we can choose $\mu$ on $[0, q)$ such that

$$\mu(q') + q' W^{(q')}(q')[\mu] = 0 \quad \text{for all } q' < q. \tag{5.30}$$

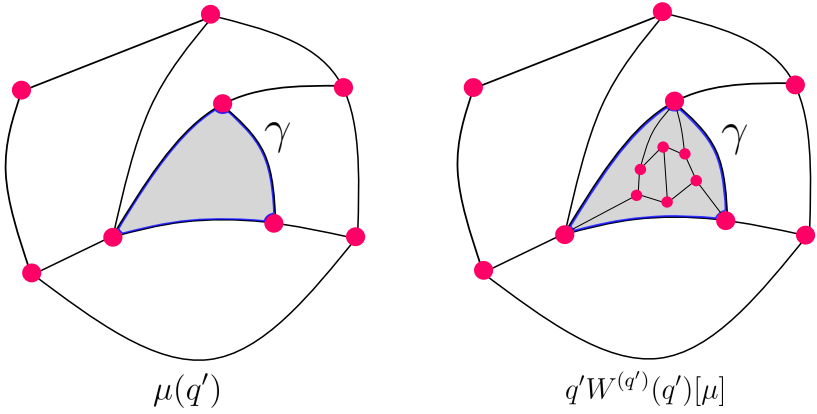

**Figure 5.3.:** A pair of metric maps that are identical outside of their shortest cycle $\gamma$ (shown in blue) of length $q'$ but differ on the inside of $\gamma$. The two cases are **Left**: $\gamma$ is the contour of a face of length $q'$ and **Right**: the interior of $\gamma$ is a $q'$-irreducible metric map of boundary length $q'$.

Now suppose $\ell < q$. Clearly, all metric maps with boundary length $\ell$ are $q$-reducible, and the same analysis applies to show that their contribution to $W^{(0)}(\ell)[\mu]$ cancels. The only difference is that now $\gamma$ can correspond to the boundary curve (and therefore $q' = \ell$), but by our definition of metric maps having at least three faces, there is no companion map that has the interior of $\gamma$ empty. Hence, $W^{mm}(\ell)[\mu] = W^{(\ell)}(\ell)[\mu]$ and therefore (5.30) implies that

$$W^{mm}(\ell)[\mu] = -\frac{\mu(\ell)}{\ell} \quad \text{for all } \ell < q. \tag{5.31}$$

Our first task is thus to show that this equation admits a solution, which a priori is not clear since the left-hand side depends on the weight $\mu$ in a highly non-linear fashion. Next, we need to show that this solution also satisfies (5.30).

## 5.3.2. Half-tight cylinders

Before discussing the solution to (5.31), we need to introduce an important decomposition of metric maps with two boundaries. The volume generating function of metric maps with two boundaries of lengths $\ell, p$, also known as the cylinder function, is given by

$$W^{mm}(\ell, p)[\mu] := \frac{\delta^2 F_0^{mm}}{\delta\mu(\ell)\mu(p)} \tag{5.32}$$

$$= \sum_{k=1}^{\infty} \frac{1}{k!} \int_q^{\infty} d\ell_1 \mu(\ell_1) \cdots \int_q^{\infty} d\ell_k \mu(\ell_k) V_{0,k+2}^{mm}(\ell, p, \ell_1 \ldots, \ell_k) \tag{5.33}$$

$$= \frac{1}{2} \int_0^{E_\mu} du \, I_0 \left(\ell \sqrt{u}\right) I_0 \left(p \sqrt{u}\right). \tag{5.34}$$

We say a metric map with two boundaries is half-tight if the contour of the second boundary (of length $p$) is the shortest curve separating the two boundaries. Let $V_{0,k+2}^{ht}(p_1, p_2, \ell_1 \ldots, \ell_k) < V_{0,k+2}^{mm}(p_1, p_2, \ell_1 \ldots, \ell_k)$ be the volume of this restricted set of metric maps and $H(\ell, p)[\mu]$ its generating function,

$$H(\ell, p)[\mu] = \sum_{k=1}^{\infty} \frac{1}{k!} \int_0^{\infty} d\ell_1 \mu(\ell_1) \cdots \int_0^{\infty} d\ell_k \mu(\ell_k) V_{0,k+2}^{ht}(\ell, p, \ell_1 \ldots, \ell_k). \tag{5.35}$$

Following [33, 44] one can relate $H(\ell, p)[\mu]$ to the cylinder function using the following canonical decomposition. Consider a metric map $\mathfrak{m}$ with two boundaries of lengths $\ell$ and $p$ such that $\ell > p$, then there exists a shortest curve $\gamma$ separating the two boundaries. We will assume this curve is unique because that is the generic situation. If $\gamma$ is the contour of the second boundary (and thus $\gamma$ has length $p$), then $\mathfrak{m}$ is a half-tight cylinder. Otherwise, if $\gamma$ has length $y < p$, cutting $\mathfrak{m}$ along $\gamma$ will result in a pair of half-tight cylinders, one with boundaries of lengths $\ell$ and $y$ and the other with boundaries of lengths $p$ and $y$. Vice versa, one may check that gluing the tight boundaries of two half-tight cylinders results in a shortest separating curve on a metric map with two boundaries (see Figure 5.4). Since there is a twisting freedom (of size $y$) in the gluing of the half-tight cylinders, this construction results in the identity

$$W^{mm}(\ell, p)[\mu] = H(\ell, p) + \int_0^p \mathrm{d}y \; H(\ell, y) y H(p, y). \qquad (\ell > p). \qquad (5.36)$$

One may check (see [44, Section 2]) that this uniquely determines $H(\ell, p)$ in terms of the cylinder function. Since the expression for $W^{mm}(\ell, p)[\mu]$ in (5.34) is identical (up to factors of 2) to the corresponding cylinder function of Weil-Petersson volumes, we may directly apply [44, Lemma 11] to deduce that

$$H(\ell, p)[\mu] = \sqrt{\frac{E_\mu}{\ell^2 - p^2}} I_1 \left( \sqrt{\ell^2 - p^2} \sqrt{E_\mu} \right) \mathbf{1}_{\{\ell \geq p\}} \qquad (5.37)$$

$$= -\frac{1}{p} \frac{\partial}{\partial p} I_0 \left( \sqrt{\ell^2 - p^2} \sqrt{E_\mu} \right) \mathbf{1}_{\{\ell \geq p\}}. \qquad (5.38)$$

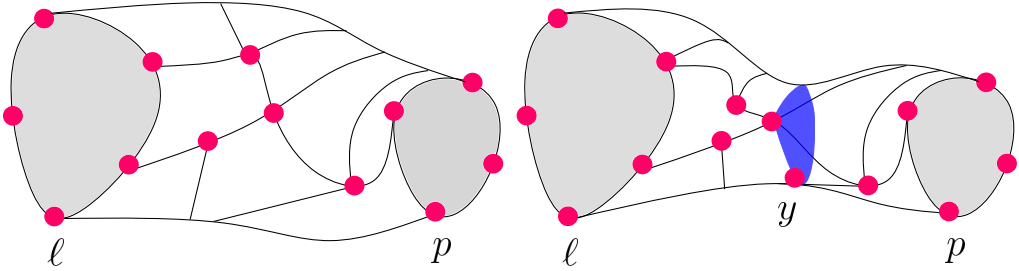

**Figure 5.4.: Left**: A tight cylinder with marked faces of lengths $\ell$ and $p$. **Right**: A non-tight cylinder can be decomposed into two tight cylinders by cutting along the shortest cycle $\gamma$ (in blue) of length $y$ connecting the two marked faces (in grey) of lengths $\ell$ and $p$.

### 5.3.3. Solving the substitution equation

We will now use the cylinder decomposition formula (5.36) to solve the substitution equation (5.31). The first thing we need to do is to relate the disk function $W^{mm}(\ell)$ to the cylinder function $W^{mm}(\ell, p)$. This will help us because the cylinder function $W^{mm}(\ell, p)$ and half-tight cylinder function $H(\ell, p)$ only depend on $\mu$ through the value $E_\mu$, while the disk function $W^{mm}(\ell)$ depends on $\mu$ in a more complicated fashion.

The identity we need is

$$W^{mm}(\ell)[\mu] = \int_0^\infty W^{mm}(\ell, p) \mu(p) \mathrm{d}p - 2 \left. \frac{\partial^2 W^{mm}(\ell, p)}{\partial p^2} \right|_{p=0}. \qquad (5.39)$$

This can be checked explicitly using the formulas (5.10) and (5.34). But there is also a simple combinatorial interpretation: the first term on the right-hand side corresponds to the generating function of metric maps with a boundary of length $\ell$ and a marked (internal) face, while the second term is half the generating function of such maps with a marked vertex[1]. The identity is then a simple consequence of Euler's formula, which in this case states that the number $|F(\mathfrak{m})|$ of internal faces and the number $|V(\mathfrak{m})|$ of vertices in a planar cubic map always satisfy $|F(\mathfrak{m})| - \frac{|V(\mathfrak{m})|}{2} = 1$.

We wish to utilize this relation (5.39) together with the decomposition (5.36) to express the disk function as a half-tight cylinder with an effective weight $\tilde{\mu}(y)$ on its tight boundary of arbitrary length $y > 0$,

$$W^{mm}(\ell)[\mu] + \frac{\mu(\ell)}{\ell} = \frac{\tilde{\mu}(\ell)}{\ell} + \int_0^\ell \mathrm{d}y \, H(\ell, y)\tilde{\mu}(y). \tag{5.40}$$

For the first term in (5.39) we observe that

$$\frac{\mu(\ell)}{\ell} + \int_0^\infty \mathrm{d}p \, W^{mm}(\ell, p)\mu(p) = \frac{\mu(\ell)}{\ell} + \int_\ell^\infty \mathrm{d}p H(p, \ell)\mu(p) + \int_0^\ell \mathrm{d}p H(\ell, p)\mu(p) \tag{5.41}$$

$$+ \int_0^\ell \mathrm{d}y H(\ell, y)y \int_y^\infty \mathrm{d}p H(p, y)\mu(p) \tag{5.42}$$

$$= \frac{U_+(\ell)}{\ell} + \int_0^\ell \mathrm{d}y \, H(\ell, y)U_+(y), \tag{5.43}$$

where

$$U_+(y) = \mu(y) + y \int_y^\infty \mathrm{d}p \, H(p, y)\mu(p) := y \, \mathbf{H}\left(\frac{\mu(p)}{p}\right)(y). \tag{5.44}$$

The definition of the operator $\mathbf{H}$ will result practical in the following section. Note that $U_+$ has an interpretation as the generating function of half-tight cylinders with a tight boundary of length $y$ and the other boundary of arbitrary length with the appropriate weight.

For the second term in (5.39), we can simply guess a solution based on the heuristic reasoning that it is of the form of the first term with a distributional

---

[1]This happens because the only contribution at quadratic order in $p$ is provided by the metric maps where the second boundary is a triangle. Assuming all vertices to be cubic, in the limit of $p \to 0$ this boundary triangle degenerates into a marked cubic vertex.

weight $\mu(p) = \delta''(0)$. The analogue of $U_+$ then is

$$U_-(y) = 2 \left. \frac{\partial^2}{\partial p^2} \sqrt{\frac{E_\mu}{p^2 - y^2}} I_1 \left( \sqrt{p^2 - y^2} \sqrt{E_\mu} \right) \right|_{p=0} + \cdots \tag{5.45}$$

$$= -\frac{2E_\mu}{y^2} J_2(y\sqrt{E_\mu}) + \cdots. \tag{5.46}$$

And one may check with a calculation that indeed

$$2 \left. \frac{\partial^2 W^{mm}(\ell, p)}{\partial p^2} \right|_{p=0} = \frac{U_-(\ell)}{\ell} + \int_0^\ell \mathrm{d}y\, H(\ell, y) U_-(y). \tag{5.47}$$

Hence (5.40) is satisfied with

$$\tilde{\mu}(y) = U_+(y) - U_-(y) \tag{5.48}$$

$$= \frac{\partial}{\partial y} \left( \frac{2\sqrt{E_\mu}}{y} J_1(y\sqrt{E_\mu}) - \int_y^\infty \mathrm{d}\ell\,\mu(\ell) I_0 \left( \sqrt{(\ell^2 - y^2)E_\mu} \right) \right) \tag{5.49}$$

$$= \frac{\partial}{\partial y} x^{(y)}(E_\mu). \tag{5.50}$$

This is useful because the operator that sends the function $\tilde{\mu}$ to the right-hand side of (5.40) is invertible and lower-triangular. Hence the substitution equation (5.31) is equivalent to the condition

$$\tilde{\mu}(\ell) = 0 \qquad \text{for all } \ell \le q. \tag{5.51}$$

Recall that $E_\mu$ was defined via $x(E_\mu) = 0$ with $x(u) = x^{(0)}(u)$ as in (5.2). This is thus also equivalent to

$$x^{(\ell)}(E_\mu) = 0 \qquad \text{for all } \ell \le q. \tag{5.52}$$

Specializing to $\ell = q$, provides a closed-form equation $x^{(q)}(E_\mu) = 0$ for $E_\mu$ since $x^{(q)}(u)$ only involves the original weight $\mu$ on $[q, \infty)$.

Once $E_\mu$ is known, $U_-(y)$ is completely determined and therefore $U_+(y)$ as well, resulting in

$$U_+(y) = \begin{cases} U_-(y) & \text{for } y \le q, \\ \mu(y) + y \int_y^\infty \mathrm{d}p\, H(p, y)\mu(p) & \text{for } y > q. \end{cases} \tag{5.53}$$

Since the operator **H** in (5.44) is invertible and upper-triangular, (5.53) has a unique solution for $\mu$ on the interval $[0, q]$.

### 5.3.4. An explicit example

In this section, we provide a concrete example to illustrate the weight substitution process outlined in the last section, specifically, in equations (5.53) and (5.44). We focus on $q$-irreducible metric maps characterized by a specific weight function

$$\mu(\ell) = c\,\ell\,e^{-\frac{\ell^2}{2t}} \qquad \text{for } \ell \geq q, \tag{5.54}$$

where $c, t > 0$ are free parameters. This choice of weight allows us to derive explicit expressions for the weight substitution. First, we note that

$$\mathbf{H}\left(\frac{\mu(p)}{p}\right)(y) = c\;e^{-\frac{y^2}{2t} + \frac{tE_\mu}{2}}. \tag{5.55}$$

We can also compute

$$\int_q^\infty \mathrm{d}\alpha\;\mu(\alpha)I_0(\sqrt{\alpha^2 - q^2}\sqrt{E_\mu}) = ct\;e^{-\frac{q^2}{2t} + \frac{tE_\mu}{2}}, \tag{5.56}$$

which provides us with an expression for $E_\mu^{(q)}$ by solving the string equation (5.17). This equation can be written as

$$\frac{2\sqrt{E_\mu}}{q}J_1\left(q\sqrt{E_\mu}\right) = ct\,e^{-\frac{q^2}{2t} + \frac{tE_\mu}{2}} \tag{5.57}$$

or equivalently (5.2), e.g. for $q = 0.4$, $c = 0.1$, $t = 1$ we have $E_\mu = 0.097092$ as a solution for both equations.

We proceed to determine the weight substitution $\mu(\ell)$ for $\ell < q$ as given by equations (5.53) and (5.44). We obtain

$$\frac{\mu(\ell)}{\ell} = \mathbf{H}^{-1}\left(\frac{U_-(y)}{y}\mathbf{1}_{\{y \leq q\}} + \frac{U_+(y)}{y}\mathbf{1}_{\{q \leq y\}}\right)(\ell) \tag{5.58}$$

$$= \ell c e^{-\frac{\ell^2}{2t}}(1 - e^{\frac{tE_\mu}{2}}) + \frac{2E_\mu J_2(\ell\sqrt{E_\mu})}{\ell} \tag{5.59}$$

$$+ \ell \int_\ell^q \mathrm{d}y\; c\; y e^{-\frac{y^2}{2t} + \frac{tE_\mu}{2}}\sqrt{\frac{E_\mu}{y^2 - \ell^2}}J_1\left(\sqrt{y^2 - \ell^2}\sqrt{E_\mu}\right) \tag{5.60}$$

$$- \ell \int_\ell^q \mathrm{d}y\frac{2E_\mu J_2(y\sqrt{E_\mu})}{y}\sqrt{\frac{E_\mu}{y^2 - \ell^2}}J_1\left(\sqrt{y^2 - \ell^2}\sqrt{E_\mu}\right). \tag{5.61}$$

This equation provides a detailed expression for the weight $\mu(\ell)$ for $\ell < q$, while the weight for $q < \ell$ is (5.54). The resulting weight in $[0, \infty)$ is shown in Figure 5.5. This weight now defines a metric map without irreducibility constraints.

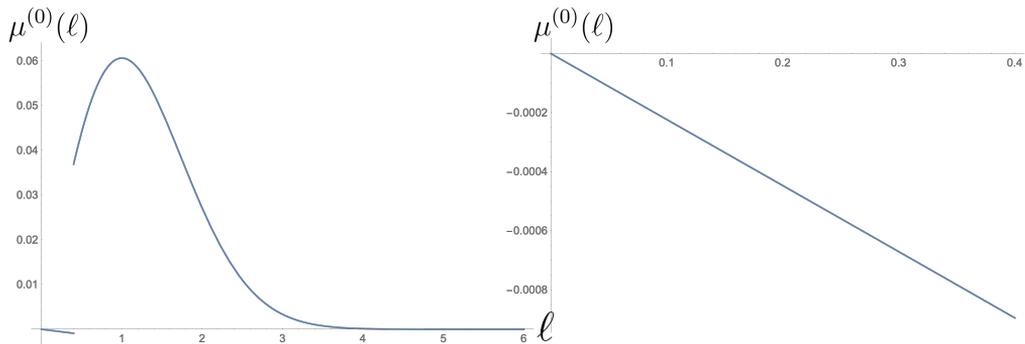

**Figure 5.5.: Left**: Full weight on $[0, \infty)$. **Right**: weight substitution for $\ell < q$. Both graphs are shown for $q = 0.4$ and $E_\mu = 0.097092$.

We aim to check how the times change with this weight substitution. For this purpose, we compute the times for $q$-irreducible metric maps $t_k^{(q)}[\mu]$ given in (5.20) for the weight (5.54)

$$t_k^{(q)}[\mu] = \frac{2^{-k} t^{k+1} \Gamma(k+1)}{k!} c e^{-\frac{\beta^2}{2t}} \tag{5.62}$$

To understand how these times change when we remove the irreducibility constraint, we need to compute the metric map times $t_k^{(0)}$ given by (4.8) for the new weight: (5.61) for $\ell < q$ and (5.54) for $\ell \geq q$. These times are given by

$$t_k^{(0)}[\mu] = \frac{4^{-k}}{k!} \int_0^q \mathrm{d}\ell \mu(\ell)\ell^{2k} + \frac{4^{-k}}{k!} \int_q^\infty \mathrm{d}\ell \mu(\ell)\ell^{2k}. \tag{5.63}$$

We show both of these times in Figure 5.6.

The main observation we want to make here is that the genus-0 weight substitution we constructed to go from $q$-irreducible metric maps to metric maps has effects on the times and therefore, on the generating function (5.14). We would like to make this effect more precise and understand the weight substitution not only at the computational level but at the level of the integrable structure that comes with the intersection theory of moduli space described in Section 5.1.1.

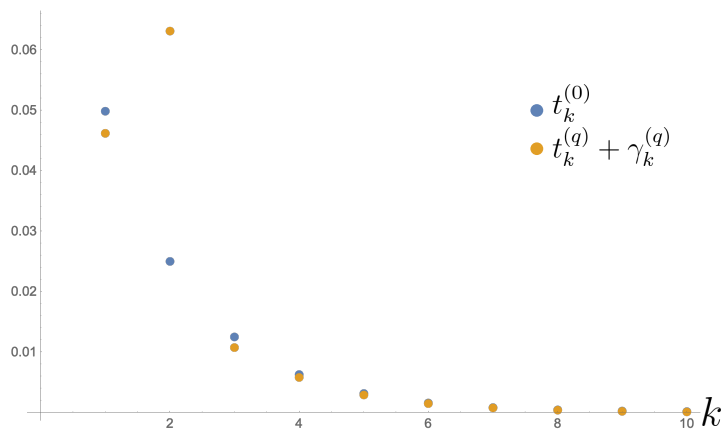

**Figure 5.6.:** Times for $q$-irreducible metric maps in yellow and times after the weight substitution (5.61) in blue, for $q = 0.4$ and $E_\mu = 0.097092$.

## 5.4. Moments

As illustrated in Section 5.3.4's example, the weight substitution has a very intricate analytical expression, and, in order to compute the times, we would need to substitute it into the expressions for the times (4.8). This approach can become highly complex and may not offer a straightforward understanding of how the irreducibility constraint removal impacts the times.

As we have seen for metric maps, irreducible metric maps and hyperbolic surfaces, the generating functions can be constructed from the moments through the function $\mathcal{P}_g$ (4.15). Therefore, an alternative path to delve into the intersection theory of $\mathcal{M}_{g,n}$ is by examining the moments of these types of surfaces. In the case of metric maps, the moments are given by (5.8) as follows

$$M_p^{(0)} = 1_{p=0} - \sum_{m=0}^{\infty} t_{m+p+1}^{(0)} \frac{E_\mu^m}{m!} \tag{5.64}$$

On the other hand, for $q$-irreducible maps, the moments were obtained in [36] by considering limits of the discrete moments arising from the enumeration of irreducible discrete maps. These are given by (5.25)

$$M_p^{(q)} = \sum_{k=0}^{p} \frac{1}{(p-k)!} \left(\frac{q}{2}\right)^{2(p-k)} \left(1_{k=0} - \sum_{m=0}^{\infty} \frac{t_{m+k+1}^{(q)} + \gamma_{m+k+1}^{(q)}}{m!} E_\mu^m\right) \tag{5.65}$$

In this section, we will prove this formula for the moments by using the weight substitution obtained in Section 5.3. To start, we need to express the moments of metric maps in terms of an expression that allows us to impose the irreducibility constraint (5.53). From (5.50), we can see that it is convenient to denote $\bar{U}_+(y)$ to the primitive of $U_+(y)$ and the same for $\bar{U}_-(y)$ and $U_-(y)$. Then (5.52) implies that for $y \leq q$, $\bar{U}_+(y) = \bar{U}_-(y)$.

Next, utilizing the identity

$$- \int_0^\infty \mathrm{d}\ell\,\ell\ \bar{U}_+(y)\ell^{2p} = \frac{1}{2} 2^{2p+1} p! \int_0^\infty \mathrm{d}\alpha \mu(\alpha) \left( \frac{\alpha}{2\sqrt{E_\mu}} \right)^{p+1} I_{p+1}\left( \alpha \sqrt{R} \right), \quad (5.66)$$

we can rewrite the moments for metric maps (without the irreducibility constraint) as

$$M_p^{(0)} = \delta_{p=0} + \frac{4^{-p}}{2p!} \int_0^\infty \ell \mathrm{d}\ell\ \bar{U}_+(\ell, E_\mu)\,\ell^{2p}. \quad (5.67)$$

In order to obtain the expression for the moments of $q$-irreducible metric maps, we want to impose the irreducibility constraint, which is equivalent to the condition (5.53) for $\ell < q$. We have

$$4^p 2p! M_p^{(0)} = - \int_0^q \ell \mathrm{d}\ell\ \bar{U}_-(\ell, E_\mu)\,\ell^{2p} - \int_q^\infty \ell \mathrm{d}\ell\ \bar{U}_+(\ell, E_\mu)\,\ell^{2p} \quad (5.68)$$

$$= \int_0^q \mathrm{d}\ell\ 2\sqrt{E_\mu} J_1(\ell\sqrt{E_\mu})\ell^{2p} + \int_q^\infty \ell \mathrm{d}\ell \int_\ell^\infty \mathrm{d}\alpha \mu(\alpha) I_0 \left( \sqrt{\alpha^2 - \ell^2}\sqrt{E_\mu} \right)\ell^{2p} \quad (5.69)$$

$$= \int_0^q \mathrm{d}\ell\ 2\sqrt{E_\mu}\ J_1(\ell\sqrt{E_\mu})\ell^{2p} \quad (5.70)$$

$$+ \int_0^\infty w \mathrm{d}w \int_w^\infty \mathrm{d}y \hat{\mu}(y) I_0 \left( \sqrt{y^2 - w^2}\sqrt{E_\mu} \right) \left( w^2 + q^2 \right)^p \quad (5.71)$$

$$= \sum_{k=0}^p \frac{4^p 2p!}{(p-k)!} \left( \frac{q}{2} \right)^{2(p-k)} \times \quad (5.72)$$

$$\left( \sum_{m=1}^\infty \frac{(-1)^{m+k+1}}{(m+k)!} \left( \frac{q}{2} \right)^{2k+2m} E_\mu^m + \int_0^\infty \ell \mathrm{d}\ell\ \bar{U}_+(\ell, E_\mu)\,\ell^{2k} \right) \quad (5.73)$$

$$\quad (5.74)$$

where $\hat{\mu}$ is given by

$$\hat{\mu}(y) = \frac{y}{\sqrt{y^2 + q^2}} \mu \left( \sqrt{y^2 + q^2} \right). \quad (5.75)$$

Therefore, we have that

$$M_p^{(0)}\big|_{\text{irred}} = \sum_{k=0}^{p} \frac{1}{(p-k)!} \left(\frac{q}{2}\right)^{2(p-k)} \left(\sum_{m=1}^{\infty} \frac{Y_m^{(q)}}{m!} E_\mu^m + \int_0^\infty \ell\, \mathrm{d}\ell\, \bar{U}_+(\ell, E_\mu)\, \ell^{2k}\right). \quad (5.76)$$

Let us review what we did in this section. We started with the expression for the moments of metric maps and expressed them in terms of the generating function $U_+$ and $U_-$. In this way, we were able to incorporate the irreducibility constraint/ weight substitution (5.53). This results in (5.76) which is a promising result that resembles the expression for the moments of $q$-irreducible metric maps found in [36]. Additionally, we observe that in order to get an expression in terms of $\bar{U}_+$, resembling the expression for $M_p^{(0)}$, the original weight $\mu$ and its argument had to be rescaled. While the full interpretation of these effects is not entirely clear at this point, it is instructive to look at how this redefinition of the weight $\hat{\mu}$ relates to the case of hyperbolic surfaces.

## 5.5. Relation to hyperbolic surfaces

In this section, we aim to gain a deeper understanding of our results by examining the relationship between volumes of irreducible metric maps and Weil-Petersson volumes.

Let us start by considering the disk function of $2\pi$-irreducible metric maps (2.18). To establish a connection between this disk function and the disk function of hyperbolic surfaces (4.18), we perform the change of variables $b \longrightarrow \sqrt{b^2 + 4\pi^2}$ and $\ell_i \longrightarrow \sqrt{\ell_i^2 + 4\pi^2}$, this is explicitly given by

$$W_\mu^{(2\pi)}(b) = \sum_{k=2}^{\infty} \frac{1}{k!} \int_{2\pi}^{\infty} \mathrm{d}\ell_1 \mu(\ell_1) ... \int_{2\pi}^{\infty} \mathrm{d}\ell_k \mu(\ell_k) V_{0,k+1}^{(2\pi)}(b, \ell_1, ..., \ell_k) \quad (5.77)$$

$$W_\mu^{(2\pi)}\left(\sqrt{b^2 + 4\pi^2}\right) = \sum_{k=2}^{\infty} \frac{1}{k!} \int_0^\infty \mathrm{d}\ell_1 \frac{\ell_1 \mu\left(\sqrt{\ell_1^2 + 4\pi^2}\right)}{\sqrt{\ell_1^2 + 4\pi^2}} ... \int_0^\infty \mathrm{d}\ell_k \frac{\ell_k \mu\left(\sqrt{\ell_k^2 + 4\pi^2}\right)}{\sqrt{\ell_k^2 + 4\pi^2}} \quad (5.78)$$

$$\times V_{0,k+1}^{(2\pi)}\left(\sqrt{b^2 + 4\pi^2}, \sqrt{\ell_1^2 + 4\pi^2}, ..., \sqrt{\ell_k^2 + 4\pi^2}\right). \quad (5.79)$$

Looking at the disk function for hyperbolic surfaces (4.18) and applying the same argument as we did for metric maps, that is: for $b > 2\pi$, the disk function should remain invariant, we have that

$$W_\mu^{(2\pi)}(b) = W_{\hat\mu}^{WP}\left(\sqrt{b^2 - 4\pi^2}\right) \tag{5.80}$$

if and only if

$$V_{0,k+1}^{(2\pi)}(b, \ell_1, ..., \ell_k) = 2^{2-(k+1)} V_{0,k+1}^{WP}\left(\sqrt{b^2 - 4\pi^2}, \sqrt{\ell_1^2 - 4\pi^2}, ..., \sqrt{\ell_k^2 - 4\pi^2}\right) \tag{5.81}$$

and $\hat\mu$ is given by (5.75). We can immediately check that this expression for the volumes is the same as (5.26) obtained in [36].

Furthermore, this implies that, for the disk functions to coincide for $b > 2\pi$, a length substitution $\ell_i \longrightarrow \sqrt{\ell_i^2 - 4\pi^2}$ must be accompanied by a corresponding transformation of the weight that is consistent with (5.75).

## 5.5.1. Potential relation with JT gravity

At this stage, it is reasonable to ask if this relation between the volumes of $2\pi$-irreducible metric maps and WP volumes has any interesting application in JT gravity. At the level of the disk function we impose for the disk functions to be equal for $b > 2\pi$, nevertheless, the disk function is not an observable/physical quantity per se in JT gravity. The quantity that contributed to the partition function (4.40) is the trumpet transform of the disk function. We observe that

$$\int_0^\infty b\,db \frac{e^{-b^2/4\beta}}{2\sqrt{\pi\beta}} W_{\hat\mu}^{WP}(b) = e^{\frac{\pi^2}{\beta}} \int_{2\pi}^\infty b\,db \frac{e^{-b^2/4\beta}}{2\sqrt{\pi\beta}} W_\mu^{(q)}(b). \tag{5.82}$$

Since the extra exponential in the right-hand side does not affect the scaling of that partition function at first order in $\beta$, this suggests that the partition function of JT gravity with defects characterized by $\hat\mu$ behaves in the same way as the partition function of 'Irreducible gravity' or topological gravity on a $2\pi$-irreducible metric map background with defect of weight $\mu$. However, there are obviously corrections at higher order in $\beta$ and these may bring potential generalizations of JT gravity. Conversely, the rich framework of JT gravity and hyperbolic surfaces, such as topological recursion, could provide new tools for studying irreducible metric maps.

As a last note, we note that one might think that the trumpet is tightly connected to JT gravity and it cannot be generalized to topological gravity with arbitrary background times $\gamma_k$. However, in [123] it was shown that the trumpet partition function can be written in terms of the Liouville wavefunction which gives a universal definition of the trumpet in arbitrary backgrounds.

## 5.6. Discussion

In this chapter, we focused on understanding the concept of irreducible metric maps, its relation to topological gravity via metric maps, and its potential relevance in the context of JT Gravity. To this end, we re-derived results on genus-0 $q$-irreducible metric maps in a different way than [36], by explicitly disallowing cycles of length $\ell < q$. We found a weight substitution that assigns negative weights to forbidden configurations so that the disk function remains invariant for cycles of length $\ell > q$. In this way, we constructed a mapping between metric maps and $q$-irreducible metric maps.

Aiming to better understand the role of this relation between reducible and irreducible metric maps with the intersection theory of $\mathcal{M}_{g,n}$, we studied how the moments of metric maps transform under our weight substitution (5.53). We (almost) re-derived the results for moments of $q$-irreducible metric maps obtained by studying limits of enumeration of irreducible maps. In this analysis, we discovered that there is another weight transformation (5.75) one needs to do to match the structure of the moments.

Moreover, in [36], it was proven that the volumes of $2\pi$-irreducible metric maps match the Weil-Petersson volumes for genus 0 and 1 up to a rescaling of the length of the faces/boundaries. In Section 5.5, we re-derived this relation for $g = 0$ revealing the extra insight that the weights need to be transformed exactly in the same form to be compatible with the expression for the moments, that is, accordingly to (5.75).

A noteworthy comment to make is that even though irreducible metric maps have connections with metric maps and their volumes to Weil-Petersson volumes, irreducible metric maps do not adhere to the framework of topological recursion. This makes it highly non-trivial to obtain results at higher genus, for example, the weight substitution (5.53) then needs to be solved explicitly for each genus. However, the fact that the generating functions for irreducible maps for all genus were previously derived using combinatorial constructions in the discrete, strongly suggests that the weight substitution (5.53) can likely be extended in a manner that is not overly intricate, although it may not be as straightforward as with the topological recursion method.

Continuing the discussion on topological recursion, recent studies on hyperbolic surfaces with tight boundaries [44] used the tight cylinder decomposition to prove topological recursion for a more general family of hyperbolic surfaces than it was originally done in [120]. In that case, it was useful to study how the times $t_k$ transform. Some preliminary results suggest that the same properties of such transformation in the times (see [44, 72]) are true for irreducible metric maps. In that context, this relation was used to map the generating function of tight Weil-Petersson volumes to the generating function of regular Weil-Petersson volumes. It is reasonable to anticipate that a similar computational approach could be applicable in the context of irreducible metric maps. However, we have not pursued this line of investigation in our current work, and it remains an intriguing avenue for future research.

In the context of JT Gravity, (5.82) suggests that the partition function of JT gravity with defects characterized by $\hat{\mu}$ exhibits similar behavior to that of 'Irreducible gravity' or topological gravity on a $2\pi$-irreducible metric map background with defects of weight $\mu$. However, there are higher-order corrections in $\beta$ that are expected to introduce variations, potentially leading to generalized models of JT gravity. In this regard, it is also an interesting question if there is any other instance of 2-dimensional gravity where irreducible metric maps naturally arise.

In conclusion, our exploration of the relations of irreducible metric maps with metric maps and hyperbolic surfaces has revealed intriguing connections at the level of genus $g$ generating functions and promising avenues for future research, such as generalizations of JT gravity. Nonetheless, we are yet to establish a bijective correspondence between irreducible metric maps and hyperbolic surfaces. Additionally, we lack a more profound comprehension of hyperbolic surfaces within the framework of discrete geometry, along with a link between hyperbolic surfaces and metric maps beyond the Kontsevich matrix model. We want to remark that there is ongoing research investigating combinatorial constructions of hyperbolic surfaces in terms of trees [43] which have potential to shed light on these regards.

# 6

# CONCLUSIONS AND OUTLOOK

The quest for a quantum theory of gravity is intrinsically linked to the search for an object that replaces the classical notion of spacetime. As highlighted by General Relativity, gravity is described by the geometrical properties of a manifold, spacetime. The question then becomes what is this geometry in the quantum regime? what are its properties? is this even a manifold? Addressing this question through the lens of statistical physics prompts us to think about the space of all possible spacetime configurations—the ensemble, so to speak. In a manner analogous to how one approaches the quantization of the trajectory of a point particle by summing over all possible paths in a path integral, the computation of the quantum gravity path integral can be thought of as a sum over all possible spacetimes that satisfy certain boundary conditions. The question becomes even more precise when one considers the Euclidean path integral and can make sense of a sum over all possible geometries with a well-defined probability per configuration. In this sense, random geometries appear as a cornerstone in the study of quantum gravity.

In this thesis, we embarked on the exploration of applications of random geometry to quantum gravity, particularly driven by different approaches to quantum gravity such as Asymptotic Safety, Dynamical Triangulations, JT gravity and Topological gravity.

# What did we learn from random geometries about the structure of quantum spacetime?

First, the asymptotic safety scenario for quantum gravity led us to look for scale-invariant random geometries. To this end, we utilized a random geometry construction in two dimensions called Mating of trees, which encompasses seamlessly with Liouville Quantum Gravity, a two-dimensional QG model. However, our universe is described by a four-dimensional manifold, so a natural step is to generalize this construction of scale-invariant random geometries to higher dimensions. To this end, similar to discrete approaches like Dynamical Triangulations, we introduced a sequence of discrete $d$-dimensional metric spaces denoted as $G_n^C$, known as $d$-dimensional Mated-CRT maps, associated with correlated Brownian excursions in $d$ dimensions. Our hypothesis is that these metric spaces converge to non-trivial continuous random metric spaces as $n \to \infty$, inheriting their scaling properties from the underlying Brownian excursions. By implementing this construction numerically, we demonstrated the scaling properties of mated-CRT maps for $d = 2$ by making Hausdorff dimension measurements that align with one of the two predictions in the literature. Additionally, our implementation allowed us to explore regimes that had been challenging with other approaches. Most importantly, we investigated the mating of trees construction in $d = 3$, where there were no previous examples of non-trivial scale-invariant random geometries, and studied its scaling properties. We found a full family of scale-invariant random geometries characterized by two critical exponents: Hausdorff dimension and string susceptibility. This unveils new universality classes of random geometries based on correlated CRTs. These geometries hold potential for understanding quantum spacetime, although many questions remain unanswered. The central question of whether the newfound scale-invariant random geometries represent any form of spacetime geometry, particularly with manifold topology, remains open.

After our exploration of three dimensions, our journey brought us back to the domain of 2-dimensional quantum gravity, specifically within the context of JT gravity in negatively curved spacetimes. We highlighted the strong relation between JT gravity and a well-defined notion of random hyperbolic surfaces. In this exploration of random hyperbolic surfaces, we encountered many elements from analytic combinatorics used to enumerate discrete maps, such as the string equation. Our main contribution was the finding of criti-cal regimes for hyperbolic surfaces inspired by the random map case. These critical regimes are obtained by fine-tuning weights associated with geodesic boundaries which are associated with matter and branes. In particular, we found so-called non-generic criticality in the system together with phase transitions. Throughout the phase transitions, these models exhibit a continuous shift in the density of states behavior, transitioning from $\rho_0(E) \sim \sqrt{E - E_0}$ to $\rho_0(E) \sim (E - E_0)^{3/2}$. These two regimes have been observed in JT gravity coupled to branes, however, this is the first time a non-analytical behavior of the density of states ($\rho_0(E) \sim (E - E_0)^{\alpha - 1}$ with $3/2 < \alpha < 5/2$) appears in JT gravity. We identified that this transition corresponds to a proper phase transition, occurring when the expected number of geodesic boundaries diverges. Our research also raised questions regarding the stability of critical regimes, non-perturbative properties, and the possibility of discrete JT gravity spectra. Additionally, we suggested that the exploration of non-generic critical phases holds potential implications for the Page curve and black hole evaporation in the context of quantum gravity.

The last topic we covered in this thesis was that of irreducible metric maps and their relation to topological gravity. Topological gravity is the study of the intersection theory of the moduli space of Riemann surfaces. First, we focused on how topological gravity brings the fields of random hyperbolic surfaces and another random geometry: random metric maps, to common ground. Then, based on previous results that found relations between the volumes of hyperbolic surfaces and the volumes of irreducible metric maps, we explored the latter. Interestingly, irreducible metric maps possess a connection with metric maps too. In this thesis, we introduced a novel algorithm to transition from genus-0 $q$-irreducible metric maps to genus-0 metric maps. This is done by explicitly disallowing metric maps with cycles of length smaller than $q$. This results in a weight substitution that generalizes results in discrete irreducible maps. We explored how moments of metric maps transform under this weight substitution and uncovered another weight transformation necessary for compatibility with moment expressions. When we explored the significance of this last weight substitution, we found that it is exactly the weight transformation one needs to apply to derive the known relation between volumes of hyperbolic surfaces and volumes of irreducible metric maps. The implications of our findings suggest that the partition function of JT gravity with defects characterized by a weight $\hat{\mu}$ behaves similarly to 'Irreducible gravity' or topological gravity on a $2\pi$-irreducible metric map background with defects of weight $\mu$. Nonetheless, higher-order corrections are expected to introduce variations, potentially leading to generalized JT gravity models. Furthermore, we highlighted that irreducible metric maps do not adhere to the topological recursion framework, posing challenges in our extending results to higher genus. While we made substantial progress, questions about bijective correspondences and deeper connections between hyperbolic surfaces and discrete maps remain.

The interplay between random geometry and quantum gravity holds promise for understanding fundamental quantum spacetime structures. This thesis has established a common ground between current topics in both fields, however, the journey is far from over. Questions like: What is the role of random geometries in continuous 4-dimensional Lorentzian quantum gravity? is random geometry even useful when quantum gravity is coupled to matter in higher dimensions? remain open for exploration. The tools and insights developed in this work provide a robust foundation for future research, offering promising avenues to deepen our understanding of quantum spacetime and its fundamental properties.

# A

# IRREDUCIBILITY IN A MATRIX MODEL

In this appendix, we explore the connection between matrix models and irreducible maps. The central question we address is whether we can incorporate the irreducibility constraint into a matrix model by restricting the existence of faces with lengths less than a certain threshold, denoted as $\mathfrak{q}$. Additionally, we investigate whether this constraint can be mapped to a weight substitution at the level of the couplings in the matrix model potential.

Consider a matrix model with a quartic potential, as defined by the action

$$I_{MM} = \frac{1}{2}\text{Tr}(A^2) + \frac{\mu_2^{(4)}}{2}\text{Tr}(A^2) + \frac{\mu_4^{(4)}}{4}\text{Tr}(A^4), \tag{A.1}$$

We aim to relate this matrix model to the properties of $\mathfrak{q}$-irreducible maps. To establish this connection, we need to relate the matrix model parameters to the irreducibility constraint via the loop operator

$$W^{(\mathfrak{q})}(L) = \lim_{N\to\infty}\left\langle\frac{1}{N}\text{Tr}A^{2L}\right\rangle = \begin{cases} \text{Cat}(L) & L < \mathfrak{q} \\ \text{Cat}(L) + \mu_{\mathfrak{q}} & L = \mathfrak{q} \end{cases}. \tag{A.2}$$

Specifically, we seek to find expressions for $\mu_2^{(4)}$ and $\mu_4^{(4)}$ that account for the irreducibility condition imposed on maps. In this case, we will study 4-irreducible maps. Therefore, we need a weight substitution such that

$$W^{(2)}(L)[q_4] = W^{(0)}(L)[\mu_2^{(4)}[q_4], \mu_4^{(4)}[q_4]] \tag{A.3}$$

It is well-known how to compute the loop expectation values (A.2) for even potentials

$$\lim_{N \to \infty} \left\langle \frac{1}{N} \mathrm{Tr} A^{2L} \right\rangle = \int_{-a}^{a} \mathrm{d}z \; z^{2L} \rho(z) \tag{A.4}$$

with

$$\rho(z) = -\frac{1}{2\pi} \sqrt{4P(z) - V'^2(z)} = -\frac{1}{2\pi} \left( (1 - \mu_2^{(4)}) + \mu_4^{(4)} \left( \frac{a^2}{2} + z^2 \right) \right) \sqrt{a^2 - z^2} \tag{A.5}$$

with

$$a^2 = \frac{2}{3\mu_4^{(4)}} \left( \sqrt{(1 - \mu_2^{(4)})^2 - 12\mu_4^{(4)}} - (1 - \mu_2^{(4)}) \right) \tag{A.6}$$

Using these expressions and imposing (A.2), we obtain the system of equations

$$\frac{1}{16} \sqrt{\frac{1}{a}} a^{9/2} \left( -a^2 \mu_4^{(4)} + \mu_2^{(4)} - 1 \right) = 2 \tag{A.7}$$

$$-\frac{1}{256} \sqrt{\frac{1}{a}} a^{13/2} \left( 9a^2 \mu_4^{(4)} - 8\mu_2^{(4)} + 8 \right) = 2 + q_4. \tag{A.8}$$

The solution to this system is shown in Figure A.1. Remarkably, we notice that similarly to the example for $q$-irreducible metric maps in Section 5.3.4, the weight substitution yields a non-combinatorial weight.

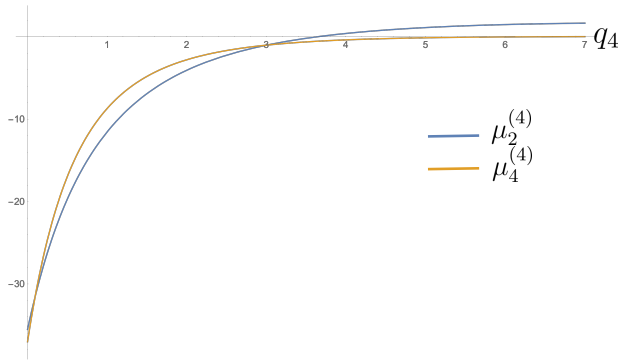

**Figure A.1.:** Weight substitution to go from 4-irreducible maps with faces of degree 4 and weights $q_4$ to 2-irreducible maps with faces of degrees 2 and 4 and weights $\mu_2^{(4)}$ and $\mu_4^{(4)}$, respectively.

# B

# Research data management

This thesis research has been carried out under the institute research data management policy of the Institute for Mathematics, Astrophysics and Particle Physics, as documented in
https://www.ru.nl/publish/pages/1029522/imapp_rdm_policy_1.pdf
The code used in Chapter 3 as well as the data and manual for its implementation are available in [39].

# Summary

How do space and time look at very high energies? In the past century, two very successful, yet very different, theories were developed: Quantum Mechanics, which describes elementary particles, and General Relativity, which describes how space and time deform due to the presence of energy and matter. Each of these theories allows us to describe most of the phenomena in our universe. For example, quantum mechanics and its generalization, quantum field theory, allow us to predict the results of experiments that involve colliding particles at very high energies in the LHC. On the other hand, General Relativity allowed us to predict the shadow observed in the first image of a black hole. Nevertheless, we can find in nature physical systems that seem to require both of these descriptions, such as the universe's very early stages and black holes. This brings us to one of the main problems in contemporary physics: how to construct a quantum theory of gravity.

The problem we find when trying to quantize gravity in a similar way to other forces of nature is that it results in a perturbatively non-renormalizable theory, this means that this theory does not help us to make predictions at the energies of the universe's very early stages and the center of black holes. This challenges us to look into the more fundamental aspects of this theory, such as the concept of spacetime itself: are space and time continuous or discrete? are causal relations between events fundamental? was our universe always four-dimensional?

In this thesis, we approach this problem by mixing elements of both gravity and quantum mechanics straightforwardly: the quantization of a classical system is the sum over all its possible *random* configurations and the configurations of gravity are given by the *geometry* of spacetime. Therefore, *random geometry* is a natural setting for studying at least certain aspects of quantum gravity!

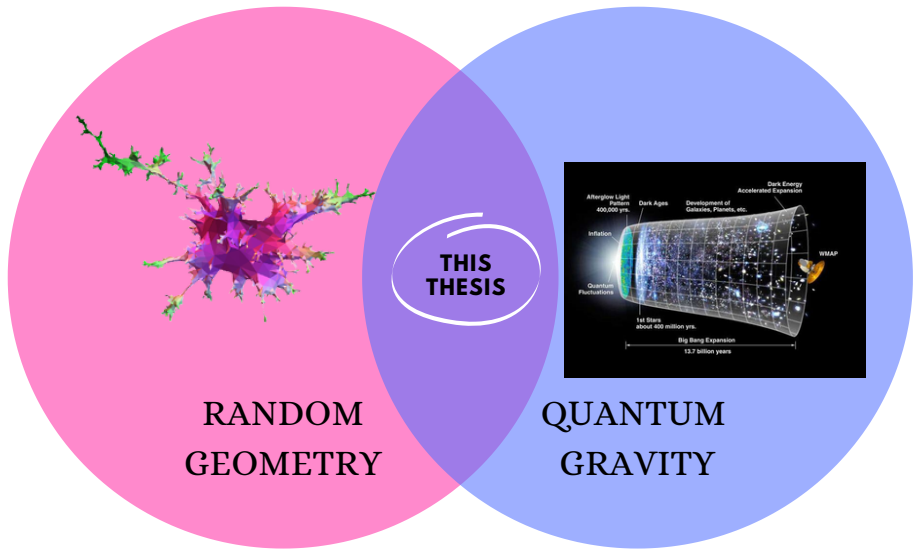

Random geometry is a field of Mathematics that studies the question: given the space of all geometries if I pick one randomly what is the likelihood that this geometry has property $X$. For example, consider a bag filled with an equal number of balls of volumes $\{1, 2, 3, 4, 5, 6\}$, then the likelihood of choosing randomly a ball of volume 1 is 1/6. Likewise, one can ask: what is the most likely volume to pick? In this case, all volumes are equally likely, then we say that they are *uniformly distributed.* In this example, geometry is characterized only by the volume of spheres. Still, one can imagine a bag filled with objects of all kinds of shapes and other properties characterizing their geometry.

In the context of quantum gravity, random geometries in quantum gravity are spacetime itself, and their geometrical properties translate into physical ones. In this thesis, we navigate through three seemingly different types of random geometries: scale-invariant random geometries and the Brownian sphere, random hyperbolic surfaces, and random irreducible metric maps, their relation to three quantum gravity approaches: Asymptotic safety, JT gravity and topological gravity, and uncover new relations between them.

# Acknowledgements

*Para Mello, Daisy, Maya, Tony y Remi*

I want to thank my daily supervisor Timothy Budd for all the support and patience during these four years. From teaching me the basics of combinatorics and Monte Carlo simulations to the long sessions of learning JT gravity together and breaking our heads with irreducible metric maps. Our collaboration highly impacted my current view on quantum gravity. Likewise, I want to thank my promotor Renate Loll for her guidance, advice, and the chance to teach General Relativity (even with short notice). My PhD was enriched by long discussions with Bart Zonneveld and Pjotr Koster in the HEP kitchen, with the random geometry group during our online sessions, and with Clifford Johnson during my stay at USC.

# Bibliography

[1] M. Aganagic, R. Dijkgraaf, A. Klemm, M. Marino, and C. Vafa, *Topological strings and integrable hierarchies*, Commun. Math. Phys., 261 (2006), pp. 451–516.

[2] D. Aldous, *The Continuum Random Tree. I*, The Annals of Probability, 19 (1991), p. 1 – 28.

[3] J. Ambjorn, *Lattice Quantum Gravity: EDT and CDT*, 9 2022.

[4] J. Ambjørn, T. Budd, and Y. Makeenko, *Generalized multicritical one-matrix models*, Nucl. Phys. B, 913 (2016), pp. 357–380.

[5] J. Ambjørn, L. Chekhov, and Y. Makeenko, *Perturbed generalized multicritical one-matrix models*, Nucl. Phys. B, 928 (2018), pp. 1–20.

[6] J. Ambjørn, G. Czelusta, J. Gizbert-Studnicki, A. Görlich, J. Jurkiewicz, and D. Németh, *The higher-order phase transition in toroidal CDT*, Journal of High Energy Physics, 30 (2020), pp. 1–18.

[7] J. Ambjørn, J. Gizbert-Studnicki, A. Görlich, J. Jurkiewicz, N. Klitgaard, and R. Loll, *Characteristics of the new phase in CDT*, The European Physical Journal C, 77 (2017), pp. 1–17.

[8] J. Ambjorn and J. Jurkiewicz, *Four-dimensional simplicial quantum gravity*, Phys. Lett. B, 278 (1992), pp. 42–50.

[9] J. Ambjørn and R. Loll, *Non-perturbative lorentzian quantum gravity, causality and topology change*, Nuclear Physics B, 536 (1998), pp. 407–434.

[10] J. Ambjorn, R. Loll, J. L. Nielsen, and J. Rolf, *Euclidean and Lorentzian quantum gravity: Lessons from two-dimensions*, Chaos Solitons Fractals, 10 (1999), pp. 177–195.

[11] J. Ambjørn and S. Varsted, *Three-dimensional simplicial quantum gravity*, Nuclear Physics B, 373 (1992), pp. 557–577.

[12] J. Ambjørn, B. Durhuus, and J. Fröhlich, *Diseases of triangulated random surface models, and possible cures*, Nuclear Physics B, 257 (1985), pp. 433–449.

[13] M. Ang, *Comparison of discrete and continuum Liouville first passage percolation*, Electron. Commun. Probab., 24 (2019), pp. 1–12.

[14] M. Ang, G. Remy, and X. Sun, *FZZ formula of boundary Liouville CFT via conformal welding*, arXiv preprint: 2104.09478, (2021).

[15] O. Angel, *Scaling of percolation on infinite planar maps, I*, arXiv preprint: math/0501006, (2005).

[16] J. Aru, N. Holden, E. Powell, and X. Sun, *Mating of trees for critical Liouville quantum gravity*, arXiv preprint: 2109.00275, (2021).

[17] J. Aru, Y. Huang, and X. Sun, *Two perspectives of the 2D unit area quantum sphere and their equivalence*, Comm. Math. Phys., 356 (2017), pp. 261–283.

[18] A. Ashtekar and E. Bianchi, *A short review of loop quantum gravity*, Rept. Prog. Phys., 84 (2021), p. 042001.

[19] R. Bañuelos and R. G. Smits, *Brownian motion in cones*, Probab. Theory Related Fields, 108 (1997), pp. 299–319.

[20] T. Banks, W. Fischler, S. H. Shenker, and L. Susskind, *M theory as a matrix model: A Conjecture*, Phys. Rev. D, 55 (1997), pp. 5112–5128.

[21] J. Barkley and T. Budd, *Precision measurements of Hausdorff dimensions in two-dimensional quantum gravity*, Classical and Quantum Gravity, 36 (2019), p. 244001.

[22] J. D. Bekenstein, *Black holes and entropy*, Phys. Rev. D, 7 (1973), pp. 2333–2346.

[23] O. BERNARDI, *Bijective counting of Kreweras walks and loopless triangulations*, Journal of Combinatorial Theory, Series A, 114 (2007), pp. 931–956.

[24] O. BERNARDI, N. CURIEN, AND G. MIERMONT, *A boltzmann approach to percolation on random triangulations*, Canadian Journal of Mathematics, 71 (2019), p. 1–43.

[25] O. BERNARDI, N. HOLDEN, AND X. SUN, *Percolation on triangulations: a bijective path to Liouville quantum gravity*, arXiv preprint: 1807.01684, (2018).

[26] A. BLOMMAERT, L. V. ILIESIU, AND J. KRUTHOFF, *Gravity factorized*, JHEP, 09 (2022), p. 080.

[27] A. BLOMMAERT AND J. KRUTHOFF, *Gravity without averaging*, SciPost Phys., 12 (2022), p. 073.

[28] A. BLOMMAERT, T. G. MERTENS, AND H. VERSCHELDE, *Eigenbranes in Jackiw-Teitelboim gravity*, JHEP, 02 (2021), p. 168.

[29] B. BOGOSEL, V. PERROLLAZ, K. RASCHEL, AND A. TROTIGNON, *3d positive lattice walks and spherical triangles*, J. Comb. Theory, Ser. A, 172 (2020), p. 105189.

[30] G. BOROT, J. BOUTTIER, AND E. GUITTER, *A recursive approach to the O(n) model on random maps via nested loops*, J. Phys. A, 45 (2012), p. 045002.

[31] G. BOROT, J. BOUTTIER, AND E. GUITTER, *More on the O(n) model on random maps via nested loops: loops with bending energy*, Journal of Physics A: Mathematical and Theoretical, 45 (2012), p. 275206.

[32] J. BOUTTIER AND E. GUITTER, *On irreducible maps and slices*, Combinatorics, Probability and Computing, 23 (2013), pp. 914 – 972.

[33] J. BOUTTIER, E. GUITTER, AND G. MIERMONT, *Bijective enumeration of planar bipartite maps with three tight boundaries, or how to slice pairs of pants*, Annales Henri Lebesgue, 5 (2022), pp. 1035–1110.

[34] T. BUDD, *The peeling process on random planar maps coupled to an O(n) loop model (with an appendix by Linxiao Chen)*, arXiv preprint: 1809.02012, (2018).

[35] ——, *On polynomials counting essentially irreducible maps*, Electron. J. Comb., 29 (2020).

[36] ——, *Irreducible metric maps and weil–petersson volumes*, Communications in Mathematical Physics, 394 (2022), pp. 887–917.

[37] ——, *Lessons from the Mathematics of Two-Dimensional Euclidean Quantum Gravity*, arXiv preprint: 2212.03031, (2022).

[38] T. BUDD AND A. CASTRO, *Irreducible metric maps, topological gravity and JT gravity*, to appear.

[39] ——, *Data and code for Scale-invariant random geometry from mating of trees: a numerical study*, 2022.

[40] ——, *Scale-invariant random geometry from mating of trees: A numerical study*, Phys. Rev. D, 107 (2023), p. 026010.

[41] T. BUDD AND P. KOSTER, *Universality classes of 2D hyperbolic Riemannian manifolds*, to appear.

[42] T. BUDD AND L. LIONNI, *A family of triangulated 3-spheres constructed from trees*, arXiv preprint:2203.16105, (2022).

[43] T. BUDD, T. MEEUSEN, AND B. ZONNEVELD, *A combinatorial approach to random hyperbolic surfaces*, to appear.

[44] T. BUDD AND B. ZONNEVELD, *Topological recursion of the Weil-Petersson volumes of hyperbolic surfaces with tight boundaries*, arXiv preprint: 2307.04708, (2023).

[45] G. CALCAGNI, D. ORITI, AND J. THÜRIGEN, *Spectral dimension of quantum geometries*, Class. Quant. Grav., 31 (2014), p. 135014.

[46] A. CASTRO, *Critical JT gravity*, JHEP, 08 (2023), p. 036.

[47] A. CASTRO AND T. KOSLOWSKI, *Renormalization Group Approach to the Continuum Limit of Matrix Models of Quantum Gravity with Preferred Foliation*, Front. in Phys., 9 (2021), p. 114.

[48] R. CORI AND B. VAUQUELIN, *Planar maps are well labeled trees*, Canadian J. Math., 33 (1981), pp. 1023–1042.

[49] N. Curien, *Peeling random planar maps*, Saint-Flour lecture notes, (2019).

[50] N. Curien, T. Hutchcroft, and A. Nachmias, *Geometric and spectral properties of causal maps*, Journal of the European Mathematical Society, 22 (2020), pp. 3997–4024.

[51] J. Dahne and B. Salvy, *Computation of tight enclosures for Laplacian eigenvalues*, SIAM Journal on Scientific Computing, 42 (2020), pp. A3210–A3232.

[52] F. David, *A model of random surfaces with non-trivial critical behaviour*, Nuclear Physics B, 257 (1985), pp. 543–576.

[53] F. David, A. Kupiainen, R. Rhodes, and V. Vargas, *Liouville quantum gravity on the Riemann sphere*, Comm. Math. Phys., 342 (2016), pp. 869–907.

[54] P. Deligne and D. Mumford, *The irreducibility of the space of curves of given genus*, Publications mathématiques de l'IHÉS, 36 (1969).

[55] D. Denisov, V. Wachtel, et al., *Random walks in cones*, Annals of Probability, 43 (2015), pp. 992–1044.

[56] P. Di Francesco, P. H. Ginsparg, and J. Zinn-Justin, *2-D Gravity and random matrices*, Phys. Rept., 254 (1995), pp. 1–133.

[57] R. Dijkgraaf and E. Witten, *Developments in topological gravity*, International Journal of Modern Physics A, (2018).

[58] J. Ding, J. Dubedat, and E. Gwynne, *Introduction to the liouville quantum gravity metric*, arXiv preprint: 2109.01252, (2021).

[59] J. Ding and S. Goswami, *Upper bounds on Liouville First-Passage Percolation and Watabiki's prediction*, Communications on Pure and Applied Mathematics, 72 (2019), pp. 2331–2384.

[60] J. Ding and E. Gwynne, *The fractal dimension of Liouville quantum gravity: universality, monotonicity, and bounds*, Commun. Math. Phys., 374 (2019), pp. 1877–1934.

[61] N. Do, *The asymptotic weil-petersson form and intersection theory on M_{g,n}*, arXiv preprint: 1010.4126, (2010).

[62] J. L. Doob and J. Doob, *Classical potential theory and its probabilistic counterpart*, vol. 549, Springer, 1984.

[63] B. Duplantier, J. Miller, and S. Sheffield, *Liouville quantum gravity as a mating of trees*, Astérisque, (2021), pp. viii+258.

[64] A. Eichhorn, *An asymptotically safe guide to quantum gravity and matter*, Front. Astron. Space Sci., 5 (2019), p. 47.

[65] A. Eichhorn, S. Surya, and F. Versteegen, *Spectral dimension on spatial hypersurfaces in causal set quantum gravity*, Classical and Quantum Gravity, 36 (2019), p. 235013.

[66] B. Eynard, *Counting Surfaces*, vol. 70 of Progress in Mathematical Physics, Springer, 2016.

[67] B. Eynard and C. Kristjansen, *More on the exact solution of the O(n) model on a random lattice and an investigation of the case |n| > 2*, Nucl. Phys. B, 466 (1996), pp. 463–487.

[68] B. Eynard and N. Orantin, *Weil-petersson volume of moduli spaces, Mirzakhani's recursion and matrix models*, arXiv preprint: 0705.3600, (2007).

[69] V. Fateev, A. B. Zamolodchikov, and A. B. Zamolodchikov, *Boundary Liouville field theory I. Boundary state and boundary two–point function*, arXiv preprint: hep-th/0001012, (2000).

[70] W. Feller, *An introduction to probability theory and its applications. Vol. II.*, Second edition, John Wiley & Sons Inc., New York, 1971.

[71] F. Ferrari, *D-Brane Probes in the Matrix Model*, Nucl. Phys. B, 880 (2014), pp. 290–320.

[72] P. Flajolet and R. Sedgewick, *Analytic Combinatorics*, Cambridge University Press, 2009.

[73] S. Forste, H. Jockers, J. Kames-King, and A. Kanargias, *Deformations of JT gravity via topological gravity and applications*, JHEP, 11 (2021), p. 154.

[74] J.-F. L. Gall, *The topological structure of scaling limits of large planar maps*, Invent. Math., 169 (2007), pp. 621–670.

[75] J.-F. L. Gall and G. Miermont, *Scaling limits of random planar maps with large faces*, The Annals of Probability, 39 (2011), pp. 1 – 69.

[76] ——, *Scaling limits of random trees and planar maps*, Probability and statistical physics in two and more dimensions, 15 (2012), pp. 155–211.

[77] P. Gao, D. L. Jafferis, and D. K. Kolchmeyer, *An effective matrix model for dynamical end of the world branes in Jackiw-Teitelboim gravity*, JHEP, 01 (2022), p. 038.

[78] G. Gibbons and S. Hawking, *Euclidean Quantum Gravity*, World Scientific, 1993.

[79] E. Gwynne, N. Holden, and X. Sun, *Mating of trees for random planar maps and Liouville quantum gravity: a survey*, arXiv preprint: 1910.04713, (2019).

[80] ——, *Joint scaling limit of site percolation on random triangulations in the metric and peanosphere sense*, Electronic Journal of Probability, 26 (2021), pp. 1–58.

[81] E. Gwynne and J. Miller, *Existence and uniqueness of the Liouville quantum gravity metric for $\gamma \in (0, 2)$*, Invent. Math., 223 (2021), pp. 213–333.

[82] ——, *Random walk on random planar maps: spectral dimension, resistance and displacement*, Ann. Probab., 49 (2021), pp. 1097–1128.

[83] E. Gwynne, J. Miller, and S. Sheffield, *The Tutte embedding of the mated-CRT map converges to Liouville quantum gravity*, The Annals of Probability, 49 (2021), pp. 1677–1717.

[84] E. Gwynne and J. Pfeffer, *Bounds for distances and geodesic dimension in Liouville first passage percolation*, Electron. Commun. Probab., 24 (2019), pp. 1–12.

[85] ——, *KPZ formulas for the Liouville quantum gravity metric*, arXiv preprint: 1905.11790, (2019).

[86] D. Harlow et al., *TF1 Snowmass Report: Quantum gravity, string theory, and black holes*, arXiv preprint: 2210.01737, (2022).

[87] S. W. Hawking, *Particle Creation by Black Holes*, Commun. Math. Phys., 43 (1975), pp. 199–220. [Erratum: Commun.Math.Phys. 46, 206 (1976)].

[88] G. Hooft, *A planar diagram theory for strong interactions*, Nuclear Physics B, 72 (1974), pp. 461–473.

[89] P. Horava, *Spectral Dimension of the Universe in Quantum Gravity at a Lifshitz Point*, Phys. Rev. Lett., 102 (2009), p. 161301.

[90] C. Itzykson and J. B. Zuber, *Combinatorics of the modular group. 2. The Kontsevich integrals*, Int. J. Mod. Phys. A, 7 (1992), pp. 5661–5705.

[91] R. Jackiw, *Lower Dimensional Gravity*, Nucl. Phys. B, 252 (1985), pp. 343–356.

[92] S. Jain and S. D. Mathur, *World sheet geometry and baby universes in 2-D quantum gravity*, Phys. Lett. B, 286 (1992), pp. 239–246.

[93] C. V. Johnson, *Nonperturbative Jackiw-Teitelboim gravity*, Phys. Rev. D, 101 (2020), p. 106023.

[94] ——, *Consistency Conditions for Non-Perturbative Completions of JT Gravity*, arXiv preprint: 2112.00766, (2021).

[95] ——, *The Microstate Physics of JT Gravity and Supergravity*, arXiv preprint: 2201.11942, (2022).

[96] C. V. Johnson and F. Rosso, *Solving Puzzles in Deformed JT Gravity: Phase Transitions and Non-Perturbative Effects*, JHEP, 04 (2021), p. 030.

[97] R. Kaufmann, Y. Manin, and D. Zagier, *Higher Weil-Petersson volumes of moduli spaces of stable n-pointed curves*, Communications in Mathematical Physics, 181 (1996), pp. 763–787.

[98] R. Kenyon, J. Miller, S. Sheffield, and D. B. Wilson, *Bipolar orientations on planar maps and* $SLE_{12}$, Ann. Probab., 47 (2019), pp. 1240–1269.

[99]  A. KITAEV, *A simple model of quantum holography*.

[100] V. G. KNIZHNIK, A. M. POLYAKOV, AND A. B. ZAMOLODCHIKOV, *Fractal Structure of 2D Quantum Gravity*, Mod. Phys. Lett. A, 3 (1988), p. 819.

[101] M. KONTSEVICH, *Intersection theory on the moduli space of curves and the matrix Airy function*, Commun. Math. Phys., 147 (1992), pp. 1–23.

[102] G. KREWERAS, *Sur une classe de problèmes de dénombrement liés au treillis des partitions des entiers*, Cahiers du Bureau universitaire de recherche opérationnelle Série Recherche, 6 (1965), pp. 9–107.

[103] Y. LI, X. SUN, AND S. S. WATSON, *Schnyder woods, SLE$_{16}$, and Liouville quantum gravity*, arXiv preprint: 1705.05573, (2017).

[104] R. LOLL, *Discrete approaches to quantum gravity in four-dimensions*, Living Rev. Rel., 1 (1998), p. 13.

[105] ——, *Quantum gravity from causal dynamical triangulations: a review*, Classical and Quantum Gravity, 37 (2019), p. 013002.

[106] R. LOLL, *Quantum Curvature as Key to the Quantum Universe*, 6 2023.

[107] J. MALDACENA AND D. STANFORD, *Remarks on the Sachdev-Ye-Kitaev model*, Physical Review D, 94 (2016), p. 106002.

[108] J. MALDACENA, D. STANFORD, AND Z. YANG, *Conformal symmetry and its breaking in two dimensional Nearly Anti-de-Sitter space*, PTEP, 2016 (2016), p. 12C104.

[109] J. M. MALDACENA, *The Large N limit of superconformal field theories and supergravity*, Adv. Theor. Math. Phys., 2 (1998), pp. 231–252.

[110] Y. I. MANIN AND P. ZOGRAF, *Invertible cohomological field theories and weil-petersson volumes*, Annales de l'Institut Fourier, 50 (1999), pp. 519–535.

[111] J.-F. MARCKERT AND G. MIERMONT, *Invariance principles for random bipartite planar maps*, The Annals of Probability, 35 (2007), pp. 1642 – 1705.

[112] J.-F. MARCKERT AND A. MOKKADEM, *Limit of normalized quadrangulations: the Brownian map*, Ann. Probab., 34 (2006), pp. 2144–2202.

[113] H. Maxfield and G. J. Turiaci, *The path integral of 3D gravity near extremality; or, JT gravity with defects as a matrix integral*, JHEP, 01 (2021), p. 118.

[114] T. G. Mertens and G. J. Turiaci, *Solvable models of quantum black holes: a review on Jackiw–Teitelboim gravity*, Living Rev. Rel., 26 (2023), p. 4.

[115] G. Miermont, *On the sphericity of scaling limits of random planar quadrangulations*, Electronic Communications in Probability, 13 (2008), pp. 248–257.

[116] G. Miermont, *The Brownian map is the scaling limit of uniform random plane quadrangulations*, Acta Math., 210 (2013), pp. 319–401.

[117] J. Miller and S. Sheffield, *Imaginary geometry IV: interior rays, whole-plane reversibility, and space-filling trees*, Probab. Theory Related Fields, 169 (2017), pp. 729–869.

[118] ——, *Liouville quantum gravity spheres as matings of finite-diameter trees*, Ann. Inst. Henri Poincaré Probab. Stat., 55 (2019), pp. 1712–1750.

[119] ——, *Liouville quantum gravity and the Brownian map I: the* QLE(8/3, 0) *metric*, Invent. Math., 219 (2020), pp. 75–152.

[120] M. Mirzakhani, *Weil-petersson volumes and intersection theory on the moduli space of curves*, Journal of The American Mathematical Society, 20 (2007), pp. 1–23.

[121] R. C. Mullin, *On the enumeration of tree-rooted maps*, Canadian Journal of Mathematics, 19 (1967), pp. 174–183.

[122] K. Okuyama and K. Sakai, *JT gravity, KdV equations and macroscopic loop operators*, JHEP, 01 (2020), p. 156.

[123] ——, *FZZT branes in JT gravity and topological gravity*, JHEP, 09 (2021), p. 191.

[124] ——, *Page curve from dynamical branes in JT gravity*, JHEP, 02 (2022), p. 087.

[125] D. ORITI, *The Group field theory approach to quantum gravity*, Approaches to Quantum Gravity: Toward a New Understanding of Space, Time and Matter, (2006), pp. 310–331.

[126] G. PENINGTON, S. H. SHENKER, D. STANFORD, AND Z. YANG, *Replica wormholes and the black hole interior*, JHEP, 03 (2022), p. 205.

[127] T. REGGE, *GENERAL RELATIVITY WITHOUT COORDINATES*, Nuovo Cim., 19 (1961), pp. 558–571.

[128] M. REUTER, *Nonperturbative evolution equation for quantum gravity*, Physical Review D, 57 (1998), p. 971.

[129] M. REUTER AND F. SAUERESSIG, *Fractal space-times under the microscope: A Renormalization Group view on Monte Carlo data*, JHEP, 12 (2011), p. 012.

[130] ——, *Quantum Gravity and the Functional Renormalization Group: The Road towards Asymptotic Safety*, Cambridge University Press, 1 2019.

[131] D. REVUZ AND M. YOR, *Continuous Martingales and Brownian Motion*, Grundlehren der mathematischen Wissenschaften, Springer Berlin Heidelberg, 2004.

[132] P. SAAD, S. H. SHENKER, AND D. STANFORD, *JT gravity as a matrix integral*, arXiv preprint: 1903.11115, (2019).

[133] S. SACHDEV AND J. YE, *Gapless spin-fluid ground state in a random quantum Heisenberg magnet.*, Physical Review Letters, 70 21 (1992), pp. 3339–3342.

[134] F. SAUERESSIG, *The Functional Renormalization Group in Quantum Gravity*, arXiv preprint: 2302.14152, (2023).

[135] G. SCHAEFFER, *Conjugaison d'arbres et cartes combinatoires aléatoires*, PhD thesis, Bordeaux 1, 1998.

[136] S. SHEFFIELD, *Quantum gravity and inventory accumulation*, Ann. Probab., 44 (2016), pp. 3804–3848.

[137] D. STANFORD AND E. WITTEN, *JT gravity and the ensembles of random matrix theory*, Adv. Theor. Math. Phys., 24 (2020), pp. 1475–1680.

[138] S. Surya, *The causal set approach to quantum gravity*, Living Rev. Rel., 22 (2019), p. 5.

[139] C. Teitelboim, *Gravitation and Hamiltonian Structure in Two Space-Time Dimensions*, Phys. Lett. B, 126 (1983), pp. 41–45.

[140] G. J. Turiaci, M. Usatyuk, and W. W. Weng, *2D dilaton-gravity, deformations of the minimal string, and matrix models*, Class. Quant. Grav., 38 (2021), p. 204001.

[141] W. T. Tutte, *On the enumeration of planar maps*, Bulletin of the American Mathematical Society, 74 (1968), pp. 64–74.

[142] H. Walden and R. Kellogg, *Numerical determination of the fundamental eigenvalue for the Laplace operator on a spherical domain*, Journal of Engineering Mathematics, 11 (1977), pp. 299–318.

[143] Y. Watabiki, *Analytic study of fractal structure of quantized surface in two-dimensional quantum gravity*, Prog. Theor. Phys. Suppl., 114 (1993), pp. 1–17.

[144] S. Weinberg, *Ultraviolet divergences in quantum theories of gravitation.*, in General Relativity: An Einstein centenary survey, S. W. Hawking and W. Israel, eds., Jan. 1979, pp. 790–831.

[145] W. Werner and E. Powell, *Lecture notes on the gaussian free field*, arXiv preprint: 2004.04720, (2020).

[146] C. Wetterich, *Exact evolution equation for the effective potential*, Phys. Lett., B301 (1993), pp. 90–94.

[147] E. Witten, *Two-dimensional gravity and intersection theory on moduli space*, Surveys Diff. Geom., 1 (1991), pp. 243–310.

[148] ——, *Anti-de Sitter space and holography*, Adv. Theor. Math. Phys., 2 (1998), pp. 253–291.

[149] ——, *Deformations of JT Gravity and Phase Transitions*, arXiv preprint: 2006.03494, (2020).

[150] ——, *Matrix Models and Deformations of JT Gravity*, Proc. Roy. Soc. Lond. A, 476 (2020), p. 20200582.

[151] ——, *A Note On Complex Spacetime Metrics*, arXiv preprint: 2111.06514, (2021).

[152] ——, *Volumes And Random Matrices*, Quart. J. Math. Oxford Ser., 72 (2021), pp. 701–716.

[153] S. WOLPERT, *On the weil-petersson geometry of the moduli space of curves*, American Journal of Mathematics, 107 (1985), pp. 969–997.